\documentclass[12pt]{article}
\usepackage[dvips]{graphicx}
\usepackage{amssymb}
\usepackage{amsmath}

\def\beq{\begin{equation}}\def\eeq{\end{equation}}
\def\bea{\begin{eqnarray}}\def\eea{\end{eqnarray}}

\begin{document}
 
\title{Non-linear corrections to Lagrangians predicted by causal set theory: Flat space bosonic toy model}
 
\author{Roman Sverdlov
Institute of Mathematical Sciences ,
\\IV Cross Road, CIT Campus, Taramani, Chennai, 600 113, Tamil Nadu, India} 
\date{January 27, 2012}
\maketitle
 
\begin{abstract}
\noindent  A while ago a proposal have been made regarding Klein Gordon and Maxwell Lagrangians for causal set theory. These Lagrangian densities are based on the statistical analysis of the behavior of field on a sample of points taken throughout some "small" region of spacetime. However, in order for that sample to be statistically reliable, a lower bound on the size of that region needs to be imposed. This results in "unwanted contributions" from higher order derivatives to the Lagrangian density, as well as non-trivial curvature effects on the latter. It turns out that both gravitational and non-gravitational effects end up being highly non-linear. In the previous papers we were focused on leading order terms, which allowed us to neglect these nonlinearities. We would now like to go to the next order and investigate them. In the current paper we will exclusively focus on the effects of higher order derivatives in the flat-space toy model. The gravitational effects will be studied in another paper which is currently in preparation. Both papers are restricted to bosonic fields, although the issue probably generalizes to fermions once Grassmann numbers are dealt with in appropriate manner. 
\end{abstract}

\subsection*{1. Introduction}

A causal set, originally proposed by Rafael Sorkin, is a model of spacetime that replaces coordinate system with lightcone causal relations. After all, any geometrical information that we "know" about the spacetime we have actually "learned" by analyzing the signals that our eyes receive. These signals, in term, are constrained to propagate only between "causally related" pairs of points. In light of lack of circular causality, it is clear that causal relations form a partial ordering $\prec$, and a signal can propagate from point $a$ to point $b$ if and only if $a \prec b$.  Thus, the information that we learn first hand is precisely the specific structure of that partial ordering. Later on, we "analyze" this structure and "infer" that there is some coordinate system that "created" it. The contention of causal set theory is to "unlearn" any information we might have inferred (including the coordinate system) and view the partial ordering as the one and only geometry available. Accordingly, some models have been proposed that attempt to describe Lagrangians (\cite{SverdlovBombelli} and \cite{BombelliSverdlov}) and propagators (\cite{Johnston1} and \cite{Johnston2}) in a coordinate-independent fashion. 

In principle, there are discrete structures for which such procedure is straightforward. For example, if we assume cubic lattice, the "nearest causal neighbor" will identify a diagonal of any given "cube". It is relatively easy to count such "diagonals" in order to reconstruct the information we had from coordinate system. At the same time, however, cubic structure would lead to "preferred directions" (such as edges of the cubes). In order to make the structure more covariant-looking, we need to replace cubic structure with Poisson distribution of point. However, in case of Poisson distribution, we expect a lot of random variation of distances between neighboring points which makes them no longer reliable. In order to "recover" continuum quantities we are forced to look at "statistically large" sample of points, so that by the law of large numbers the random fluctuations cancel out. That sample of points, of course, is assumed to lie within a "very small" region of spacetime, so that the linear approximation still holds to high accuracy, thus allowing us to compute Lagrangian densities. At the same time, the size of this region is assumed to be several magnitudes larger than the discretization scale. 

The fact that the above-described region is finite leads to some unwanted effects. In particular, in order to compute Lagrangian density, which is defined in terms of derivatives, we need some kind of statistical analysis of the behavior of relevant fields over the "sample of points". The statistical analysis has to be designed in such a way that the main contribution comes from the first derivatives of the fields involved. However, the input of the analysis consists of the values of the fields at various sample points. Due to the "finite" size of the region, the latter includes higher derivatives. If we were to take direct neighbors (as we would have done in the case of cubic lattice) we would be able to dismiss higher order derivatives by simply saying that they are not well defined since there is nothing "between" the neighbors. In causal set context, however, we are taking a "large sample of points" which means that we are no longer able to dismiss higher order derivatives in the above way. 

In case of curved spacetime, we will also have curvature effects in addition to the above. Again, if we use some regular lattice to model the curved spacetime, we can argue that curvature is not defined ''between'' the neighboring points and, therefore, is of no consequence. In case of causal set, however, the lack of regular structure forces us to take statistically large sample of points. This ultimately implies that the curvature \emph{will} in fact have some finite effect within that sample. At the same time, we are using that \emph{finite} sample as a discretized ''infinitesimal'' region where we are supposed to ''take derivatives''. Thus, curvature is ''not supposed'' to have any effect on that region; yet it will. Intuitively, this means that the very ''tangent plane'' is now ''curved''. This, of course, is logically separate from the effects of curvature we would expect in ''usual'' cases. Therefore, the continuum limit of resulting effects might be very different as well,  and one needs to perform explicit causal set calculations to find out what these effects might be. 

In the previous papers, we were focused on leading order terms which allowed us to neglect both the contributions of higher order derivatives as well as contributions of curvature. We would now like to consider higher order terms and analyze these contributions. This attempt is partly motivated by the results of recent OPERA experiment regarding neutrino speed. It is conceivable that the non-linear interactions of a ''very weak'' field (neutrino) with ''much larger'' background field (earth's gravity as well as its magnetic field, the atmosphere, and so forth) might result in the latter setting a ''preferred frame'' for the propagation of the former. In that ''preferred frame'' the ''weaker fields'' can, conceivably, be either subluminal or superluminal. 

However, no attempt to compute the deviation of speed of propagation was made in this paper. Furthermore, the very existence of such ''speed'' is highly questionable. In fact, non-linearities imply that we can no longer appeal to the concept of eigenfunction. If we "insist" in doing Fourier decomposition of our solutions, we will likely conclude that any given Fourier component "gives birth" to a number of other components moving with different, and seemingly unrelated, velocities. One can attempt to argue that the rate of this process is very small and, therefore, negligible; at the same time, one can also credit the ''very large'' fields of the earth as an exclusive reason that the non-linear interaction with the latter can not be similarly neglected. Again, however, in order to make such claims one needs to do a large amount of research that has not been done in this paper. 

The exclusive goal of this paper is simply to explore the non-linear effects that the causal set theory will predict, independently of OPERA experiment or any other ''outside'' motivations we might have. One reason we are independently interested in non-linearities is that it is perhaps the only tangible way of using causal set theory for the purposes of making new predictions (whatever these ''new predictions'' might ''happen'' to be). As we have stated earlier, causal set theory does not assume any regular structure. This makes it very difficult to make analytic calculations without going back to the continuum limit. Thus, in order for causal set theory to make new predictions, one has to think of some deviations from ''traditional'' theories that persist in the continuum as well. Non-linear effects we just described meet this criteria. On a down side, however, such effects might imply that the resulting theories are no longer renormalizable. In principle, we can still attempt to compute non-renormalizable theories since the discreteness scale sets up a ''preferred'' value of ultraviolet cutoff. Again, however, the fruitfulness of such calculations is questionable. 

In this paper we will restrict ourselves to finding the non-linear effects on Lagrangian density, without actually ''using'' that Lagrangian density for any calculations. We do, however, plan to return to the above issues in future work. Apart from this, the paper at hand is restricted to the flat space toy model. In other words, we are exclusively focused on the effects of higher order derivatives of Klein Gordon and Maxwell fields in flat space context. The effects of curvature will be studied in \cite{Gravity}, and the generalization to fermions will be studied in \cite{Fermions}, which is likewise in preparation. Again, however, both of these papers are, likewise, restricted exclusively to Lagrangian densities and they make no attempt to compute any Feynmann diagrams or tackle any specific problem (such as OPERA experiment). But hopefully these papers might give some direction to either the author, or other readers of this paper, to work in future. 

\subsection*{2. Geometry and fields on a causal set: a brief review}

We consider a partially ordered set $(S, \prec_S)$, where $S$ is a discrete set. The main reason we write $\prec_S$ rather than simply $\prec$ is to distinguish it from $\prec_{\cal M}$ which is to be defined on a smooth manifold $\cal M$.The relation $\prec_S$ satisfies the axioms of partial ordering: if $a \prec_S b$ and $b \prec_S c$ then $a \prec_S c$, and there is \emph{no} point $a$ satisfying $a \prec a$. The relation $\prec$ is called \emph{causal relation}. Physically, $a \prec b$ holds if and only if we can travel from point $a$ to point $b$ without going faster than the speed of light. The transitivity of causal relation follows from the fact that we can travel from $a$ to $c$ by "first" traveling from $a$ to $b$ and "then" traveling from $b$ to $c$. The fact that none of the points satisfy $a \prec a$ is equivalent to the statement that there are no closed causal loops. If, for example, we could travel from $a$ to $b$ and then come back to $a$, this would mean that $a \prec b$ and $b \prec a$ both hold; by transitivity, this would imply $a \prec a$, which is "forbidden". 

Two points $a$ and $b$ are said to be \emph{direct neighbors} if they are causally related and there is no other point that is causally "between them". If $a \prec_S b$ holds, then $a \prec_S^* b$ holds if and only if $a$ and $b$ are direct neighbors: 
\beq a \prec_S^* b \Longleftrightarrow ((a \prec_S b) \wedge \not\exists c (a \prec_S c \prec_S b)) \eeq
A fundamental structure of our theory is Alexandrov set, $\alpha (p, q)$, which consists of all points $r$ satisfying $p \preceq_S r \preceq_S q$:
\beq \alpha_S (p, q) = \{r \vert p \preceq_S r \preceq_S q \} \eeq
where $\preceq_S$ is defined by 
\beq a \preceq_S b \Longleftrightarrow [(a \prec_S b) \vee (a=b)] \eeq
 Geometrically, this looks like a compact region of spacetime bounded by two light cones: "future" light cone of $p$ and "past" light cone of $q$. This set defines a "preferred frame"; namely, the "local" $t$-axis coincides with the geodesic passing from $p$ to $q$. It is obvious that $p$ and $q$ are "direct neighbors" if and only if they happen to be the only elements of the Alexandrov set that they form:  
\beq a \prec_S^* b \Longleftrightarrow [(a \prec_S b) \wedge (\alpha_S (a, b) = \{a, b \})] \eeq
In general, however, we would be interested in Alexandrov sets that are non-empty and, in fact, contain large enough number of points to be statistically relevant. 

We observe that in case of Minkowski space the distance between two timelike separated points is equal to the length of the "longest path" connecting them (which, in flat case, happens to be straight line). The fact that that path is the "longest" rather than the "shortest" is due to the minus signs in Minkowskian metric. Therefore, we will \emph{define} a discretized distance on a causal set to be the length of the "longest" possible chain of points $p \prec r_1 \prec \cdots \prec r_{n-1} \prec q$, where the "length" of the above chain is identified with $n$:
\beq \tau_S (p, q) = \xi \max \{n  \vert \exists r_1, \cdots, r_{n-1} \in S \colon p \prec_S r_1 \prec \cdots \prec_S r_{n-1} \prec q \}, \label{taus} \eeq
It is commonly assumed that $\xi$ coincides with Plank scale; but in the opinion of the author such doesn't have to be the case. It should be noticed that the two points are "direct neighbors" if and only if the distance between them is $\xi$:
\beq a \prec_S^* b \Longleftrightarrow [(a \prec_S b) \wedge (\tau_S (a, b) = \xi)] \eeq
A sequence of points is said to be a \emph{geodesic} if the "longest" path connecting any of its points happen to be the segment of that sequence itself:
\beq \{\cdots, a_{-n-1}, a_{-n}, \cdots, a_m, a_{m+1}, \cdots \} \; {\rm is \; geodesic} \; \Longleftrightarrow \nonumber \eeq
\beq \Longleftrightarrow \forall k< l \not\exists b_1, \cdots, b_{l-k} (a_k \prec_S b_1 \prec_S \cdots \prec_S b_{l-k} \prec a_l) \label{geodesics} \eeq
It should be noticed that the path $a_k \prec_S b_1 \prec_S \cdots \prec_S b_{l-k-1} \prec_S a_l$ is "allowed" and, at the same time, it has "the same" length as the "segment" in question. The only thing we claim is that there is no path "longer" than the latter. This choice is crucial since we don't want the presence of two same-length paths to prevent us from drawing a "geodesic". At the same time, in order to preserve existence, we sacrifice uniqueness (although we could restore uniqueness by "taking a union" of all possible geodesics). 

Since we are planning to routinely compare the causal set $S$ to a manifold $\cal M$, it is important to formally define the corresponding notions in $\cal M$. The \emph{timelike curve} on $\cal M$ is defined in a coordinate based way:
\beq \gamma \colon \mathbb{R} \rightarrow {\cal M} \; {\rm is \; timelike \; curve} \; \Longleftrightarrow \; \forall t \Big( g_{\mu \nu} (\gamma (t)) \frac{d \gamma^{\mu}}{dt} \frac{d \gamma^{\nu}}{dt} \geq 0 \Big) \eeq
The causal relation on a manifold $\cal M$ is $\prec_{\cal M}$ (while causal relation on a causal set $S$ is $\prec_S$). The relation $x^{\mu} \prec y^{\mu}$ holds  if and only if $x^{\mu}$ and $y^{\mu}$ are connected by at least one timelike curve:
\beq x \prec_{\cal M} y \; \Longleftrightarrow \; \exists \gamma \colon [0, 1] \rightarrow {\cal M}  \Big(\gamma^{\mu} (0) = x^{\mu} \; , \; \gamma^{\mu} (1) = y^{\mu} \; ; \; g_{\mu \nu} (\gamma (\tau)) \frac{d \gamma^{\mu}}{d \tau} \frac{d \gamma^{\nu}}{d \tau} = 1, \Big) \label{precM} \eeq
If $\gamma \colon \mathbb{R} \rightarrow \cal M$ is \emph{timelike}, then its \emph{length} is given by 
\beq \tau_{\cal M} (\gamma; \tau_1, \tau_2) = \int_{\tau_1}^{\tau_2} \sqrt{g_{\mu \nu} (\tau) \frac{d \gamma^{\mu}}{d \tau} \frac{d \gamma^{\nu}}{d \tau}} d \tau \eeq
The \emph{distance} between $x^{\mu} \in \cal M$ and $y^{\mu} \in \cal M$ is the length of the longest possible timelike, future-directed, curve that connects the two points:
\beq \tau (x^{\mu}, y^{\mu}) = \sup \Big\{ \tau(\gamma; \tau_1, \tau_2) \Big\vert \gamma (\tau_1) = x^{\mu}  ;  \gamma (\tau_2) = y^{\mu} ;  \frac{d \gamma^{\mu}}{d \tau} > 0 ; g_{\mu \nu} (\tau) \frac{d \gamma^{\mu} }{d \tau} \frac{d \gamma^{\nu} }{d \tau} > 0 \Big\} \eeq
Finally, we will define an Alexandrov set on $\cal M$ to be 
\beq \alpha_{\cal M} (x^{\mu}, y^{\mu}) = \{z^{\mu} \in {\cal M} \vert x^{\mu} \prec_{\cal M} z^{\mu} \prec_{\cal M} y^{\mu} \} \eeq

If $\cal M$ is a Lorentzian manifold, then the function $f \colon S \rightarrow \cal M$ is said to be an \emph{embeding} of $S$ into $\cal M$ if it respects causal structure. In other words, 
\beq p \prec_S q \Rightarrow f(p) \prec_{\cal M} f(q) \label{respectcausal}\eeq
The set $S$ is said to \emph{approximate} $\cal M$ (which we write as $S \approx \cal M$ if $f(S)$ "looks like" a Poisson scattering on $\cal M$ with density so large that any "small" region of $\cal M$ has "very large" number of points. Such set can be produced by "starting" from $\cal M$, performing the Poisson process on $\cal M$ to obtain a set $T \subset \cal M$, and finally identifying the set $T$ with $f(S)$ for some other "abstract" set $S$. Then the causal relation $\prec_S$ can be "read off" from $\prec_{\cal M}$ together with Equation \ref{respectcausal}. Such sets have been widely studies and it has been confirmed that distances, indeed, approximate what we would expect them to be once the number of points is statistically large:
\beq S \approx {\cal M} \Longrightarrow \tau_S (p, q) \approx k_d \tau_{\cal M} (f(p), f(q)) \eeq
where $d$ is the dimensionality of $\cal M$, and $k_d$ is a coefficient appropriate for that specific dimensionality. If we stick to one specific dimensionality it is possible to set $k_d$ to $1$ by appropriately scaling $\xi$. 

However, if one insists on viewing $\prec_S$ as fundamental rather than $\prec_{\cal M}$, then one is "not allowed" to "start off" from $\cal M$. Instead, one is hard pressed to formulate $S$-based "physics laws" that would "force" $S$ to approximate $\cal M$ on a sufficiently large scale. This is one of the big problems of causal set theory that is still unresolved. In principle, one can try to do a "short cut" by saying that the "physics law" is simply a constraint that "forbid" any $S$ that would not "approximate" at least one sufficiently smooth manifold. But then the question is how can one rigorously formulate such a constraint? Since $S$ is fundamental rather than $\cal M$, that would force us to find an intrinsic property of $S$ through which we can judge whether or not $S$ is "manifold-like". In principle, there should be such property: after all, we have "learned" that our universe is manifoldlike by "analyzing" the set of rays that hit our eyes (in other words, causal relations). At the same time, at least so far, the attempts to specify such property had not been successful. Furthermore, even if one does it successfully, it would be "pointless" if the constructions involved are too complicated. After all, they would be even "less" natural than the coordinate system we were trying to get rid of.  

For the purposes of this paper, we will not attempt to address the questions in the above paragraph. Instead, we will simply assume that $S \approx \cal M$, without specifying the reason. Our only purpose is to work out the Lagrangian densities on already-given causal set, while leaving the question of its origin aside. 

Let us now proceed to define sample fields on a causal set as well as their Lagrangians. For the purposes of this paper, we will limit ourselves to Klein Gordon and electromagnetic fields, which we will define as $\phi_S \colon S \rightarrow \mathbb{C}$ and $a_S \colon S \times S \rightarrow \mathbb{R}$, respectively. It is further assumed that they correspond to well behaved differential functions $\phi_{\cal M} \colon {\cal M} \rightarrow \mathbb{C}$ and $a_{\cal M} \colon {\cal M} \times {\cal M} \rightarrow \mathbb{R}$ on $\cal M$:
\beq \phi_{\cal M} (f(p)) = \phi_S (p) \; ; \; a_{\cal M} (f(p), f(q)) = a_S (p, q) \eeq
The "scalar field" on a manifold $\cal M$ is identified with $\phi_{\cal M}$ itself, while the electromagnetic field is identified with $A \colon {\cal M} \rightarrow T \cal M$ (where $T \cal M$ is a "tangent bundle" of $\cal M$) and it is assumed that
\beq a_{\cal M} (x, y) = \exp \Big( i \; \int_{\gamma (x, y)} g_{\mu \nu} A^{\mu} (z) dz^{\mu}  \Big) \eeq
where $\gamma (x, y)$ is a geodesic connecting $x$ and $y$. The geodesic $\gamma$ is defined in terms of ordinary manifold-based way,
\beq \frac{d \gamma^{\rho}}{d \tau} + \Gamma^{\rho}_{\mu \nu} \frac{dx^{\mu}}{d \tau} \frac{dx^{\nu}}{d \tau} \label{geodesicssmooth}\eeq
\emph{as opposed to} Equation \ref{geodesics}. After all, Equation \ref{geodesics} is a definition of geodesics on a \emph{discrete} set $S$, whereas Equation \ref{geodesicssmooth} continues to be the definition in continuum case of $\cal M$. As we said previously, this paper is focused on flat space, while curvature effects are postponed to \cite{Gravity}. The reason we are including Christoffel's symbols is simply because it might benefit the reader who will be interested to read other papers, and it doesn't require a lot of extra labor on our part. For the purposes of this paper we can assume that $\Gamma^{\gamma}_{\alpha \beta} =0$.

In the above expression, we have written $A^{\mu}$ instead of $A_{\cal M}^{\mu}$ because $A^{\mu}$ does not have $S$-counterpart. On a manifold $\cal M$, we define a \emph{Lagrangian density} ${\cal L}_{\cal M} \colon \{ \phi_{\cal M} \} \times \{A^{\mu} \} \times {\cal M} \rightarrow \mathbb{R}$ in a usual way, 
\beq {\cal L} (\phi, A^{\mu} ; x^{\mu}) = F^{\mu \nu} F_{\mu \nu} + {\cal D}^{\mu} \phi^* {\cal D}_{\mu} \phi \eeq
where 
\beq F^{\mu \nu} = \partial^{\mu} A^{\nu} - \partial^{\nu} A^{\mu} \; ; \; {\cal D}^{\mu} \phi = \partial^{\mu} \phi + ieA^{\mu} \phi \eeq
Our goal is to define a corresponding Lagrangian density ${\cal L}_S \colon \{\phi_S \} \times \{ a_S \} \times S \rightarrow \mathbb{R}$ such that 
\beq {\cal L}_S (\phi_S, a_S ; p) \approx {\cal L}_{\cal M} (\phi_{\cal M}, A^{\mu} ; f(p)) \label{projectionlagrangian} \eeq
We are approaching that goal by focusing our attention on $f(S)$, but, at the same time, attempt to re-express ${\cal L}_{\cal M}$ in such a way that it does not appeal to anything containing Lorentzian indexes or anything else that lacks $S$-counterpart. This will allow us to "rewrite" our final answer in terms of $S$ rather than $f(S)$. 

Let us now go ahead and define ${\cal L}_S$, in a way that meets the criteria of Equation \ref{projectionlagrangian}. We will define a small scale $\tau$. This scale should be "small enough" in order for the well behaved fields to be approximately linear ($\tau \ll 1$) but, at the same time, it should be "large enough" to contain statistically reliable sample of points ($\tau \gg \xi$); thus, 
\beq \xi \ll \tau \ll 1 \eeq
The criteria of "successful" definition of ${\cal L}_S$ is that it \emph{approximates} ${\cal L}_{\cal M}$ on "our" scale; that is, 
\beq {\cal L}_S = {\cal L}_{\cal M}(1 + 0 (\tau)) \eeq
Producing such ${\cal L}_S$ is the goal of the current chapter. Then, in the next chapter, we will "take" ${\cal L}_S$ produced in this chapter \emph{without changing it at all}, and evaluate higher order terms it would produce. We will find that 
\beq {\cal L}_S = {\cal L}_{\cal M} + \delta {\cal L}_{\cal M} + 0 (\tau^3) \eeq
The "small variation" $\delta {\cal L}_{\cal M}$ is the ultimate goal of this paper. As explained in the introduction, this goal is motivated by a possibility that $\delta {\cal L}_{\cal M}$ might be responsible (or at least contribute to) the deviation of the speed of neutrino from the speed of light (although, as far as this paper is concerned, we are not considering the neutrino field). 

\subsection*{3. First order Lagrangian density for scalar fields}

One proposal of ${\cal L}_S$ has been made in \cite{SverdlovBombelli}. However, at the time of writing of that paper, we were not interested in finding $\delta {\cal L}_{\cal M}$. As a result, we made some "sloppy" moves that would make $\delta {\cal L}_{\cal M}$ much larger than it should be. The main example of this sort of thing is that we assumed that the "neighborhood" of a given point lies "to the future" of that point, instead of assuming that the point is at the center of the neighborhood. In the former case, we obtain 
\beq {\cal L}_S = {\cal L}_{\cal M} + 0 (\tau) = {\cal L}_{\cal M} + \delta {\cal L}_{\cal M} + 0 (\tau^2) \; ; \; \delta {\cal L}_{\cal M} = 0 (\tau) \eeq
while at the latter case we obtain
\beq {\cal L}_S = {\cal L}_{\cal M} + 0 (\tau^2) = {\cal L}_{\cal M} + \delta {\cal L}_{\cal M} + 0 (\tau^3) \; ; \; \delta {\cal L}_{\cal M} = 0 (\tau^2) \eeq
While the $\delta {\cal L}_{\cal M} = 0 (\tau)$ would still be "formally okay", it is clear that the option that produces $\delta {\cal L}_{\cal M} = 0 (\tau^2)$ is by far more reasonable. Thus, in this section we will redo \cite{SverdlovBombelli} while moving the point in question to the center of the neighborhood, and then in the next section we will proceed to finding the $0 (\tau^2)$ correction. 

According to our model, the field has \emph{two} Lagrangians rather than one. In particular, the Lagrangians are ${\cal L}_t$ and ${\cal L}_s$ (where letters "t" and "s" stand for "timelike" and "spacelike") defined as 
\beq {\cal L}_s (\phi; r) = \frac{1}{2}  (\partial^{\mu} \phi \partial_{\mu} \phi) (-1 + sgn (\partial^{\mu} \phi \partial_{\mu} \phi)) \label{WhySpacelike}\eeq
\beq {\cal L}_t (\phi; r) = \frac{1}{2}  (\partial^{\mu} \phi \partial_{\mu} \phi) (1 + sgn (\partial^{\mu} \phi \partial_{\mu} \phi)) \label{WhyTimelike}\eeq
where "sgn" is a "signum" function defined as 
\beq sgn (x) = \left\{
	\begin{array}{ll}
		-1 & {\rm If} \; x < 0 \\
                         0 & {\rm if} \; x=0 \\
		+1 &  {\rm if} \; x>0
	\end{array}
\right.
\label{SignumScalar} \eeq
and we are using $(+,-,-,-)$ metric convention. Thus, the "spacelike" Lagrangian returns zero if the gradient of $\phi$ is timelike at a given point and it returns the Lagrangian density "with the wrong sign" if the gradient of $\phi$ is spacelike. On the other hand, timelike Lagrangian returns zero if gradient is spacelike, and it returns Lagrangian density with the correct sign if the gradient is timelike. After we have obtained ${\cal L}_s$ and ${\cal L}_t$, we will then define $\cal L$ by superimposing the two:
\beq {\cal L} = {\cal L}_t - {\cal L}_s \eeq
where the minus sign is meant to "correct" the "sign error" inside of ${\cal L}_s$.  It is easy to see that substitution of ${\cal L}_t$ and ${\cal L}_s$ into the above equation results in
\beq {\cal L} = \partial^{\mu} \phi \partial_{\mu} \phi \eeq
as expected.  Now, each ${\cal L}_s$ and ${\cal L}_t$ need to be defined for arbitrary causal set.Since a general causal set is not manifold-like, we are not allowed to refer to coordinate system the way we do when we write partial derivatives. At the same time, however, we will have to show that \emph{in the special case} of manifold-like causal set the "general" definition reduces to coordinate-based one. Therefore, we will propose the following construction. We will define a \emph{Lagrangian generator} to be 
\beq {\cal K} (\phi; s_1, s_2) = (\phi (s_2) - \phi (s_1))^2 \label{AbstractGenerator}\eeq
We will then define a \emph{Lagrangian generators} ${\cal J}_s$ and ${\cal J}_t$ to be 
\beq {\cal J}_s (\phi; p, q) = \max \{ (\phi (s_1) - \phi (s_2))^2 \vert p \prec^* s_1 \prec^* q \; , \; p \prec^* s_2 \prec^* q \} \label{AbstractPreLagrangianSpacelike}\eeq 
\beq {\cal J}_t (\phi; p, q) = \max \{ (\phi (s_1) - \phi (s_2))^2 \vert s_1 = p \; , \; s_2 = q \}  \eeq 
Thus, of course, ${\cal J}_t$ simplifies to 
\beq {\cal J}_t (\phi; p, q) = (\phi (q) - \phi (p))^2 \label{AbstractPreLagrangianTimelike} \eeq
while ${\cal J}_s (\phi; p, q)$ would require a little bit more work. Finally, we will define \emph{Lagrangian densities} at $r$ as 
\beq {\cal L}_s (\phi; r) = \min \{ {\cal J}_s (p, q) \vert p \prec r \prec q \; , \; \tau (p, r) = \tau (r, q) = \tau \} \label{AbstractLagrangianSpacelike}\eeq
\beq {\cal L}_t (\phi; r) = \min \{ {\cal J}_t (p, q) \vert p \prec r \prec q \; , \; \tau (p, r) = \tau (r, q) = \tau \} \label{AbstractLagrangianTimelike}\eeq
Then \emph{after} the Lagrangians have been defined by the above method they are manually subtracted:
\beq {\cal L} = {\cal L}_t - {\cal L}_s \label{AbstractLagrangian}\eeq
Strictly speaking, the definition of Lagrangian is given by Equations \ref{AbstractGenerator}, \ref{AbstractPreLagrangianTimelike}, \ref{AbstractPreLagrangianSpacelike}, \ref{AbstractLagrangianTimelike}, \ref{AbstractLagrangianSpacelike} and \ref{AbstractLagrangian}. None of these equations have any reference to coordinate system; thus, the Lagrangian is defined for abstract causal set. 

Let us now show that \emph{in a special case} where the causal set \emph{happens} to be manifoldlike (defined in terms of embedding $f \colon S \rightarrow \cal M$, we would, in fact, obtain Equations \ref{WhySpacelike} and \ref{WhyTimelike}. Geometrically, the fact that $p \prec^* s_1$ implies that the Lorentzian distance between $f(p)$ and $f(s_1)$ is $0$. In other words, $f(s_1)$ lies in the light cone of $f(p)$. The fact that we have $p \prec^* s_1$ as opposed to $s_1 \prec^* p$ means that $f(s_1)$ lies in the \emph{future} light cone of $f(p)$. Similarly, the fact that $s_1 \prec^* q$ implise that $s_1$ lies in the \emph{past} light cone of $q$. Geometrically, it is easy to see that these two conditions imply that $s_1$ lies on the surface of the "equator" of the Alexandrov set $\alpha (p, q)$. In other words, if we define our coordinate system in such a way that 
\beq f(p) = (- \tau, 0, 0, 0) \; , \; f(q)=(\tau,0,0,0) \eeq
then the $t$-coordinate of $s_1$ is zero, 
\beq s_1 = (0, x_1, y_1, z_1) \eeq
and its space coordinates satisfy
\beq x_1^2 + y_1^2 + z_1^2 = \tau^2 \eeq
The same, of course, is true for $s_2$:
\beq s_2 = (0, x_2, y_2, z_2) \; , \; x_2^2 +y_2^2 + z_2^2 = \tau^2 \eeq
Let us now consider the situation in which the gradient of $\phi$ is spacelike. We can select a coordinate system in which the spacelike part of gradient of $\phi$ points in $z$-direction. Thus, 
\beq \partial_x \phi = \partial_y \phi =0 \eeq
Now, up to linear order, the Lagrangian generator is 
\beq {\cal K}_S (\phi; p, q, s_1, s_2) = ((s_2^{\mu} - s_1^{\mu}) \partial_{\mu} \phi)^2 \eeq
which, in our coordinate system, becomes 
\beq {\cal K}_S (\phi; p, q, s_1, s_2) = ((z_2- z_1) \partial_z \phi)^2 \eeq
Therefore, it is being maximized by the choice of points $s_1$ and $s_2$ that lie on $z$-axis. The maximization of "Lagrangian generator" $\cal K$ is identified with "pre-Lagrangian" ${\cal J}$:
\beq \partial_x \phi = \partial_y \phi = 0 \Longrightarrow {\cal J} = 4 \tau^2 \partial_z \phi^2 \eeq
This generalizes to rotationally covariant (but \emph{not} Lorentz covariant) expression 
\beq {\cal J} = 4 \tau^2 \vert \vec{\nabla} \phi \vert^2 \eeq
Now, we would like to "minimize" $\cal J$. Since the gradient of $\phi_{\cal M}$ is spacelike, there is a frame in which its time component is zero. It is easy to see that if we will \emph{first} start from that frame and then perform Lorentz boost \emph{away} from that frame, then $\vert \vec{\nabla} \phi \vert$ will increase. This means that the frame in which the time component is zero is the one where minimization of $\cal J$ occurs. The minimum of $\cal J$ is identified with $\cal L$. Thus,
\beq \partial_0 \phi = 0 \Longrightarrow {\cal L}_s = 4 \tau^2 \vert \vec{\nabla} \phi \vert^2 \eeq
This generalizes to a Lorentz covariant  expression
\beq {\cal L}_s = - 4 \tau^2 \partial^{\mu} \phi \partial_{\mu} \phi \label{SpacelikeSpacelike}\eeq
where the minus sign comes from $(+, -,-,-)$ metric convention. The above was derived for the case where the gradient is spacelike. Now, if the gradient happens to be timelike then ${\cal L}_S$ is still formally defined. This time, however, it is equal to zero. After all, if the gradient is timelike, we can select Alexandrov set in such a way that $\partial^{\mu} \phi$ points in a direction parallel to $q^{\mu} - p^{\mu}$ and, therefore, perpendicular to equator. This would imply that for \emph{arbitrary} $s_1$ and $s_2$ lying on the equator, $\phi (s_2) - \phi (s_1) = 0$. Thus, for arbitrary $s_1$ and $s_2$ on the equator ${\cal K}_s (\phi; s_1, s_2) = 0$. This means that "maximum" over "all possible" ${\cal K}_s$ "throughout equator" is $0$ as well. Thus, ${\cal J}_S$ is zero. Now, since all of the expressions involve squaring, none of them are negative. Therefore, the fact that ${\cal J}_S$ is zero "at least once" means that the minimum of ${\cal J}_S$ is zero as well. Finally, since ${\cal L}_s$ is identified with a minimum of ${\cal J}_S$, this implies that 
\beq \partial^{\mu} \phi \partial_{\mu} \phi > 0 \Longrightarrow {\cal L}_s (\phi; r) = 0 \label{SpacelikeTimelike}\eeq
Finally, putting together Equation \ref{SpacelikeSpacelike} and \ref{SpacelikeTimelike} gives us
\beq {\cal L}_s = 2 \tau^2 (\partial^{\mu} \phi \partial_{\mu} \phi)(-1 + sgn(\partial^{\mu} \phi \partial_{\mu} \phi) \label{SpacelikeMichigan}\eeq
Let us now move on to ${\cal L}_t$. Again, we have to do it for two cases: the gradient of $\phi$ being timelike and spacelike (and in the former case we will get well known Lagrangian density while in the latter case we will get zero). Let us start from the case of $\partial^{\mu} \phi$ being timelike. By definition,  the points $s_1$ and $s_2$ are "constrained" to coincide with $p$ and $q$ respectively. Thus, the "maximum" over $s_1$ and $s_2$ trivially reduces to the corresponding expression over $p$ and $q$. Thus, 
\beq {\cal J}_t (\phi; p, q) = {\cal K}_t (\phi; p, q) = (\phi (q) - \phi (p))^2 \eeq
If we assume that $\phi$ is linear, this becomes 
\beq {\cal J}_t (\phi; p, q) = ((f^{\mu} (q) - f^{\mu} (p))\partial_{\mu} \phi)^2 \eeq
We will choose coordinate system in such a way that the gradient points along $t$-axis. In this case the above expression becomes 
\beq {\cal J}_t (\phi; p, q) = (f^0 (q) - f^0 (p))^2 (\partial_0 \phi)^2 \eeq
In order to minimize ${\cal J}_t$, we have to minimize $f^0 (q) - f^0 (p)$. Since the constraint of minimization is $\tau (p, r) =\tau (r, q) = \tau$, it is easy to see that the minimum is achieved when $f^{\mu} (q) - f^{\mu} (p)$ is parallel to $t$-axis; that is,
\beq f^{\mu} (q) - f^{\mu} (r) = f^{\mu} (r) - f^{\mu} (p) = \tau \delta^{\mu}_0 \eeq
We now substitute the above into ${\cal J}_t$ and identify the result with ${\cal L}_t$: 
\beq {\cal L}_t = 4 \tau^2 (\partial_0 \phi)^2 \eeq
Since the above equation was derived in a frame where $\partial_k \phi = 0$, this generalizes to a Lorentz covariant expression 
\beq {\cal L}_t = 4 \tau^2 \partial^{\mu} \phi \partial_{\mu} \phi \label{TimelikeTimelike} \eeq
Let us now assume that $\partial^{\mu} \phi$ is spacelike. In this case, it is possible to select Alexandrov set in such a way that its axes point perpendicularly to the gradient of $\phi$. This will immediately imply that ${\cal K}_t=0$ and ${\cal J}_t =0$. Since ${\cal K}_t$ and ${\cal J}_t$ involve only quadratic expressions, they are both non-negative. Thus, the fact that ${\cal J}_t$ coincides with zero \emph{at least once} implies that the "minimum" of ${\cal J}_t$ is zero. Now, since the minimum of ${\cal J}_t$ is identified with ${\cal L}_t$, this implies that 
\beq \partial^{\mu} \phi \partial_{\mu} \phi < 0 \Longrightarrow {\cal L}_t (\phi; r) = 0 \label{TimelikeSpacelike}\eeq
Puting together Equations \ref{TimelikeTimelike} and \ref{TimelikeSpacelike}, we obtain
\beq {\cal L}_t = 2 \tau^2 (\partial^{\mu} \phi \partial_{\mu} \phi)(1 + sgn(\partial^{\mu} \phi \partial_{\mu} \phi) \label{TimelikeMichigan}\eeq
Finally, combining the Equations \ref{SpacelikeMichigan} and \ref{TimelikeMichigan} we obtain
\beq {\cal L} = {\cal L}_t - {\cal L}_s = 2 \tau^2 \partial^{\mu} \phi \partial_{\mu} \phi \eeq
The moral of the story is that, due to the lack of reference to coordinates in the original definitions of Lagrangians, we are bound to obtain relativistically covariant result. At the same time, however, we can have "relativistically covariant" things that are "not observed in nature", such as $sgn (\partial^{\mu} \phi \partial_{\mu} \phi)$. In order to "get rid" of these things we need to find "by hand" an appropriate linear combination of Lagrangians that would cancel unwanted quantities. Similar situation will arise in electromagnetic case. We will have two different Lagrangians: one is "electric" and the other is "magnetic". Both will be expressed in Lorentz covariant form. But we would have "unwanted" contraction $\epsilon_{\alpha \beta \gamma \delta} F^{\alpha \beta} F^{\gamma \delta}$. That contraction, however, would get canceled when we find appropriate linear combination of "electric" and "magnetic" Lagrangians. 

\subsection*{4.Higher order correction for scalar field}

Let us now compute higher order corrections to the scalar field Lagrangian. In the previous section we have found out that the Lagrangian is of $0 (\tau^2)$. Let us now agree to the next order we are looking at. Let us denote by $\phi_{\cal M}$ a "linear" approximation and by $\phi'_{\cal M}$ its modification. If $f(s_1)$ and $f(s_2)$ maximizes $(\phi{\cal M} (f(s_2)) - \phi_{\cal M} (f(s_1)))^2$, the points $f(s_1')$ and $f(s_2')$ maximize $(\phi'_{\cal M} (f(s'_2)) - \phi'_{\cal M} (f(s'_1))$. Then the ''corrected'' expression will take the form
\beq \phi'_{\cal M} (f(s'_2)) - \phi'_{\cal M} (f(s'_1)) = [\phi_{\cal M} (f(s_2)) - \phi_{\cal M} (f(s_1))] + [(\phi'_{\cal M} (f(s_2)) - \phi_{\cal M} (f(s_2))]  - \nonumber \eeq
\beq  - [\phi'_{\cal M} (f(s_1)) - \phi_{\cal M} (f(s_1))] + [\phi'_{\cal M} (f(s'_2)) - \phi'_{\cal M} (f(s_2))] - [(\phi'_{\cal M} (f(s'_1)) - \phi'_{\cal M} (f(s_1))]  \label{ScalarModification}\eeq
Now, up to the first order, $\psi_{\cal M}$ and $\psi'_{\cal M}$ agree. Therefore, when we look at $\psi'_{\cal M} (s) - \psi_{\cal M} (s)$, we are referring to the second order or higher. Now, we recall from previous section that the points $f(s_1)$ and $f(s_2)$ are "exactly opposite" to each other relative to the center:
\beq f(s_1) = - f(s_2) \eeq
This statement is no longer true for $f(s_1')$ and $f(s_2')$, but it is still true for $f(s_1)$ and $f(s_2)$. This means that the even order terms in $\phi'_{\cal M} (f(s_2)) - \phi_{\cal M} (f(s_2))$ and $\phi'_{\cal M} (f(s_1)) - \phi_{\cal M} (f(s_1))$ will match. Since in the above expression one is being subtracted from the other, they will cancel. Therefore, the contributions from second and third term on the right hand side will come from $0 (\tau^3)$ as opposed to $0 (\tau^2)$:
\beq [(\phi'_{\cal M} (f(s_2)) - \phi_{\cal M} (f(s_2))]  - [\phi'_{\cal M} (f(s_1)) - \phi_{\cal M} (f(s_1))] = 0 (\tau^3) \eeq
As far as the last two terms of Equation \ref{ScalarModification} are concerned, the symmetry arguments no longer apply since the displacement between we know that displacement betweein ''primed'' and ''unprimed'' points broke that symmetry. Now, the displacement that we have just mentioned is "much smaller" than the size of the Alexandrov set. Since the size of Alexandrov set is of $0 (\tau)$, the displacement is of $0 (\tau^2)$. Therefore \emph{if the derivatives were "finite"} this would have, in fact, produced $0 (\tau^2)$ difference. However, we recall from the previous section that we have chosen a coordinate system in such a way that 
\beq \frac{\partial \phi_{\cal M}}{\partial x} \Big\vert_0 = \frac{\partial \phi_{\cal M}}{\partial y} \Big\vert_0 = 0 \eeq
Since primed and un-primed points are located in the equator of the Alexandrov set rather than the origin, $\partial_x \phi_{\cal M}$ and $\partial_y \phi_{\cal M}$ is no longer zero at these points. However, due to the fact that their displacement from the origin is $\tau$, we know that the $x$- and $y$- derivatives are of $0 (\tau)$ as well: 
\beq \frac{\partial \phi_{\cal M}}{\partial x} \Big\vert_s =  0 (\tau) \; , \; \frac{\partial \phi_{\cal M}}{\partial y} \Big\vert_s = 0 (\tau) \eeq
As a result, the effect of these derivatives upon $0 (\tau^2)$ displacement between primed and un-primed points leads to the field difference of the order of $0 (\tau) \times 0 (\tau^2) = 0 (\tau^3)$. On the other hand, if we consider $z$-derivative, then it is, in fact, "finite":
\beq \frac{\partial \phi_{\cal M}}{\partial z} = 0 (1) \eeq
 After all, the "smallness" of $x$- and $y$-derivatives comes from the assumption that they are zero at the origin; we made no such assumption regarding $z$-derivative. At the same time, however, the $z$-displacement of "primed" points relative to un-primed is of $0 (\tau^4)$ rather than $0 (\tau^3)$. This is due to the fact that, as was explained in the previous section, the un-primed points lie on $z$-axis. After all, constraint that all of the points are "on the surface" of the equator thus implies that 
\beq z' = \sqrt{\tau^2 - x^2 - y^2} = \tau \sqrt{1 - \frac{x^2 + y^2}{\tau^2}} \eeq
Now, since we already agree that 
\beq x= 0 (\tau^2) \; , \; y= 0 (\tau^2), \eeq
 we also know that 
\beq \frac{x^2 + y^2}{\tau^2} = 0 (\tau^2) \eeq
Therefore
\beq z' = \tau \sqrt{1 - \frac{x^2 + y^2}{\tau^2}} = \tau \Big(1- \frac{1}{2} \frac{x^2 + y^2}{\tau^2} + 0 (\tau^4) \Big) = \tau (1+ 0 (\tau^2)) = \tau + 0 (\tau^3) \eeq
Thus, the assumption that 
\beq z = \tau \eeq
implies that 
\beq z' - z = 0 (\tau^3) \eeq
At the same time, as mentioned earlier, the $z$-derivative is of $0(1)$. Thus, the contribution of $z$-derivative is $0 (1) \times 0 (\tau^3) = 0 (\tau^3)$. This means that all three derivatives contribute $0 (\tau^3)$ towards the last two terms of Equation \ref{ScalarModification}, even though the $0 (\tau^3)$ arises for different reasons (in case of $x$ and $y$ it arises as $0 (\tau) \times 0 (\tau^2) = 0 (\tau^3)$ and in case of $z$ it arises as $0 (1) \times 0 (\tau^3) = 0 (\tau^3)$):
\beq \phi'_{\cal M} (f(s'_1)) - \phi'_{\cal M} (f(s_1)) = \frac{\partial \phi}{\partial x} \Big\vert_{f(s_1)} \delta x + \frac{\partial \phi}{\partial y} \Big\vert_{f(s_1)} \delta y + \frac{\partial \phi}{\partial z} \Big\vert_{f(s_1)} \delta z + 0 ((\delta x)^2) = \nonumber \eeq
\beq = 0 (\tau) \times 0 (\tau^2) + 0 (\tau) \times 0 (\tau^2) + 0 (1) \times 0 (\tau^3) + 0 (\tau^4) = 0 (\tau^3) \eeq
Now, if $\phi$ is being modified by $0 (\tau^3)$, then the kinetic term of the Lagrangian will be modified by $0 (\tau^4)$. After all, if we alter we alter
\beq \phi (s_2) \rightarrow \phi (s_2) + \delta \phi \eeq
the Lagrangian generator is being altered according to 
\beq (\phi (s_2) - \phi (s_1))^2 \rightarrow (\phi (s_2) - \phi (s_1))^2 + (\phi (s_2) - \phi (s_1)) \delta \phi \eeq
The fact that 
\beq \phi (s_2) - \phi (s_1) = 0 (\tau) \; , \; \delta \phi = 0 (\tau^3) \eeq
implies that 
\beq  (\phi (s_2) - \phi (s_1)) \delta \phi = 0 (\tau) \times 0 (\tau^3) = 0 (\tau^4) \eeq
It should be emphasized though that $0 (\tau^4)$ corrections towards the Lagrangian density were produced from $0 (\tau^3)$ corrections to scalar field. This means that $0 (\tau^4)$ corrections to scalar field will have an effect of $0 (\tau^5)$ or higher. Thus, we only need to compute the scalar field up to $0 (\tau^3)$. The expression for $\phi$ up to that order is 
\beq \phi_{\cal M} (f(s')) = \phi ({f(s)}) + \frac{\partial \phi}{\partial x} \Big\vert_{f(s)} x + \frac{\partial \phi}{\partial y} \Big\vert_{f(s)} y - \frac{\partial \phi}{\partial z} \Big\vert_{f(s)} (\tau - \sqrt{\tau^2 - x^2 -y^2}) \nonumber \eeq
\beq + \frac{1}{2} \frac{\partial^2 \phi}{\partial x^2} \Big\vert_{f(s)} x^2 + \frac{1}{2} \frac{\partial^2 \phi}{\partial y^2} \Big\vert_{f(s)} y^2 + \frac{\partial^2 \phi}{\partial x \partial y} \Big\vert_{f(s)} xy \eeq
In the above expression, we didn't have $\partial^2 \phi/ \partial x \partial z$, $\partial^2 \phi/ \partial y \partial z$ and $\partial^2 \phi / \partial z^2$. The reason is that the coefficient in front of $\partial \phi/ \partial z$ is of $0 (\tau^3)$; thus we would expect that the coefficients next to higher order derivatives that happen to include $z$ would be of $0 (\tau^4)$ or higher. Since, as mentioned before, we are computing $\phi$ up to $0 (\tau^3)$, we can throw these terms away. 

Now, in order to find the displacement of point $s'$ relative to $s$, we have to find the extremum of $\phi_{\cal M} (f(s'))$ given above. As usual, we find the extremum by taking the derivatives with respect to $x$ and $y$ and equating them with zero (we don't need to do the derivative with respect to $z$ since we have re-expressed $z$ as a function of $x$ and $y$). Thus, 
\beq \frac{\partial \phi}{\partial x} \Big\vert_{f(s)} - \frac{x}{\sqrt{\tau^2 - x^2 - y^2}} \frac{\partial \phi}{\partial z} \Big\vert_{f(s)} + \frac{\partial^2 \phi}{\partial x^2} \Big\vert_{f(s)} x + \frac{\partial^2 \phi}{\partial x \partial y} \Big\vert_{f(s)} y + 0 (\tau^2) = 0 \label{System1a} \eeq 
\beq \frac{\partial \phi}{\partial y} \Big\vert_{f(s)} - \frac{y}{\sqrt{\tau^2 - x^2 - y^2}} \frac{\partial \phi}{\partial z} \Big\vert_{f(s)} + \frac{\partial^2 \phi}{\partial y^2} \Big\vert_{f(s)} y + \frac{\partial^2 \phi}{\partial x \partial y} \Big\vert_{f(s)} x + 0 (\tau^2)= 0 \label{System1b} \eeq 
The above derivatives are computed up to $0 (\tau^2)$ rather than $0 (\tau^3)$ because the derivative of $0 (\tau^3)$ is of $0 (\tau^2)$. This, however, does not change the fact that the values of extreme points of two functions differing by $0 (\tau^3)$ from each other still differ by $0 (\tau^3)$ (one can construct a proof to this effect even without reference to derivatives). Therefore, we trust ourselves that if we start off from functions defined up to $0 (\tau^3)$ and follow our noses, we will obtain an $0 (\tau^3)$ accuracy at the end, even if we will be "running into" $0 (\tau^2)$ along the way. Now, if we multiply the above expressions by $\tau$, we obtain 
\beq \tau \frac{\partial \phi}{\partial x} \Big\vert_{f(s)} - \frac{x}{\sqrt{1 - \frac{x^2 + y^2}{\tau^2}}} \frac{\partial \phi}{\partial z} \Big\vert_{f(s)} + \frac{\partial^2 \phi}{\partial x^2} \Big\vert_{f(s)} x \tau + \frac{\partial^2 \phi}{\partial x \partial y} \Big\vert_{f(s)} y \tau + 0 (\tau^3) = 0 \label{System2a} \eeq 
\beq \tau \frac{\partial \phi}{\partial y} \Big\vert_{f(s)} - \frac{y}{\sqrt{1 - \frac{x^2 + y^2}{\tau^2}}} \frac{\partial \phi}{\partial z} \Big\vert_{f(s)} + \frac{\partial^2 \phi}{\partial y^2} \Big\vert_{f(s)} y \tau + \frac{\partial^2 \phi}{\partial x \partial y} \Big\vert_{f(s)} x \tau + 0 (\tau^3)= 0 \label{System2b} \eeq 
In light of the fact that $x$ and $y$ are of $0 (\tau^2)$, we know that 
\beq \frac{x^2 + y^2}{\tau^2} = 0 (\tau^2)  \eeq
This means that the second terms in the above two expressions are equal to $x + 0 (\tau^3)$ and $y + 0 (\tau^3)$, respectively. Furthermore, again from the fact that $x$ and $y$ are of $0 (\tau^2)$, we know that $x \tau$ and $y \tau$ are of $0 (\tau^3)$. In other words, the last two terms in Equations \ref{System2a} and \ref{System2b} are of $0 (\tau^3)$. \emph{But}, as we mentioned earlier, the increment of $\phi$ would be a \emph{product} of the $x$- and $y$- displacements by $\partial_x \phi$ and $\partial_y \phi$ which, itself, is of $0 (\tau)$. Thus, in order to know $\phi$ up to $0 (\tau^3)$ we only need to know $x$ and $y$ up to $0 (\tau^2)$. Therefore, we throw away all of the $\tau^3$ terms thus simplifying the above expressions: 
\beq \tau \frac{\partial \phi}{\partial x} \Big\vert_{f(s)} - x \frac{\partial \phi}{\partial z} \Big\vert_{f(s)} +  0 (\tau^3) = 0 \label{System3a} \eeq 
\beq \tau \frac{\partial \phi}{\partial y} \Big\vert_{f(s)} - y \frac{\partial \phi}{\partial z} \Big\vert_{f(s)} + 0 (\tau^3)= 0 \label{System3b} \eeq 
Now, since the $x$- and $y$- derivatives at the origin are \emph{exactly} zero, their values at $f(s)$ are given by 
\beq \frac{\partial \phi}{\partial x} \Big\vert_{f(s)} = \frac{\partial^2 \phi}{\partial x \partial z} \Big\vert_0 \tau \; , \; \frac{\partial \phi}{\partial y} \Big\vert_{f(s)} = \frac{\partial^2 \phi}{\partial y \partial z} \Big\vert_0 \tau \label{MixedDerivatives} \eeq
The above was computed to $0 (\tau)$ because in Equations \ref{System3a} and \ref{System3b} these derivatives are multiplied by $\tau$ which would turn $0 (\tau)$ into $0 (\tau^2)$, and $0 (\tau^2)$ is the order up to which we are doing our calculation. By substituting Equation \ref{MixedDerivatives} into Equations \ref{System3a} and \ref{System3b}, we obtain 
\beq \tau^2 \frac{\partial^2 \phi}{\partial x \partial z} \Big\vert_0 - x \frac{\partial \phi}{\partial z} \Big\vert_{f(s)} +  0 (\tau^3) = 0 \label{System4a} \eeq 
\beq \tau^2 \frac{\partial^2 \phi}{\partial y \partial z} \Big\vert_0 - y \frac{\partial \phi}{\partial z} \Big\vert_{f(s)} + 0 (\tau^3)= 0 \label{System4b} \eeq 
Now, the only reasons $\partial \phi/ \partial x$ and $\partial \phi/ \partial y$ are "small" is that coordinate system is chosen in such a way that the gradient of $\phi$ is parallel to $z$-axis at the origin. This implies that $\partial \phi/ \partial z$ is \emph{large}. Thereore, we immediately obtain the expression up to $0 (\tau^2)$ for $x$ and $y$:
\beq x = \tau^2 \frac{\partial^2 \phi/ \partial x \partial z}{\partial \phi/ \partial z} + 0 (\tau^3) \; , \; y = \tau^2 \frac{\partial^2 \phi/ \partial y \partial z}{\partial \phi/ \partial z} + 0 (\tau^3) \label{Approximatexy}\eeq
where we have dropped the indications of points at which the partial derivatives are being evaluated since that would lead to $0 (\tau) \times 0 (\tau^2) = 0 (\tau^3)$ effect which we are ignoring. We can now use these $x$ and $y$ to compute $z$:
\beq z = \sqrt{\tau^2 - x^2 - y^2} = \tau \sqrt{1- \tau^2 \Big(\Big(\frac{\partial^2 \phi/ \partial x \partial z}{\partial \phi/ \partial z} \Big)^2 + \Big(\frac{\partial^2 \phi/ \partial y \partial z}{\partial \phi/ \partial z}\Big)^2 \Big)} = \nonumber \eeq
\beq = \tau - \frac{\tau^3}{2} \Big(\Big(\frac{\partial^2 \phi/ \partial x \partial z}{\partial \phi/ \partial z} \Big)^2 + \Big(\frac{\partial^2 \phi/ \partial y \partial z}{\partial \phi/ \partial z}\Big)^2 \Big) \label{Approximatez} \eeq
Thus, we would like to compute the value of $\phi_{\cal M} (s'_2)$, up to $0 (\tau^3)$. Since the deviation of $z$ is of $0 (\tau^3)$, we only need the first derivative with respect to $z$ up to finite order. On the other hand, since the deviations of $x$ and $y$ are of $0 (\tau^2)$, we need to know respective first derivatives up to $0 (\tau)$. \emph{But}, as stated earlier, the $x$- and $y$- derivatives \emph{are} of $0 (\tau)$ to begin with! Thus, again, we only need a leading order terms of first derivatives, just for a different reason. Therefore, we don't need any second derivatives at all. Thus, we use
\beq \phi_{\cal M} (f(s')) - \phi_{\cal M} (f(s)) = x \frac{\partial \phi_{\cal M}}{\partial x} \Big\vert_{f(s)}+ y \frac{\partial \phi_{\cal M}}{\partial y} \Big\vert_{f(s)} + (z- \tau) \frac{\partial \phi_{\cal M}}{\partial z} \Big\vert_{f(s)}+ 0 (\tau^4) \eeq
By substituting the Equations \ref{Approximatexy} and \ref{Approximatez} into the above, we obtain
\beq \phi_{\cal M} (f(s')) - \phi_{\cal M} (f(s)) = \tau^2 \frac{\partial^2 \phi_{\cal M} / \partial x \partial z}{\partial \phi_{\cal M} / \partial z} \frac{\partial \phi_{\cal M}}{\partial x}\Big\vert_{f(s)} + \tau^2 \frac{\partial^2 \phi_{\cal M} / \partial y \partial z}{\partial \phi_{\cal M}/ \partial z} \frac{\partial \phi_{\cal M}}{\partial y} \Big\vert_{f(s)}+ \nonumber \eeq
\beq  - \frac{\tau^3}{2} \Big(\Big(\frac{\partial^2 \phi_{\cal M}/ \partial x \partial z}{\partial \phi_{\cal M}/ \partial z} \Big)^2 + \Big(\frac{\partial^2 \phi_{\cal M}/ \partial y \partial z}{\partial \phi_{\cal M}/ \partial z}\Big)^2 \Big) \frac{\partial \phi_{\cal M}}{\partial z}\Big\vert_{f(s)} + 0 (\tau^4) \eeq
Now, by substituting Equation \ref{MixedDerivatives} into the $\partial/ \partial x$ and $\partial/ \partial y$ terms above, we obtain
\beq \phi_{\cal M} (f(s')) - \phi_{\cal M} (f(s)) = \tau^3 \frac{\partial^2 \phi_{\cal M} / \partial x \partial z}{\partial \phi_{\cal M} / \partial z} \frac{\partial^2 \phi_{\cal M}}{\partial x \partial z}\Big\vert_{f(s)} + \tau^3 \frac{\partial^2 \phi_{\cal M} / \partial y \partial z}{\partial \phi_{\cal M}/ \partial z} \frac{\partial^2 \phi_{\cal M}}{\partial y \partial z} \Big\vert_{f(s)}+ \nonumber \eeq
\beq  - \frac{\tau^3}{2} \Big(\Big(\frac{\partial^2 \phi_{\cal M}/ \partial x \partial z}{\partial \phi_{\cal M}/ \partial z} \Big)^2 + \Big(\frac{\partial^2 \phi_{\cal M}/ \partial y \partial z}{\partial \phi_{\cal M}/ \partial z}\Big)^2 \Big) \frac{\partial \phi_{\cal M}}{\partial z}\Big\vert_{f(s)} + 0 (\tau^4) \eeq
Now, from the inspection of the above equation, one can see that the first and second terms match the third and fourth term, respectively, except for the factor of $-1/2$ that the latter two terms are multiplied by. Therefore, the above expression simplifies to 
\beq \phi_{\cal M} (f(s')) - \phi_{\cal M} (f(s)) = \frac{\tau^3}{2} \frac{(\partial^2 \phi_{\cal M} / \partial x \partial z)^2}{\partial \phi_{\cal M} / \partial z} \Big\vert_{f(s)} + \frac{\tau^3}{2} \frac{(\partial^2 \phi_{\cal M} / \partial y \partial z)^2}{\partial \phi_{\cal M}/ \partial z}  \Big\vert_{f(s)} \eeq
Now, our eventual goal is to look at shifts of \emph{two different} points: namely, a shift from $s_1$ to $s'_1$ and a shift from $s_2$ to $s'_2$. As we recall, we were assuming that $s_1$ is located at $(0, 0, 0, - \tau)$ and $s_2$ is located at $(0, 0, 0, + \tau)$. Our previous calculations were assuming that $z = + \tau$ and, therefore, the results apply to $s_2$.  If we are to use $z= - \tau$ for $f(s_1)$ the signs of odd-order terms will be reversed while the signs of even-order terms will stay the same. Since the above expression does not have $0 (\tau^2)$, the leading order correction is of $0 (\tau^3)$ and, therefore, comes with the reversed sign:
\beq \phi_{\cal M} (f(s'_1)) - \phi_{\cal M} (f(s_1)) = \frac{\tau^3}{2} \frac{(\partial^2 \phi_{\cal M} / \partial x \partial z)^2}{\partial \phi_{\cal M} / \partial z} \Big\vert_{f(s_1)} - \frac{\tau^3}{2} \frac{(\partial^2 \phi_{\cal M} / \partial y \partial z)^2}{\partial \phi_{\cal M}/ \partial z}  \Big\vert_{f(s_1)} \label{ExpressionFors1}\eeq
On the other hand, since $s_2$ is located at $z= + \tau$, the expression for $s_2$ does not change sign. Thus, 
\beq \phi_{\cal M} (f(s'_2)) - \phi_{\cal M} (f(s_2)) = \frac{\tau^3}{2} \frac{(\partial^2 \phi_{\cal M} / \partial x \partial z)^2}{\partial \phi_{\cal M} / \partial z} \Big\vert_{f(s_2)} + \frac{\tau^3}{2} \frac{(\partial^2 \phi_{\cal M} / \partial y \partial z)^2}{\partial \phi_{\cal M}/ \partial z} \Big\vert_{f(s_2)} \label{ExpressionFors2} \eeq
If we now subtract Equation \ref{ExpressionFors1} from Equation \ref{ExpressionFors2}, and move $f(s_1)$ and $f(s_2)$ to the right hand side while keeping $f(s'_1)$ and $f(s'_2)$ at the left, we obtain
\beq \phi_{\cal M} (f(s_2')) - \phi_{\cal M} (f(s_1')) = \phi_{\cal M} (f(s_2)) - \phi_{\cal M} (f(s_1)) + \nonumber \eeq 
\beq + \frac{\tau^3}{2} \frac{(\partial^2 \phi_{\cal M} / \partial x \partial z)^2}{\partial \phi_{\cal M} / \partial z} \Big\vert_{f(s_2)} + \frac{\tau^3}{2} \frac{(\partial^2 \phi_{\cal M} / \partial y \partial z)^2}{\partial \phi_{\cal M}/ \partial z}  \Big\vert_{f(s_2)}  + \eeq
\beq + \frac{\tau^3}{2} \frac{(\partial^2 \phi_{\cal M} / \partial x \partial z)^2}{\partial \phi_{\cal M} / \partial z} \Big\vert_{f(s_1)} + \frac{\tau^3}{2} \frac{(\partial^2 \phi_{\cal M} / \partial y \partial z)^2}{\partial \phi_{\cal M}/ \partial z}  \Big\vert_{f(s_1)} \nonumber \eeq
Now, the difference between the derivatives (of any order) evaluated at $s_1$ and $s_2$ is of $0 (\tau)$. Since the derivatives are being multiplied by $\tau^3$, this $0 (\tau)$ difference will become $0 (\tau) \times 0 (\tau^3) = 0 (\tau^4)$ which we can ignore. Therefore, we can replace the derivatives "at $f(s_1)$" and "at $f(s_2)$" with derivatives "at the origin". This will allow us to combine these for terms into two terms: 
\beq \phi_{\cal M} (f(s_2')) - \phi_{\cal M} (f(s_1')) = \phi_{\cal M} (f(s_2)) - \phi_{\cal M} (f(s_1)) + \nonumber \eeq
\beq + \tau^3  \frac{(\partial^2 \phi_{\cal M} / \partial x \partial z)^2}{\partial \phi_{\cal M} / \partial z} \Big\vert_0 + \tau^3\frac{(\partial^2 \phi_{\cal M} / \partial y \partial z)^2}{\partial \phi_{\cal M}/ \partial z}  \Big\vert_0  \eeq
Now, the first two terms on the right hand side have yet more higher order derivatives hidden in them. Namely,
\beq \phi_{\cal M} (f(s_2)) - \phi_{\cal M} (f(s_1)) = 2 \tau \frac{\partial \phi_{\cal M}}{\partial z} \Big\vert_0 + \frac{4 \tau^3}{3}  \frac{\partial^3 \phi_{\cal M}}{\partial z^3} \Big\vert_0 \eeq
Thus, the final expression becomes 
\beq \phi_{\cal M} (f(s_2')) - \phi_{\cal M} (f(s_1')) = 2 \tau \frac{\partial \phi_{\cal M}}{\partial z} \Big\vert_0 + \frac{4 \tau^3}{3}  \frac{\partial^3 \phi_{\cal M}}{\partial z^3} \Big\vert_0 + \nonumber \eeq
\beq + \tau^3  \frac{(\partial^2 \phi_{\cal M} / \partial x \partial z)^2}{\partial \phi_{\cal M} / \partial z} + \tau^3\frac{(\partial^2 \phi_{\cal M} / \partial y \partial z)^2}{\partial \phi_{\cal M}/ \partial z}  \Big\vert_0 \eeq
Now in order to get Lagrangian we need to square the above expression. Up to $0 (\tau^4)$ terms,  the latter is
\beq {\cal J}_s = (\phi_{\cal M} (f(s_2')) - \phi_{\cal M} (f(s_1')))^2 = 4 \tau^2 \Big(\frac{\partial \phi_{\cal M}}{\partial z}  \Big\vert_0 \Big)^2 + \label{SpacelikeScalarCorrectionxy}\eeq
\beq +  \frac{16 \tau^4}{3}  \frac{\partial^3 \phi_{\cal M}}{\partial z^3} \Big\vert_0 \frac{\partial \phi_{\cal M}}{\partial z} \Big\vert_0  + 2 \tau^4  \Big(\frac{\partial^2 \phi_{\cal M}}{\partial x \partial z}\Big)^2 \Big\vert_0 + 2 \tau^4  \Big(\frac{\partial^2 \phi_{\cal M}}{\partial y \partial z}\Big)^2 \Big\vert_0 \nonumber \eeq
So far we have found a way of maximizing $(\phi_{\cal M} (f (s_2')) - \phi_{\cal M} (f (s_1')))^2$ for a \emph{fixed} Alexandrov set. The next step in the prescription outlined in the previous section is to look at the different possible Alexandrov sets, do the "maximization" within each one, and then select the one Alexandrov set that would \emph{minimize} the "maximum". Now, the original Alexandrov set was selected in such a way that the gradient of $\phi_{\cal M}$ is parallel to $z$-axis at the origin. This implies that, upon "very small" coordinate rotation, the first term on the right hand side varies quadratically. On the other hand, the last three terms on the right hand side vary linearly. In other words, the equation behaves as 
\beq {\cal J}_s = {\cal J}_0 + a \tau^2 \theta^2 + b \tau^4 \theta \label{AvoidMinimizationTrick} \eeq
where $\theta$ is the angle of rotation. In order to minimize the above we have to find the place where derivative is zero with respect to $\theta$:
\beq 0 = \frac{d}{d \theta} (a \tau^2 \theta^2 + b \tau^4 \theta) = 2a \tau^2 \theta + b \tau^4 \eeq
This implies that 
\beq \theta = \frac{b \tau^4}{2a \tau^2} = \frac{b \tau^2}{2a} \eeq
By substituting it back into Equation \ref{AvoidMinimizationTrick}, we obtain
\beq {\cal J}_s = {\cal J}_0 + \frac{b^2 \tau^6}{4a} +  \frac{b^2 \tau^6}{2a} = {\cal J}_0 + \frac{3b^2 \tau^6}{4a} \label{AvoidMinimizationSuccess} \eeq
Since we are computing Lagrangian density only up to $0 (\tau^4)$, this means that we are free to ignore the effect of "minimization" part. In other words, we don't have to rotate Alexandrov set from its original position, and simply copy the result of "maximization" (Equation \ref{SpacelikeScalarCorrectionxy}) for $\cal J$ as our answer for $\cal L$. Since we are not going to need the Alexandrov set any more, we can think of it as "pointwise" Lagrangian where the "point" in which Lagrangian is evaluated is what \emph{used to be} the origin. Thus, we will rewrite Equation \ref{SpacelikeScalarCorrectionxy} while dropping "at the origin" signs: 
\beq {\cal L}_s = 4 \tau^2 \Big(\frac{\partial \phi_{\cal M}}{\partial z}  \Big)^2 +  \frac{16 \tau^4}{3}  \frac{\partial^3 \phi_{\cal M}}{\partial z^3}  \frac{\partial \phi_{\cal M}}{\partial z} + 2 \tau^4  \Big(\frac{\partial^2 \phi_{\cal M}}{\partial x \partial z}\Big)^2  + 2 \tau^4  \Big(\frac{\partial^2 \phi_{\cal M}}{\partial y \partial z}\Big)^2 \label{ScalarSpacelikeLxy} \eeq
Now, the above expression is true \emph{only} in the coordinate system where 
\beq \frac{\partial \phi_{\cal M}}{\partial x} = \frac{\partial \phi_{\cal M}}{\partial y} = 0 \eeq
We can, however, use the fact that $x$- and $y$- derivatives are zero in our particular frame in order to replace $z$-derivatives with covariant expressions: 
\beq \partial_z \lambda = \frac{\partial_i \phi \partial_i \lambda}{\sqrt{\partial_j \phi \partial_j \phi}} \; , \; \partial_z^2 \lambda = \frac{\partial_i \phi \partial_j \phi \partial_i \partial_j \lambda}{\partial_j \phi \partial_j \phi} \; , \; \partial_z^3 \lambda = \frac{\partial_i \phi \partial_j \phi \partial_k \phi \partial_i \partial_j \partial_k \lambda}{(\partial_l \phi \partial_l \phi)^{3/2}} \label{zDerivativesAreCovariant}\eeq
Basically, we have replaced $z$-derivative with a contraction to gradient of $\phi_{\cal M}$ since the latter is supposed to be parallel with $z$-axis. If the amplitude of gradient of $\phi_{\cal M}$ is something other than unity, this would pick unwanted scalar factors. In order to "get rid" of them, we divide the contraction by the appropriate power of the amplitude of gradient of $\phi$. 

Apart from "getting rid" of $z$-derivative, we have to also "get rid" of $x$ and $y$. In order to do it, we add and subtract $(\partial_z \partial_z \phi_{\cal M})^2$:
\beq {\cal L}_s = 4 \tau^2 (\partial_z \phi_{\cal M})^2 + \frac{16 \tau^4}{3} \partial^3_z \phi_{\cal M} \partial_z \phi_{\cal M} + 2 \tau^4 \partial_k \partial_z \phi_{\cal M} \partial_k \partial_z \phi_{\cal M} - 2 \tau^4 (\partial_z \partial_z \phi_{\cal M})^2 \eeq
This reduces the task of "getting rid" of $x$ and $y$ to the task of "getting rid" of $z$. We then use Equation \ref{zDerivativesAreCovariant} to do the latter, obtaining
\beq {\cal L}_s = 4 \tau^2 \partial_k \phi_{\cal M} \partial_k \phi_{\cal M} +  \frac{16 \tau^4}{3}  \frac{\partial_i \phi_{\cal M} \partial_j \phi_{\cal M} \partial_k \phi_{\cal M} \partial_i \partial_j \partial_k \phi_{\cal M}}{\partial_l \phi_{\cal M} \partial_l \phi_{\cal M}} + \nonumber \eeq
\beq + 2 \tau^4  \frac{\partial_i \phi_{\cal M} \partial_j \phi_{\cal M} \partial_i \partial_k \phi_{\cal M} \partial_j \partial_k \phi_{\cal M} }{\partial_l \phi_{\cal M} \partial_l \phi_{\cal M}} - 2 \tau^4 \Big(\frac{\partial_i \phi_{\cal M} \partial_j \phi_{\cal M} \partial_i \partial_j \phi_{\cal M} }{\partial_k \phi_{\cal M} \partial_k \phi_{\cal M}}\Big)^2 \label{ScalarSpacelikeLxyz} \eeq
Finally, we would like to rewrite it in Lorentz covariant way. We recall from previous section that we originally chose the frame in which 
\beq \partial_0 \phi = 0 \eeq
This seem to suggest that we can blindly replace Latin indexes with Greek ones. Of course, we still have non-zero values of second derivatives involving time:
\beq \partial_0 \partial_{\mu} \neq 0 \eeq
However, by quick inspection of Equation \ref{ScalarSpacelikeLxyz}, we see that any second derivative that involves any given index is always coupled to the first derivative with respect to that index. Thus, if we were to replace the non-covariant indexes with covariant ones, the "unwanted" $\partial_{\mu} \partial_0 \phi$ terms will be multiplied by $\partial_0 \phi$ and, therefore, sent to zero. The only possible concern we might still have is that, due to the perturbations, the axis of Alexandrov set has been rotated which would result in non-zero value of time derivative. However, in the argument we have made in Equations \ref{AvoidMinimizationTrick} and \ref{AvoidMinimizationSuccess}, the effect of such rotation is of $0 (\tau^6)$ and, therefore, negligible as far as the precision of the calculations at hand is concerned. Thus, at the expense of extra $0 (\tau^6)$ error, we will agree \emph{not} to rotate the axis of Alexandrov set which will, in turn, allow us to use $\partial_0 \phi = 0$. Thus, we will go ahead and rewrite Equation \ref{ScalarSpacelikeLxyz} in a covariant form:
\beq {\cal L}_s = - 4 \tau^2 \partial^{\mu} \phi_{\cal M} \partial_{\mu} \phi_{\cal M} +  \frac{16 \tau^4}{3}  \frac{\partial^{\mu} \phi_{\cal M} \partial^{\nu} \phi_{\cal M} \partial^{\rho} \phi_{\cal M} \partial_{\mu} \partial_{\nu} \partial_{\rho} \phi_{\cal M}}{\partial^{\sigma} \phi_{\cal M} \partial_{\sigma} \phi_{\cal M}} + \nonumber \eeq
\beq + 2 \tau^4  \frac{\partial^{\mu} \phi_{\cal M} \partial^{\nu} \phi_{\cal M} \partial_{\mu} \partial^{\rho} \phi_{\cal M} \partial_{\nu} \partial_{\rho} \phi_{\cal M} }{\partial^{\sigma} \phi_{\cal M} \partial_{\sigma} \phi_{\cal M}} - 2 \tau^4 \Big(\frac{\partial^{\mu} \phi_{\cal M} \partial^{\nu} \phi_{\cal M} \partial_{\mu} \partial_{\nu} \phi_{\cal M} }{\partial^{\sigma} \phi_{\cal M} \partial_{\sigma} \phi_{\cal M}}\Big)^2 \label{ScalarSpacelikeLorentz} \eeq
where the sign change in the first term on the right hand side is due to $(+1, -1,-1,-1)$ convention. 

Now we recall that the calculation above was made under assumption that the grandient of $\phi_{\cal M}$ happened to be spacelike at a point we are interested in. After all, this is what allowed us to align the gradient of $\phi$ with $z$-axis (as opposed to $t$-axis). Let us now consider the case when gradient is timelike. Let us select Alexandrov set in such a way that $f^{\mu} (q) - f^{\mu} (p)$ is parallel to $(\partial^{\mu} \phi)(0)$, and let us select coordinate system in such a way that $t$-axis passes through $f(p)$ and $f(q)$ with origin at the middle. Thus, 
\beq f^{\mu} (p) = - \frac{\tau}{2} \delta^{\mu}_0 \; , \; f^{\mu} (q) = \frac{\tau}{2} \delta^{\mu}_0 \eeq
\beq \frac{\partial \phi}{\partial x} \Big\vert_0 = \frac{\partial \phi}{\partial y} \Big\vert_0 = \frac{\partial \phi}{\partial z} \Big\vert_0 = 0 \eeq
In this case, linear terms no longer contribute to Lagrangian generator. Now, in the previous calculation the presence of linear terms was the reason why we had to select $s_1$ and $s_2$ in the nearly-opposite directions from the origin. Therefore in the present situation this is no longer the case. Now, one consequence of the fact that $s_1$ and $s_2$ were selected in the opposite directions from the origin was lack of contribution from second order derivative terms. Thus, the next order correction was coming from third derivatives. In our present situation, since $s_1$ and $s_2$ are no longer opposite, the second order terms begin to contribute. 

Apart from that, there is yet another difference. In the previous calculation, the $0 (\tau^3)$ correction to $\phi$ lead to $0 (\tau^3) \times 0 (\tau) = 0 (\tau^4)$ correction to the Lagrangian. This time, due to lack of linear terms, $0 (\tau)$ is being replaced by $0 (\tau^2)$. Thus, the contribution of $0 (\tau^3)$ correction to $\phi$ is $0 (\tau^3) \times 0 (\tau^2) = 0 (\tau^5)$. Since we are only computing Lagrangian up to $0 (\tau^4)$, we can neglect $0 (\tau^5)$ effects on Lagrangian and, therefore, we can likewise neglect $0 (\tau^3)$ correction to $\phi$. On the other hand, $0 (\tau^2)$ correction to $\phi$ leads to $0 (\tau^2) \times 0 (\tau^2) = 0 (\tau^4)$ contribution, which we can not neglect. Thus, we will assume that $\phi$ is quadratic. 

Now, in light of lack of spacelike gradient of $\phi$, we don't have any "preferred" space coordinate the way we had before; the only "preferred" coordinate we have specified so far is $t$. Therefore, we are free to perform spacelike rotations as long as $t$ stays fixed. We will, therefore, rotate the spacelike coordinates in such a way that $3 \times 3$ matrix $D_{ij} = \partial_i \phi \partial_j \phi$ is diagonalized, and its eigenvalues are $\lambda_1$, $\lambda_2$ and $\lambda_3$. Thus,
\beq \frac{\partial^2 \phi}{\partial x^2} \Big\vert_0 = \lambda_1 \; , \;  \frac{\partial^2 \phi}{\partial y^2} \Big\vert_0 = \lambda_2 \; , \;  \frac{\partial^2 \phi}{\partial z^2} \Big\vert_0 = \lambda_3 \eeq
The value of $\phi$ at an arbitrary point on the equator is 
\beq \phi (s) = \phi (0) + \frac{\lambda_1 x^2}{2} + \frac{\lambda_2 y^2}{2} + \frac{\lambda_3 z^2}{2} \eeq
Therefore, the "maximum" and "minimum" of $\phi (s)$ are given by 
\beq \phi_{\rm max} = \phi_0 + \frac{\tau^2}{2} \max (\lambda_1, \lambda_2, \lambda_3) \; , \; \phi_{\rm min} = \phi_0 + \frac{\tau^2}{2} \max (\lambda_1, \lambda_2, \lambda_3) \eeq
This means that pre-Lagrangian is given by 
\beq {\cal J}_s = \frac{\tau^4}{4} (\max (\lambda_1, \lambda_2, \lambda_3) - \min (\lambda_1, \lambda_2, \lambda_3))^2 \eeq
Now we have to see whether or not we have to rotate the axis of Alexandrov set by a small amount in order to minimize $\cal J$. Geometrically, it is easy to see that if a function is symmetric around the origin, then the rotation away from the symmetric state would only increase the difference between maximum and minimum. Since our goal is to \emph{minimize}, the only instance where we would need to rotate is when the symmetry is already broken by odd-order terms. Now we already know that we don't have linear terms. Therefore, the only justification for rotation would be third-order terms. But we have already established that $0 (\tau^3)$ terms in $\phi$ have $0 (\tau^5)$ effect on Lagrangian which we ignore. Therefore, the effects of minimization of ${\cal J}_s$ are, likewise, of $0 (\tau^5)$ and are likewise ignored. Thus, we simply copy the expression we had for $\cal J$ into $\cal L$ without further modifications:
\beq {\cal L} = \frac{\tau^4}{4} (\max (\lambda_1, \lambda_2, \lambda_3) - \min (\lambda_1, \lambda_2, \lambda_3))^2 \eeq
One should note that the above expression is not covariant. After all, we have only \emph{three} eigenvalues rather than four, since we are referring to only \emph{spacelike} components of second derivative, $\partial_i \partial_j \phi$. We can obtain a covariant expression by replacing $t$ axis with $\partial^{\mu} \phi/ \vert \partial^{\mu} \phi \vert$. We thus define the following tensor:
\beq D_{\mu \nu} = \partial_{\mu} \partial_{\nu} \phi - \frac{\partial^{\mu} \phi \partial_{\mu} \partial_{\nu} \phi }{\sqrt{\partial^{\rho} \partial_{\rho} \phi}} - \frac{\partial^{\nu} \phi \partial_{\mu} \partial_{\nu} \phi }{\sqrt{\partial^{\rho} \partial_{\rho} \phi}} + \frac{\partial^{\mu} \phi \partial^{\nu} \phi \partial_{\mu} \partial_{\nu} \phi }{\partial^{\rho} \partial_{\rho} \phi}\eeq
In this case $D_{\mu \nu}$ will have \emph{four} eigenvalues; but one of them will be zero. Thus, we would have to take a maximum and minimum out of $\{\lambda_1, \lambda_2, \lambda_3, \lambda_4 \} \setminus \{0 \}$. In order to make ourselves completely safe for the situations in which the "zero" eigenvalue will deviate from $0$, we will instead do $\{\lambda_1, \lambda_2, \lambda_3, \lambda_4 \} \setminus (- \epsilon, \epsilon)$. Thus, we have 
\beq {\cal L}_s = \frac{\tau^4}{4} (\max (\{\lambda_1, \lambda_2, \lambda_3, \lambda_4 \} \setminus (- \epsilon, \epsilon)) - \min (\{\lambda_1, \lambda_2, \lambda_3, \lambda_4 \} \setminus (- \epsilon, \epsilon) ))^2 \label{NotExactZero}\eeq
Thus, if we bring together the case of spacelike and timelike gradient, we obtain
\beq {\cal L}_s = (-1 + sgn( \partial^{\mu} \phi \partial_{\mu} \phi)) \Big[ 2 \tau^2 \partial^{\mu} \phi_{\cal M} \partial_{\mu} \phi_{\cal M} -  \frac{8 \tau^4}{3}  \frac{\partial^{\mu} \phi_{\cal M} \partial^{\nu} \phi_{\cal M} \partial^{\rho} \phi_{\cal M} \partial_{\mu} \partial_{\nu} \partial_{\rho} \phi_{\cal M}}{\partial^{\sigma} \phi_{\cal M} \partial_{\sigma} \phi_{\cal M}} - \nonumber \eeq
\beq -  \tau^4  \frac{\partial^{\mu} \phi_{\cal M} \partial^{\nu} \phi_{\cal M} \partial_{\mu} \partial^{\rho} \phi_{\cal M} \partial_{\nu} \partial_{\rho} \phi_{\cal M} }{\partial^{\sigma} \phi_{\cal M} \partial_{\sigma} \phi_{\cal M}} + \tau^4 \Big(\frac{\partial^{\mu} \phi_{\cal M} \partial^{\nu} \phi_{\cal M} \partial_{\mu} \partial_{\nu} \phi_{\cal M} }{\partial^{\sigma} \phi_{\cal M} \partial_{\sigma} \phi_{\cal M}}\Big)^2 \Big] + \eeq
\beq + \frac{\tau^2}{2} (1 + sgn (\partial^{\mu} \phi \partial_{\mu} \phi)) (\max (\{\lambda_1, \lambda_2, \lambda_3, \lambda_4 \} \setminus (- \epsilon, \epsilon)) - \min (\{\lambda_1, \lambda_2, \lambda_3, \lambda_4 \} \setminus (- \epsilon, \epsilon) ))^2  \nonumber \eeq
So far we have found an expression for ${\cal L}_s$. Let us now discuss ${\cal L}_t$. In this case the calculation will be considerably simpler since the two points we are looking at will simply be $p$ and $q$ so we would no longer need to consider their displacements. The pre-Lagrangian simply becomes 
\beq {\cal J}_t = (\phi (q) - \phi (p))^2 = \Big(2 \tau \frac{\partial \phi}{\partial t} \Big\vert_0 + \frac{4 \tau^3}{3} \frac{\partial^3 \phi}{\partial t^3} \Big)^2 = 4 \tau^2 \Big( \frac{\partial \phi}{\partial t} \Big)^2 + \frac{16 \tau^4}{3} \frac{\partial \phi}{\partial t} \frac{\partial^3 \phi}{\partial t^3}  \eeq
Now we need to rotate Alexandrov set in a way that the above expression is minimized. Let us start from timelike case. In case of $\phi_{\cal M}$ being linear, we have to align the axis of Alexandrov set with the gradient of the field, as we have done in the "timelike" part of the previous section. If, on the other hand, $\phi_{\cal M}$ is non-linear, we can repeating the argument similar to Equations \ref{AvoidMinimizationTrick} and \ref{AvoidMinimizationSuccess} to show that the small Lorentz transformations would only result in $0 (\tau^6)$ corrections which we don't care about. Therefore, we will identify the above expression for $\cal J$ with a Lagrangian $\cal L$:
\beq {\cal L}_t = (\phi (q) - \phi (p))^2 = \Big(2 \tau \frac{\partial \phi}{\partial t} \Big\vert_0 + \frac{4 \tau^3}{3} \frac{\partial^3 \phi}{\partial t^3} \Big)^2 = 4 \tau^2 \Big( \frac{\partial \phi}{\partial t} \Big)^2 + \frac{16 \tau^4}{3} \frac{\partial \phi}{\partial t} \frac{\partial^3 \phi}{\partial t^3}  \eeq
Finally, by using the fact that space derivatives of $\phi$ are zero, we can generalize the above to Lorentz covariant expression:
\beq {\cal L}_t = 4 \tau^2 \partial^{\mu} \phi \partial_{\mu} \phi + \frac{16 \tau^4}{3} \frac{\partial^{\mu} \phi \partial^{\nu} \phi \partial^{\rho} \phi \partial_{\mu} \partial_{\nu} \partial_{\rho} \phi}{\partial^{\sigma} \phi \partial_{\sigma} \phi}  \label{TimelikeDeviationAlmostDone} \eeq
Now let us consider the case where the gradient is spacelike. In this case, we can pick an Alexandrov set in such a way that $\phi (q) = \phi (p)$ \emph{exactly} holds, setting ${\cal L}_t$ to \emph{exact} zero:
\beq \partial^{\mu} \phi \partial_{\mu} \phi < 0 \Longrightarrow {\cal L}_t = 0 \label{ExactZero} \eeq
 This should be contrasted with the fact that ${\cal L}_s$ is \emph{not} an exact zero in case of timelike gradient (see Equation \ref{NotExactZero}). The reason for this is that in case of ${\cal L}_s$ we are looking at "a lot of" points (namely all of the points across the equator of Alexandrov set) whereas in case of ${\cal L}_t$ we are looking at only \emph{two} points (namely $p$ and $q$). In case of two points, we can manually move them in such a way as to get an exact match. On the other hand, in case of several different points, the "rigidity" of the shape prevents us from doing it: manually adjusting some of them would compromise adjustment of the others. Thus, in case of several points the best we can do is to "trust" some specified order of derivative, which would lead to higher order deviations. Anyway, Equations \ref{TimelikeDeviationAlmostDone} and \ref{ExactZero} can be summarized as 
\beq {\cal L}_t = (1 + sgn (\partial^{\mu} \phi \partial_{\mu} \phi)) \Big( 2 \tau^2 \partial^{\mu} \phi \partial_{\mu} \phi + \frac{8 \tau^4}{3} \frac{\partial^{\mu} \phi \partial^{\nu} \phi \partial^{\rho} \phi \partial_{\mu} \partial_{\nu} \partial_{\rho} \phi}{\partial^{\sigma} \phi \partial_{\sigma} \phi} \Big) \eeq  
Finally, by using 
\beq {\cal L} = {\cal L}_t - {\cal L}_s \eeq
we obtain 
\beq {\cal L}_s =4 \tau^2 \partial^{\mu} \partial_{\mu} \phi + \frac{8 \tau^4}{3} \frac{\partial^{\mu} \phi \partial^{\nu} \phi \partial^{\rho} \phi \partial_{\mu} \partial_{\nu} \partial_{\rho} \phi (1 + sgn (\partial^{\mu} \phi \partial_{\mu} \phi)) }{\partial^{\sigma} \phi \partial_{\sigma} \phi} \Big) + \nonumber \eeq  
\beq  + (-1 + sgn( \partial^{\mu} \phi \partial_{\mu} \phi)) \Big[    \frac{8 \tau^4}{3}  \frac{\partial^{\mu} \phi_{\cal M} \partial^{\nu} \phi_{\cal M} \partial^{\rho} \phi_{\cal M} \partial_{\mu} \partial_{\nu} \partial_{\rho} \phi_{\cal M}}{\partial^{\sigma} \phi_{\cal M} \partial_{\sigma} \phi_{\cal M}} +  \eeq
\beq +  \tau^4  \frac{\partial^{\mu} \phi_{\cal M} \partial^{\nu} \phi_{\cal M} \partial_{\mu} \partial^{\rho} \phi_{\cal M} \partial_{\nu} \partial_{\rho} \phi_{\cal M} }{\partial^{\sigma} \phi_{\cal M} \partial_{\sigma} \phi_{\cal M}} - \tau^4 \Big(\frac{\partial^{\mu} \phi_{\cal M} \partial^{\nu} \phi_{\cal M} \partial_{\mu} \partial_{\nu} \phi_{\cal M} }{\partial^{\sigma} \phi_{\cal M} \partial_{\sigma} \phi_{\cal M}}\Big)^2 \Big] -  \nonumber \eeq
\beq - \frac{\tau^4}{2} (1 + sgn (\partial^{\mu} \phi \partial_{\mu} \phi)) (\max (\{\lambda_1, \lambda_2, \lambda_3, \lambda_4 \} \setminus (- \epsilon, \epsilon)) - \min (\{\lambda_1, \lambda_2, \lambda_3, \lambda_4 \} \setminus (- \epsilon, \epsilon) ))^2  \nonumber \eeq
One should note that the \emph{only} $0 (\tau^2)$ term is $\partial^{\mu} \phi \partial_{\mu} \phi$, and it comes \emph{without} $sgn$. This is similar to the result of previous section when $sgn$ canceled out after we performed a subtraction ${\cal L} = {\cal L}_t - {\cal L}_s$. However, in our present situation, $sgn$ still contributes to higher order terms, which we were ignoring in the previous section. 

\subsection*{5. First order Lagrangian density for electromagnetic field}

Let us now describe electromagnetic Lagrangian. Similarly to what happened with scalar field, we will need two Lagrangian generators: a "spacelike" and a "timelike" ones. Naturally, we will call the former "magnetic" and the latter "electric". And, again, similarly to scalar case, magnetic and electric Lagrangians will each be Lorentz covariant \emph{on their own}, but we will see some unwanted terms. These terms will be perfectly covariant and the only "problem" with them is a simple fact that they were never observed in the lab. Fortunately, they will end up canceling out once the two Lagrangians are added.

In scalar case, we have identified Lagrangian generator with $(\phi (s_2) - \phi (s_1))^2$. In the current situation, we will define it to be a \emph{four}-point function,  
\beq {\cal K}_S (s_1, s_2, s_3, s_4) = (a(s_1, s_2) + a(s_2, s_3) + a(s_3, s_4) + a(s_4, s_1))^2 \eeq
Now, if there is an embedding $f \colon S \rightarrow \cal M$, then this Lagrangian generator becomes 
\beq {\cal K}_S (s_1, s_2, s_3, s_4) = (a_S (s_1, s_2) + a_S (s_2, s_3) + a_S (s_3, s_4) + a_S (s_4, s_1))^2 = \nonumber \eeq
\beq = (a_{\cal M} (f(s_1), f(s_2)) + a_{\cal M} (f(s_2), f(s_3)) + a_{\cal M} (f(s_3), f(s_4)) + a_{\cal M} (f(s_4), f(s_1)))^2 =  \nonumber \eeq
\beq = \Big( \int_{f(s_1)}^{f(s_2)} A_{\mu} dx^{\mu} + \int_{f(s_2)}^{f(s_3)} A_{\mu} dx^{\mu} + \int_{f(s_3)}^{f(s_4)} A_{\mu} dx^{\mu} + \int_{f(s_4)}^{f(s_1)} A_{\mu} dx^{\mu} \Big)^2  = \eeq
\beq = \Big( \int_{{\rm Loop} (f(s_1), f(s_2), f(s_3), f(s_4)} A_{\mu} dx^{\mu} \Big)^2 = \Big(\int_{\rm Square (f(s_1, f(s_2), f(s_3), f(s_4)))} \vec{B} \cdot \vec{d \sigma} \Big)^2 = {\rm Flux}^2 \nonumber \eeq
where the "magnetic field" is taken in the reference frame in which $t$-axis is identified with the line passing through $f(p)$ and $f(q)$ and, as usual, we were able to use our knowledge of coordinate-based calculus (such as Stoke's theorem) because we were dealing with the space in which $f(s_k)$ are living in, as opposed to the space where $s_k$ do. 

Now, for any given Alexandrov set, we need to evaluate the \emph{maximum} of above expression. For our convenience, let us denote $f(s_1)$, $f(s_2)$, $f(s_3)$ and $f (s_4)$ by $A$, $B$, $C$ and $D$, respectively (in other words the lower-case letters, such as $s$, denote the elements of $S$, while the upper-case elements, such as $A$, denote the elements of $\cal M$). Thus, we are interested in positioning points $A$, $B$, $C$ and $D$ on the surface of equator of Alexandrov set in such a way that the flux through the contour $ABCD$ is maximized. Now, it is easy to see that the flux through the rectangle embedded in the sphere is less than or equal to the flux through the square embedded inside the circle, where we assume that the radius of the circle is the same as the one of the sphere (which, in our case, is $\tau$). Now, in case of embedding of rectangle inside the circle, 
\beq {\rm Area} (ABCD) = {\rm Area} (A0B) + {\rm Area} (B0C) + {\rm Area} (C0D) + {\rm Area} (D0A) \eeq
Furthermore, if we let $E$ be an intermediate point on the segment $AB$, then 
\beq {\rm Area} (A0B) = {\rm Area} (A0E) + {\rm Area} (B0E) = \frac{AE \times 0E}{2} + \frac{BE \times 0E}{2} \label{TwoTrianglesIsTriangle} \eeq
If we denote the angle $A0B$ by $\theta$, then 
\beq AE = BE= \tau \sin \frac{\theta}{2} \; , \; 0E  = \tau \cos \frac{\theta}{2} \eeq
By substituting this into Equation \ref{TwoTrianglesIsTriangle}, we obtain 
\beq {\rm Area} (A0B) = \tau^2 \sin \frac{\theta}{2} \cos \frac{\theta}{2} = \frac{\tau^2}{2} \sin \theta \eeq
Therefore, the area of the rectangle is 
\beq {\rm Area} (ABCD) = \frac{\tau^2}{2} (\sin A0B + \sin B0C + \sin C0D + \sin D0A) \eeq
Now, let us find the derivative of this area with respect to displacement of point $B$. In other words, we will replace $B$ with $B'$, and assume that the angle $B0B'$ is equal to $\epsilon$. Then 
\beq {\rm Area} (AB'CD) = \frac{\tau^2}{2} (\sin A0B' + \sin B'0C + \sin C0D + \sin D0A) = \nonumber \eeq
\beq = \frac{\tau^2}{2} (\sin (A0B + \epsilon) + \sin (B0C - \epsilon) + \sin C0D + \sin D0A) = \eeq
\beq = \frac{\tau^2}{2} (\sin A0B + \sin B0C + \sin C0D + \sin D0A + \epsilon (\cos A0B - \cos B0C) + 0 (\epsilon^2)) \nonumber \eeq
In light of the fact that $\epsilon$ can be both positive or negative, we can always pick a particular sign of $\epsilon$ that would increase the flux through the contour \emph{as long as} angles $A0B$ and $B0C$ are not equal to each other. This implies that \emph{at} the maximum $A0B = B0C$. By similar argument, we also know that $B0C=C0D$, $C0D=D0A$ and $D0A=A0B$. This means that every single angle is $\pi/2$. In such case, the area becomes 
\beq \max ({\rm Area}) = \frac{\tau^2}{2} \Big(\sin \frac{\pi}{2} + \sin \frac{\pi}{2} + \sin \frac{\pi}{2} + \sin \frac{\pi}{2} \Big) = 2 \tau^2 \eeq
This implies that the magnetic flux through the contour is $2 \tau^2 \vert \vec{B} \vert^2$. Now that we have done the maximization of the flux for a \emph{given} Alexandrov set, let us do the minimization over the collection of different Alexandrov sets. This amounts to the selection of reference frame in which $\vert \vec{B} \vert^2$ is minimized. We claim that such frame coincides with the frame in with $\vec{E}$ and $\vec{B}$ are parallel. This argument will consist of three parts:

a) Show that the frame where $\vec{E}$ and $\vec{B}$ are parallel exists to begin with

b) Show that in such frame the value of $\vert \vec{B} \vert^2$ is minimized

c) Find a covariant expression \emph{in arbitrary frame} that coincides with $\vert \vec{B} \vert^2$ in the frame specified above

Let us start with the proof of the existence of that frame. Let us start with arbitrary $\vec{E}$ and $\vec{B}$ and then rotate the frame in such a way that they end up being parallel. Without loss of generality, we can assume that $\vec{E}$-field is parallel to $x$-axis, while $\vec{B}$-field lies in $xy$-plane: 
\beq \vec{E} = (E_x, 0, 0) \; , \; \vec{B} = (B_x, B_y, 0) \eeq
When we make a Lorentz boost in $z$-direction, these fields transform according to 
\beq \vec{E}' = (\gamma (E_x - vB_y), \gamma v B_x, 0 ) \eeq
\beq \vec{B}' = (\gamma B_x, \gamma (B_y -v E_x), 0 ) \eeq
In order for $\vec{E}'$ and $\vec{B}'$ to be parallel, we need to have 
\beq \frac{E'_y}{E'_x} = \frac{B'_y}{B'_x} \eeq
or, in other words, 
\beq \frac{vB_x}{E_x - vB_y} = \frac{B_y - vE_x}{B_x} \eeq
This can be rewritten as a quadratic equation,
\beq (1 + v^2)  E_x B_y - v(E_x^2 + B_x^2+ B_y^2) = 0 \eeq
This equation solves to 
\beq v_1 = \frac{E_x^2 + B_x^2 + B_y^2 - \sqrt{(E_x^2 + B_x^2 +B_y^2)^2 - 4 E_x^2 B_y^2}}{2E_xB_y} \label{v1}\eeq
\beq v_2 = \frac{E_x^2 + B_x^2 + B_y^2 + \sqrt{(E_x^2 + B_x^2 +B_y^2)^2 - 4 E_x^2 B_y^2}}{2E_xB_y} \label{v2} \eeq
We now have to show that at least one of these solutions is physical. First of all, we note that neither of these two solutions have imaginary part.  After all, 
\beq (E_x - B_y)^2 \geq 0 \eeq
which implies that 
\beq 2E_x B_y \leq E_x^2 + B_y^2 \label{Inequality1}\eeq
and, therefore, 
\beq (E_x^2 + B_x^2 + B_y)^2 - 4E_x^2B_y^2 \geq (E_x^2 + B_y^2) - 4E_x^2 B_y^2 \geq 0 \label{NoImaginary} \eeq
This, however, is not enough. We also have to show that \emph{at least one} of these two velocities ($v_1$ or $v_2$) is between $-1$ and $1$. It turns out that $v_2$ does \emph{not} satisfy this condition. After all, 
\beq v_2 = \frac{E_x^2 + B_x^2 + B_y^2 + \sqrt{(E_x^2 + B_x^2 +B_y^2)^2 - 4 E_x^2 B_y^2}}{2E_xB_y} \geq \frac{E_x^2 + B_y^2}{2E_x B_y} \geq 1 \eeq
where Equation \ref{Inequality1} was used at the last step. Since $v_2$ does not fall between $-1$ and $1$, we \emph{must} show that $v_1$ does. In order to do it, we  first note that 
\beq \sqrt{a^2 - 2ab + b^2} = a-b \eeq
and, therefore
\beq \sqrt{a^2-b^2} \geq a-b \eeq
This, in particular, implies that 
\beq \sqrt{(E_x^2 + B_x^2 +B_y^2)^2 - 4 E_x^2 B_y^2}  \geq E_x^2 + B_x^2 +B_y^2 - 2E_xB_y \eeq
Therefore, 
\beq \frac{E_x^2 + B_x^2 + B_y^2 - \sqrt{(E_x^2 + B_x^2 +B_y^2)^2 - 4 E_x^2 B_y^2}}{2E_xB_y} \leq \nonumber \eeq
\beq \leq \frac{E_x^2 + B_x^2 + B_y^2 -E_x^2 - B_x^2 -B_y^2 +2E_xB_y }{2E_xB_y} = 1\eeq
Apart from showing that $v_1 \leq 1$ we also have to show that $v_1 \geq -1$. This part can be done by simply showing that $v_1 \geq 0$. Now, we have already shown that the expression under square root is positive (see Equation \ref{NoImaginary}). Furthermore, we trivially know that that expression is less than $(E_x^2 + B_x^2 + B_y^2)^2$. This, together with the fact that it is positive, implies that its \emph{absolute value} is smaller than $(E_x^2 + B_x^2 + B_y^2)^2$, as well. Thus, the value of square root is smaller than $E_x^2 + B_x^2 + B_y^2$. Since square root is being subtracted \emph{from} $E_x^2 + B_x^2 + B_y^2$, this shows that the result of subtraction is positive, thus implying that $v_1 \geq 0$. This, together with the previously shown result $v_1 \leq 1$ implies that $\vert v_1 \vert \leq 1$, which implies that $v_1$ is physical. 

Technically, we have to show that the \emph{strict inequality} $\vert v_1 \vert <1$ holds. This part is easy. By repeating the above arguments while inserting the assumptions $E_x \neq 0$ and $B_y \neq 0$  we would, in fact, obtain the strict inequality. If, on the other hand, we have $E_x =0$, then we have $\vec{E} =0$, which means that $\vec{E}$ and $\vec{B}$ are \emph{already} "parallel" and no boost is needed to begin with. If, instead, $B_y =0$ then, again they are parallel: both point along $x$-axis. Thus, again, no further boost is needed. In both of these cases, however, the Equation \ref{v1} would produce $0/0$. Thus, they would be "parallel" strictly due to qualitative argument we have just presented. If, on the other hand, both of the numbers are non-zero then the strict inequality will begin to hold and, therefore, we would be able to find $v_1 < 1$ that would make them parallel. 

Now that we have proven the existence of the frame where $\vec{E}$ and $\vec{B}$ are parallel, let us show that in this frame the value of $\vert \vec{B} \vert^2$ is minimized (see part "b" in the last outline). Let us \emph{start} from the frame where $\vec{E}$ and $\vec{B}$ are parallel and see how they transform. We can assume that their common direction is $x$. Without loss of generality, we can assume that we are doing boost along $z$-axis, which leads to the transformation 
\beq \vec{E}' = (\gamma E_x , \gamma v B_x, 0 ) \eeq
\beq \vec{B}' = (\gamma B_x, - \gamma v E_x, 0 ) \eeq
This implies that in the new frame 
\beq \vert \vec{B}' \vert = \gamma \sqrt{B_x^2 +v^2E_x^2} \geq \gamma B_x \geq B_x \eeq
Since in the original frame we have 
\beq \vert \vec{B} \vert = B_x \eeq
we have shown that 
\beq \vert \vec{B}' \vert \geq \vert \vec{B} \vert \eeq
Thus, in the frame where $\vec{E}$ and $\vec{B}$ are parallel the flux is, in fact, minimized. 

Let us now arrive at a covariant expression for the above. On first glance, it \emph{seems} like it is "impossible" due to $\vert \vec{B} \vert^2$ not being covariant. However, in light of the fact that the above computation was done in \emph{one specific} frame (namely, the one where $\vec{E}$ and $\vec{B}$ are parallel), it is logically possible that the "minimal flux" corresponds to covariant expression \emph{in all frames}, but that expression "matches" $\vert \vec{B} \vert^2$ \emph{only} in the frame where $\vec{E}$ and $\vec{B}$ are parallel. Let us, therefore, find a covariant expressions which $\vert \vec{E} \vert$ and $\vert \vec{B} \vert $ match in the above said frame. In order to agree with convention used in the rest of the paper, let us change coordinates and assume that the common direction $\vec{E}$ and $\vec{B}$ point to is $z$ rather than $x$. Thus, the only non-zero components of $F_{\mu \nu}$ are $F_{03}$ and $F_{12}$. Therefore, in that frame, 
\beq \vert \vec{E} \vert \vert \vec{B} \vert = F_{03} F_{12} = \frac{1}{8} \epsilon_{\alpha \beta \gamma \delta} F^{\alpha \beta} F^{\gamma \delta} \eeq
\beq \vert \vec{B} \vert^2 - \vert \vec{E} \vert^2 = F^{03} F_{03} + F^{12} F_{12} = F^{\alpha \beta} F_{\alpha \beta} \eeq
The linear combinations of the above two equations lead us to 
\beq (\vert \vec{B} \vert - \vert \vec{E} \vert)^2 = F^{\alpha \beta} F_{\alpha \beta} - \frac{1}{4} \epsilon_{\alpha \beta \gamma \delta} F^{\alpha \beta} F^{\gamma \delta} \eeq 
\beq (\vert \vec{B} \vert + \vert \vec{E} \vert)^2 = F^{\alpha \beta} F_{\alpha \beta} + \frac{1}{4} \epsilon_{\alpha \beta \gamma \delta} F^{\alpha \beta} F^{\gamma \delta} \eeq 
If we assume that $\vert \vec{B} \vert > \vert \vec{E} \vert$, then we obtain
\beq \vert \vec{B} \vert - \vert \vec{E} \vert  = \sqrt{F^{\alpha \beta} F_{\alpha \beta} - \frac{1}{4} \epsilon_{\alpha \beta \gamma \delta} F^{\alpha \beta} F^{\gamma \delta}} \eeq 
\beq \vert \vec{B} \vert + \vert \vec{E} \vert =\sqrt{F^{\alpha \beta} F_{\alpha \beta} + \frac{1}{4} \epsilon_{\alpha \beta \gamma \delta} F^{\alpha \beta} F^{\gamma \delta}} \eeq 
which implies that 
\beq \vert \vec{B} \vert = \frac{1}{2} \Bigg(\sqrt{F^{\alpha \beta} F_{\alpha \beta} + \frac{1}{4} \epsilon_{\alpha \beta \gamma \delta} F^{\alpha \beta} F^{\gamma \delta}} + \sqrt{F^{\alpha \beta} F_{\alpha \beta} - \frac{1}{4} \epsilon_{\alpha \beta \gamma \delta} F^{\alpha \beta} F^{\gamma \delta}}  \Bigg) \eeq
\beq \vert \vec{E} \vert = \frac{1}{2} \Bigg(\sqrt{F^{\alpha \beta} F_{\alpha \beta} + \frac{1}{4} \epsilon_{\alpha \beta \gamma \delta} F^{\alpha \beta} F^{\gamma \delta}} - \sqrt{F^{\alpha \beta} F_{\alpha \beta} - \frac{1}{4} \epsilon_{\alpha \beta \gamma \delta} F^{\alpha \beta} F^{\gamma \delta}}  \Bigg) \eeq
If, on the other hand, we assume that $\vert \vec{B} \vert < \vert \vec{E} \vert$, then we obtain
\beq \vert \vec{E} \vert - \vert \vec{B} \vert  = \sqrt{F^{\alpha \beta} F_{\alpha \beta} - \frac{1}{4} \epsilon_{\alpha \beta \gamma \delta} F^{\alpha \beta} F^{\gamma \delta}} \eeq 
\beq \vert \vec{E} \vert + \vert \vec{B} \vert =\sqrt{F^{\alpha \beta} F_{\alpha \beta} + \frac{1}{4} \epsilon_{\alpha \beta \gamma \delta} F^{\alpha \beta} F^{\gamma \delta}} \eeq 
which implies that 
\beq \vert \vec{B} \vert = \frac{1}{2} \Bigg(\sqrt{F^{\alpha \beta} F_{\alpha \beta} + \frac{1}{4} \epsilon_{\alpha \beta \gamma \delta} F^{\alpha \beta} F^{\gamma \delta}} - \sqrt{F^{\alpha \beta} F_{\alpha \beta} - \frac{1}{4} \epsilon_{\alpha \beta \gamma \delta} F^{\alpha \beta} F^{\gamma \delta}}  \Bigg) \eeq
\beq \vert \vec{E} \vert = \frac{1}{2} \Bigg(\sqrt{F^{\alpha \beta} F_{\alpha \beta} + \frac{1}{4} \epsilon_{\alpha \beta \gamma \delta} F^{\alpha \beta} F^{\gamma \delta}} + \sqrt{F^{\alpha \beta} F_{\alpha \beta} - \frac{1}{4} \epsilon_{\alpha \beta \gamma \delta} F^{\alpha \beta} F^{\gamma \delta}}  \Bigg) \eeq
Now in order to neatly write down these two cases in one equation, we introduce "signum" function defined as 
\beq sgn (x) = \left\{
	\begin{array}{ll}
		-1 & {\rm If} \; x < 0 \\
                         0 & {\rm if} \; x=0 \\
		+1 &  {\rm if} \; x>0
	\end{array}
\right.
\label{Signum} \eeq
Thus, the above equations would have to use $sgn (\vert \vec{B} \vert - \vert \vec{E}\vert$. But, of course, we do not want any references to $\vec{B}$ and $\vec{E}$ on the right hand side. What comes to our rescue is the fact that 
\beq F^{\mu \nu} F_{\mu \nu} = \vert \vec{B} \vert^2 - \vert \vec{E} \vert^2 \eeq
This means that 
\beq sgn (\vert \vec{B} \vert - \vert \vec{E} \vert) = sgn (F^{\mu \nu} F_{\mu \nu}) \eeq
We can, therefore, use the "covariant" expression $sgn (F^{\mu \nu} F_{\mu \nu}$ in order to summarize the "case by case" situation we are dealing with:
\beq \vert \vec{B} \vert = \frac{1}{2} \Bigg(\sqrt{F^{\alpha \beta} F_{\alpha \beta} + \frac{1}{4} \epsilon_{\alpha \beta \gamma \delta} F^{\alpha \beta} F^{\gamma \delta}} + (sgn(F^{\mu \nu}F_{\mu \nu})) \sqrt{F^{\alpha \beta} F_{\alpha \beta} - \frac{1}{4} \epsilon_{\alpha \beta \gamma \delta} F^{\alpha \beta} F^{\gamma \delta}}  \Bigg) \eeq
\beq \vert \vec{E} \vert = \frac{1}{2} \Bigg(\sqrt{F^{\alpha \beta} F_{\alpha \beta} + \frac{1}{4} \epsilon_{\alpha \beta \gamma \delta} F^{\alpha \beta} F^{\gamma \delta}} -(sgn(F^{\mu \nu}F_{\mu \nu})) \sqrt{F^{\alpha \beta} F_{\alpha \beta} - \frac{1}{4} \epsilon_{\alpha \beta \gamma \delta} F^{\alpha \beta} F^{\gamma \delta}}  \Bigg) \eeq
Squaring the above two equations implies
\beq \vert \vec{B} \vert^2 = \frac{1}{2} \Bigg(F^{\alpha \beta} F_{\alpha \beta} + (sgn(F^{\mu \nu}F_{\mu \nu})) \sqrt{(F^{\alpha \beta} F_{\alpha \beta})^2 - \frac{1}{16} (\epsilon_{\alpha \beta \gamma \delta} F^{\alpha \beta} F^{\gamma \delta})^2} \Bigg) \label{BisCovariant} \eeq
\beq \vert \vec{E} \vert^2 = \frac{1}{2} \Bigg(F^{\alpha \beta} F_{\alpha \beta} - (sgn(F^{\mu \nu}F_{\mu \nu})) \sqrt{(F^{\alpha \beta} F_{\alpha \beta})^2 - \frac{1}{16} (\epsilon_{\alpha \beta \gamma \delta} F^{\alpha \beta} F^{\gamma \delta})^2} \Bigg) \label{EisCovariant} \eeq
Now, as we recall, we were selecting the four points lying in the equator in our definition for Lagrangian generator. This resulted in $\vert \vec{B} \vert^2$ being the \emph{only} contributing term. As we see from the equation \ref{BisCovariant}, the absence of $\vec{E}$ does not stop us from still obtaining the relativistically covariant result. The only "problem" we encounter is the presence of $\epsilon_{\alpha \beta \gamma \delta} F^{\alpha \beta} F^{\gamma \delta}$, which is not seen in the lab. This term, however, is still covariant! This is, in fact, what one would expect. After all, our construction did not appeal to any coordinate system; thus, the result has to be covariant no matter what. 

Now, in order to get rid of $\epsilon_{\alpha \beta \gamma \delta} F^{\alpha \beta} F^{\gamma \delta}$ we introduce a \emph{separate} Lagrangian which has $\vert \vec{E} \vert^2$ \emph{alone}. As quick inspection of above equation shows, this will introduce $\epsilon_{\alpha \beta \gamma \delta} F^{\alpha \beta} F^{\gamma \delta}$ with an opposite sign, allowing for cancellation. Now, the way we introduce the $\vert \vec{E} \vert^2$-Lagrangian is by again considering the four-point sets, with the same four-point Lagrangian generator, while modifying the criteria by which we select these sets. In "magnetic" case we were constraining all four points to the equator. In the "electric" case, on the other hand, 

\subsection*{6 Corrections to magnetic Lagrangian}

Once again, we are considering a contour based on $s_1$, $s_2$, $s_3$ and $s_4$, where it is assumed that all four of these points are lying on the "equator" of $\alpha (p, q)$, where it is assumed that the point $x$ is a "midpoint" of a geodesic connecting $p$ and $q$:
\beq \tau (p, r) = \tau (r, q) = \frac{\tau (p,q)}{2} = \tau \; ; \; p \prec^* s_{1, 2, 3, 4} \prec^* q \eeq
Since we are computing pre-Lagrangian as opposed to actual Lagrangian, we are assuming that $p$ and $q$ are "fixed". Our goal is to select $s_1$, $s_2$, $s_3$ and $s_4$ in such a way that the value of 
\beq {\cal K}_B = a (s_1, s_2) + a (s_2, s_3) + a (s_3, s_4) + a (s_4, s_1) \eeq
is maximized with respect to the above constraints. Now, from section 6 we already have an "approximate" idea of where $s_1$, $s_2$, $s_3$ and $s_4$ are positioned. Our present task is to compute deviations from that configuration due to the higher order derivatives. In other words, we would like to set 
\beq s'_1 = s_1 + \delta s_1 \; ; \; s'_2 = s_2 + \delta s_2 \; ; \; s'_3 = s_3 + \delta s_3 \; ; \; s'_4 = s_4 + \delta s_4 \eeq
and then maximize it with respect to $\delta s_1$, $\delta s_2$, $\delta s_3$ and $\delta s_4$ by using the usual derivative techniques. 

For the purposes of this paper, we are only interested in lowest order correction. Let us, therefore, take a qualitative look and see which corrections we should take into account and which we should ignore. The area of the square loop is $O(\tau^2)$ which means that the flux of magnetic field through that loop is $O(\tau^2)$ as well. The displacement of our "points" has to be "small" with respect to the "size" of the Alexandrov set, $2 \tau$. Therefore, their displacement is likewise of $O (\tau^2)$. Thus, the "correction" to the flux of the magnetic field is of $O (\tau^4)$. Now, the Lagranigan generator is equal to \emph{square} of the magnetic flux. Therefore, the "original" Lagrangian generator is $(O(\tau^2))^2 = O (\tau^4)$. On the other hand, the "corrected" flux is $(O(\tau^2) + O (\tau^4))^2 = O (\tau^4) + O (\tau^6) + O (\tau^8)$. We will, therefore, "throw away" $O (\tau^8)$ term and "look for" the $O(\tau^6)$ corrections. In other words, we are seeking expression of the form $O (\tau^4) + O (\tau^6)$. 

Once again, we will select a coordinate system so that its "origin" coincides with point $r$, and $t$ axis passes through $p$ and $q$. Thus, $x$, $y$ and $z$ axes all cross the equator. In light of the fact that we assume that spacetime is flat, and also since our constraint $p \prec^* s_{1, 2, 3, 4} \prec^* q$ is highly restrictive, we know that the $t$ coordinate of these four points is \emph{exactly} zero. While the points are going to undergo small "shifts", these "shifts" will be constrained to a sphere of radius $\tau$ on $xyz$-plane. Now, since "electric field" is identified with $F_{0k}$ this immediately implies that the effect of the latter will continue to be \emph{exactly} zero. This allows us to focus exclusively on the magnetic field. Now, we will identify z-axis with the direction of the magnetic field at the origin: 
\beq \vec{B} (0) = B \hat{z} \eeq
At the same time, away from the origin, the values of $B_x$ and $B_y$ can become non-zero. However, in light of the fact that we are only concerned about the interior of the Alexandrov set, their values are of  $O (\tau)$:
\beq B_x = 0 (\tau) \; ; \; B_y = 0 (\tau) \eeq
Let us now look at the contributions of linear and quadratic terms. From antisymmetry it is clear, if we attempt to "integrate" $B_x$ and $B_y$ then the linear terms will drop out \emph{unless} we shift the contour of integration slightly. If we do shift the contour, then the effect of the first order terms will be multiplied by the \emph{modification} of area due to the shift. Since the area, itself, is of $0 (\tau^2)$, its modification is of $0 (\tau^3)$. Furthermore, since the size of Alexandrov set is of $0 (\tau)$, the linear effect on its own is of $0 (\tau)$. Thus, together, the linear effect becomes of $0 (\tau) \times 0 (\tau^3) = 0 (\tau^4)$. The quadratic term, on the other hand, is of $0 (\tau^2)$; thus it "starts out" smaller than linear term. At the same time, however, the quadratic term is no longer an odd function, thus we do \emph{not} need to shift the contour in order for its integral to be non-zero. Thus, we are multiplying the quadratic term by the \emph{unperturbed} area $0 (\tau^2)$ (as opposed to $0 (\tau^3)$ perturbation we were multiplying first order term by). Therefore, the "integral" for second order term will end up being $0 (\tau^4)$, just like it was for the first order! Thus, the "total" integral is, likewise, $0 (\tau^4)$:
\beq \int dA \; B_x = 0 (\tau^4) \; ; \; \int dA \; B_y = 0 (\tau^4) \eeq
Now, the above integrals are intentionally taken over "scalar" area element $dA$, and the integrands are "real valued" functions $B_x$ and $B_y$. Let us now replace $dA$ with $d \vec{s}$ and $B_x$ and $B_y$ with $B_x \hat{x}$ and $B_y \hat{y}$, respectively. If the contour was lying on $xy$-plane that integral would clearly be zero. Therefore, the value of that integral can be assumed to be proportional to the angle by which $s'_1$, $s'_2$, $s'_3$ and $s'_4$ have "shifted":
\beq \int d \vec{s} \cdot (B_x \hat{x} + B_y \hat{y}) = 0 \Big( \int dA B_x \Big) 0 (\angle s'_1 0 s_1) \eeq
In the above expression $B_x$ is not "better" than $B_y$, nor is $\angle s'_1 0 s_1$ any more "important" than $\angle s'_2 0 s_2$ either. We simply assumed that the "order of magnitude" won't change upon these replacements, which is why we "picked" arbitrary "examples" of the "orders of magnitude" we are looking for. Now, since points are only allowed to "shift" by "small" amounts, the angle by which the contour tilts is also of $O (\tau)$:
\beq \angle s_1 0 s'_1 = O (\tau) \; ; \; \angle s_2 O s'_2 = O (\tau) \; ; \; \angle s_3 0 s'_3 = 0 (\tau) \; ; \; \angle s_4 0 s'_4 = 0 (\tau) \eeq 
This means that the "projection" of $B_x \hat{x} + B_y \hat{y}$ onto a contour defined by $s'_1$, $s'_2$, $s'_3$ and $s'_4$ is of $0 (\tau^5)$: 
\beq \int d \vec{s} \cdot (B_x \hat{x} + B_y \hat{y}) = 0 (\tau^5) \eeq
Since our calculation is only up to $0 (\tau^4)$, this means that we can safely "throw away" $B_x$ and $B_y$ and assume that the only non-zero component of $B$ is $B_z$:
\beq B_x (t, x, y, z)= B_y (t, x, y, z) = 0 \eeq
In light of this, we will identify $B$ with $B_z$ which will allow us to "drop" index $z$:
\beq B (t, x, y, z) = B_z (t, x, y, z) \eeq
which is not to be confused with a "vector"
\beq \vec{B} (t, x, y, z) = \hat{z} B (t, x, y, z) \eeq
However, as we will soon see, $x$ and $y$ derivatives \emph{of} $B_z$ \emph{do} lead to non-negligible contribution and, therefore, we do \emph{not} ignore the latter:  
\beq \frac{\partial B_z}{\partial x} \neq 0 \; ; \; \frac{\partial B_z}{\partial y} \neq 0 \; ; \; \frac{\partial B_z}{\partial z} \neq 0 \eeq
At the same time,  since contribution of each of the above is $0 (\tau^6)$, their "products" are, in fact, negligible. This allows us to take each of the above derivatives "one at a time" and any given time assume that the derivative we are looking at is non-zero while the other two derivatives are zero.

Let us start with $\partial B_z / \partial z$. It is easy to see that if points are trying to "maximize" the flux, they will shift "upward" if $\partial B_z / \partial z$ is positive and "downward" if $\partial B_z / \partial z$ is negative. Furthermore, since we are \emph{not} neglecting $\partial B_z / \partial z$, we \emph{are} neglecting $\partial B_z / \partial x$ and $\partial B_z / \partial y$ (and we are planning to count these separately, later on). Thus, we have a symmetry around $z$ axis, which implies that the points "shift" vertically by the same amount. Let us assume that the "common" amount by which the points shift is $h$. Again, because of the symmetry around $z$-axis, we can assume that the points will \emph{not} shift horizontally (the only exception is a slight shift towards the $z$-axis in order make sure that the points "stay" on the sphere). Thus, they continue to form the square; but the length of the sides of the square changes from $a$ to $a'$ where
\beq a^2 = 2 \tau^2 \; ; \; a'^2 = 2 (\tau^2 - h^2) \eeq
Now the area of the "new" square is $a'^2$. This means that the flux through that new square is $B' a'^2$ where 
\beq B' =  B (0) + h \frac{\partial B}{\partial z} \Big\vert_0 + 0 (h^2) \eeq
Now, since $h$ is "small" compared to the size of the Alexandrov set, we know that 
\beq h = 0 (\tau^2) \eeq
This means that 
\beq 0 (h^2) = 0 (\tau^4) \eeq
Now, since the "flux" takes place through the area of $0 (\tau^2)$, the effect on flux is 
\beq \delta B = 0 (h^2) = 0 (\tau^4) \Longrightarrow \delta ({\rm flux}) = 0 (\tau^2) \times 0 (\tau^4) = 0 (\tau^6) \eeq
which makes it negligible. Thus, we have established that we can throw away the $0(h^2)$ terms and just use linear expression for $B$:
\beq B'= B(0) + h \frac{\partial B}{\partial z} \Big\vert_0 \eeq
Now, as we stated earlier, the "flux" through the shifted square is $B'a'^2$. By substituting the expressions for $B'$ and $a'$ we obtain
\beq {\rm Flux} = a'^2 B'^2 = 2 (\tau^2 - h^2) \Big( B(0) + h \frac{\partial B}{\partial h} \Big\vert_0 \Big) \eeq
Now, our previous reason for "throwing away" $0(h^2)$ terms was the fact that we were \emph{planning} to multiply them by the area which is $0(\tau^2)$. In the above expression, however, we have \emph{already} performed a multiplication by the area. Thus, no further multiplication will be performed, which means that we will \emph{keep} $0(h^2)$ terms, seeing that $0(h^2)= 0 (\tau^4)$ and we do \emph{not} neglect $0(\tau^4)$ terms. At the same time, since $0(\tau^4) = 0(h^2)$ is the highest order that we keep, we will throw away $0(h^3)$ term. Thus, we obtain
\beq {\rm Flux} = 2 \Big(\tau^2 B(0) + \tau^2 h \frac{\partial B}{\partial z} \Big\vert_0 - h^2 B(0) \Big) \eeq
Thus, in order to maximize flux, we need to find a value of $h$ such that 
\beq \frac{d}{dh} \Big(\tau^2 B(0) + \tau^2 h \frac{\partial B}{\partial z} \Big\vert_0 - h^2 B(0) \Big) = 0 \eeq
This implies that 
\beq h = \frac{\tau^2}{2B} \frac{\partial B}{\partial z} \Big\vert_0 \eeq
The fact that $h$ has the same sign as $\partial B/ \partial z$ implies that the points will be "shifted" in the direction of gradient of $B$. Indeed, this is what we intuitively expect would maximize the flux. If we now substitute the above value of $h$ into the expression for flux, we obtain 
\beq {\rm Flux} = 2B_0 \tau^2 + \frac{\tau^4}{B} \Big( \frac{\partial B}{\partial z} \Big\vert_z \Big)^2 \label{vertical} \eeq
As a result of the squaring, the increment of flux is positive, regardless of sign of $\partial B/ \partial z$. This is indeed what we expect. Both positive \emph{and} negative value of $\partial B/ \partial t$ provide extra opportunity for us to bring flux closer to what we "want" it to be. Thus, as long as we "want" to make it larger, both positive and negative values of $\partial B/ \partial z$ will do just that. 

The above calculation is made under the assumption that the only non-zero derivative is $\partial B/ \partial z$. But, as we stated previously, both $\partial B/ \partial x$ and $\partial B/ \partial y$ \emph{do} have non-negligible effect. We simply decided to compute the effects of the derivatives separately. So let us now move on to finding an effect of $\partial B/ \partial x$. Thus, we will now assume that $\partial B/ \partial z$ is equal to zero and allow both $\partial B/ \partial x$ and $\partial B/ \partial y$ to be non-zero (but, at the same time, $B$ is still parallel to $z$-axis, as always). In light of the symmetry between $x$ and $y$, we do not need to separate the effects of $\partial B / \partial x$ and $\partial B / \partial y$. Thus, we can assume that both of these derivatives are non-zero, and while $\partial B / \partial z$ is the only derivative that \emph{is} zero. 

In light of the non-zero values $\partial B / \partial x$ and $\partial B/ \partial y$, it is possible that the flux will change as we are "rotating" the "square" (with our points sitting in the corner). While the shift "away from equator" had to be "small", the rotation on the $xy$-plane can be anything from $0$ to $2 \pi$. Thus, we would like to \emph{first} make a "large" rotation of the square to "appropriate" position and only \emph{after} that make a small "shift" away from that position. The rotation will be "rigid", whereas the "shift" will involve stretching and compression of the sides of the square; that is the reason why the latter can not be absorbed into the former. 

We noticed that the only two axes that we have defined ahead of time were $t$ and $z$. In particular, $t$-axis coincides with the line passing through $p$ and $q$, while $z$-axis coincides with the direction of $\vec{B} (0)$. Thus, we are still free to select the directions of $x$ and $y$ whatever way we like. For our convenience, we will choose $x$ and $y$ in such a way that they are parallel to the edges of the square. Thus, in the $(t, x, y, z)$-notation, the coordinates of the four points are given by 
\beq s_1 = (0, - \tau, - \tau, 0) \; , \; s_2 = (0, \tau, - \tau, 0) \; , \; s_3 = (0, \tau, \tau, 0) \; , \; s_4 = (0, - \tau, \tau, 0) \eeq
Thus, \emph{instead of} rotating square, we will be "rotating" $\vec{B}$:
\beq \vec{B}' (t, x, y, z) = \vec{B} (t, x \cos \theta + y \sin \theta , - x \sin \theta + y \cos \theta, z) \eeq
Our goal is to select $\theta$ in such a way that the flux of "rotating" $\vec{B}'$ through the "fixed" square is maximized. Now, the flux through the square is given by 
\beq {\rm Flux} = \int_{-\tau}^{\tau} dx \int_{- \tau}^{\tau} dy \Big(B (0) + x \frac{\partial B}{\partial x} \Big\vert_0 + y \frac{\partial B}{\partial y} \Big\vert_0 + \nonumber \eeq
\beq + \frac{x^2}{2}  \frac{\partial^2 B}{\partial x^2} \Big\vert_0 + \frac{y^2}{2}  \frac{\partial^2 B}{\partial y^2} \Big\vert_0 + xy \frac{\partial^2 B}{\partial x \partial y} \Big\vert_0 \Big) + 0 (\tau^5) \eeq
Clearly, the terms proportional to $x$, $y$ and $xy$ are odd and, therefore, integrate to zero. Thus, our integral becomes
\beq {\rm Flux} = \int_{-\tau}^{\tau} dx \int_{- \tau}^{\tau} dy \Big(B (0)  + \frac{x^2}{2}  \frac{\partial^2 B}{\partial x^2} \Big\vert_0 + \frac{y^2}{2}  \frac{\partial^2 B}{\partial y^2} \Big\vert_0  \Big) + 0 (\tau^6) \eeq
where we have replaced $0 (\tau^5)$ with $0 (\tau^6)$ because the $0(\tau^5)$ terms are, likewise, odd and therefore also integrate to zero (but, of course, this point is simply aesthetic given that we ignore both $0(\tau^5)$ and $0(\tau^6)$). Now the above integral evaluates to
\beq {\rm Flux} = 2 \tau^2 B(0) + \frac{2}{3} \tau^4 \Big( \frac{\partial^2 B}{\partial x^2} \Big\vert_0 + \frac{\partial^2 B}{\partial y^2} \Big\vert_0 \Big) + 0 (\tau^6) \label{noshift} \eeq
The above is invariant under the rotation of $B$. However, the "very small" terms that we "neglected", \emph{regardless of how small they might be} would still result in a "finite" effect when it comes to $B$. For example, if the "small" term has the form $\tau^{100} \sin \theta$, it would still "select" $\theta= \pi/ 2$, \emph{despite} the "smallness" of $\tau^{100}$. The only thing that can "stay in the way" is a "larger" term, such as $\theta^{80}$. This, however, does not change the fact that the "largest" term, whatever that term might be, \emph{would} in fact be a determining factor of $\theta$. In the past that issue was avoided because the "largest" term happened to be $0(\tau^4)$. In the present situation, since things are symmetric up to $0(\tau^4)$, the higher order terms \emph{will} "have a final say" on what $\theta$ would be.  

The good news, however, is that the "final" effect on $\theta$ will be "unimportant". After all, the only reason we are interested in $\theta$ is that we are interested in flux. Now, from the above equation $\theta$ will have an $0(\tau^6)$ effect on flux, which means we don't care about it. In our subsequent calculations, we will pretend we will know $\theta$. However, our calculations, up to $0 (\tau^4)$, will be identical regardless of the actual value of $\theta$. Therefore, we don't have to actually "find out" what $\theta$ is, which means that we don't have to look at $0 (\tau^6)$ terms either. In fact, this pattern can be generalized to other situations. The only way "something" is "important" is when it produces $0 (\tau^4)$ terms. If such is the case, then that "something" won't be influenced by $0 (\tau^6)$ ones. On the other hand, if that "something" doesn't produce $0 (\tau^4)$ terms then it "begins" to be influenced by $0 (\tau^6)$; but then the lack of $0 (\tau^4)$ makes that "something" unimportant. Thus, $0 (\tau^6)$ terms "either" have very small influence on something "important", or they have large influence on something "unimportant". In both cases they can be neglected, just for different reasons. 

To make long story short, we have answered the question regarding "rigid rotation" of the square. In particular, yes the square will be rotated, but it is not important to find out how; so we won't. Now it is time to consider the small oscillation from its position, whatever it might be. As was stated earlier, small oscillations can not be "absorbed" into rigid rotation since they involve stretching and shrinking of sides of the square. For that same reason it turns out that the "stretching and shrinking" we are about to consider \emph{will} have the $0(\tau^4)$ effect that we are looking for, despite the fact that rigid rotation does not. Now, the displacement of each point is of $0(\tau^2)$ which means that its effect \emph{can't be larger} than $0(\tau^4)$. Therefore, the "interaction" \emph{between} different displacements will be of $0 (\tau^6)$ and can be neglected. Thus, we will consider the displacement of each point separately and then take their superposition.

As was stated earlier, the only axes that are rigidly fixed are $t$ and $z$. On the other hand, the choice of $x$ and $y$ is up to our convenience. Let us, therefore, change our choice of $xy$ coordinates and this time select them in such a way that the coordinate axes pass through the four points on the "rigid" square. Thus, in the $(t, x, y, z)$ notation,
\beq s_1 = (0, \tau, 0, 0) \; , \; s_2 = (0, 0, \tau, 0) \; , \; s_3 = (0, - \tau, 0, 0) \; , \; s_4 = (0, 0, - \tau, 0) \eeq
Now, we will consider the effect of "displacement" of $s_1$, while we will leave other points fixed. Thus, the "displacement" takes place from $s_1$ to $s'_1$, where
\beq s_1 = (0, \tau, 0, 0) \; , \; s'_1 = (0, \tau \cos \theta  ,  \tau \sin \theta, 0) \eeq
For our convenience, we will denote $\tau \sin \theta$ by $h$:
\beq h = \tau \sin \theta \eeq 
The integral over the "square" $s_1' s_2 s_3 s_4$ can be represented as a sum of the integral over a triangles $s_4 s_1' s_2$ and $s_2 s_3 s_4$. Since the only "change" we are making is going from $s_1$ to $s'_1$, the triangle $s_2 s_3 s_4$ is left unchanged. Thus, we can exclusively focus on $s'_1 s_2 s_3$. Now, as we just said, we can assume that $s'_1 s_2 s_3$ is a triangle rather than a circle. We now notice that 
\beq s'_1 s_2 s_3 = (s_1 s_2 s_3 \cup s_2 s_1 s'_1) \setminus s_4 s_1 s'_1 \eeq
This means that the "correction" to our integral takes the form
\beq \int_{s_4 s_2 s'_1} B dA - \int_{s_4 s_2 s_1} B dA = \int_{s_2 s_1 s'_1} B dA - \int_{s_4 s_1 s'_1} B dA \label{triangles} \eeq
We know that the area of the square is $0 (\tau^2)$. Therefore, the area of the two small strips we are integrating over is $0 (\tau^3)$. Thus, the terms of $0 (\tau)$ under the integral become $0 (\tau^4)$ once the integral is evaluated. Therefore, $0 (\tau)$ is the highest power we leave under the integral. This means that we can expand $B$ only up to linear terms:
\beq B = B(0) + \frac{\partial B}{\partial x} \Big\vert_0 x + \frac{\partial B}{\partial y} \Big\vert_0 y \eeq
Now, in light of the fact that $B(0)$ is "large" while linear terms are $0 (\tau)$, we will need different kinds of approximations to compute them. Thus, it would be best to compute them separately. As far as $B (0)$ integral is concerned, we simply have to look at the variation of area. It is easy to see that
\beq {\rm Area} (s_4 s_2 s_1) = \tau^2 \; ; \; {\rm Area} (s_4, s_2 s_1') = \tau^2 \cos \theta \eeq
This immediately implies that 
\beq \int_{s_4 s_2 s'_1} B (0,0) dA - \int_{s_4 s_2 s_1} B (0,0) dA = - \tau^2 (1 - \cos \theta ) \eeq
Let us now do the linear terms. From what we have just seen, the deviation between $1$ and $\cos \theta$ leads to $0 (\tau^4)$ terms when it comes to something "finite", such as $B(0,0)$. Now, in case of linear terms, we have extra $0 (\tau)$. This means that the difference between $1$ and $\cos \theta$ will now produce $0 (\tau^5)$ effect, which can be thrown away. Thus, we will replace $\cos \theta$ with $1$ by saying
\beq s_1' = (0, \tau, \tau \sin \theta, 0) \eeq
Now, the line connecting $s_2$ and $s_1$ is $t = \tau-x$. The line connecting $s_2$ and $s'_1$ is $t= \tau - x(\tau -h)/\tau$. The line connecting $s_4$ and $s_1$ is $t = x - \tau$, and the line connecting $s_4$ to $s'_1$ is $t= x(\tau+h)/\tau - \tau$. Thus, according to the Equation \ref{triangles}, the correction becomes 
\beq \int_{s_2 s_1 s'_1} (B (x,y)- B (0,0)) dA - \int_{s_4 s_1 s'_1} (B (x,y)- B (0,0))  dA  = \eeq
\beq = \int_0^{\tau} dx \int_{\tau- x}^{\tau- \frac{\tau-h}{\tau} x} dy \Big(\frac{\partial B}{\partial x} \Big\vert_0 x + \frac{\partial B}{\partial y} \Big\vert_0 y \Big) - \int_0^{\tau} dx \int_{x-\tau}^{\frac{\tau+h}{\tau}x - \tau} dy \Big( \frac{\partial B}{\partial x} \Big\vert_0 x + \frac{\partial B}{\partial y} \Big\vert_0 y \Big). \nonumber \eeq
When we compute that integral on the right hand side, we obtain 
\beq \int_{s_2 s_1 s'_1} (B (x,y)-B (0,0)) dA - \int_{s_4 s_1 s'_1} (B(x,y) - B(0,0)) dA  = \frac{h\tau}{3} \frac{\partial B}{\partial y} \Big\vert_0 (\tau - h) \eeq
And, if we apply Equation \ref{triangles} to the left hand side, we obtain 
\beq \int_{s_4 s_2 s'_1} (B(x,y) - B(0,0)) dA - \int_{s_4 s_2 s_1} (B(x,y) - B(0,0)) dA  = \frac{h\tau}{3} \frac{\partial B}{\partial y} \Big\vert_0 (\tau - h) \eeq
Now, by substituting 
\beq h = \tau \sin \theta \eeq
we obtain 
\beq \int_{s_4 s_2 s'_1} (B(x,y) - B(0,0)) dA - \int_{s_4 s_2 s_1} (B(x,y)-B(0,0)) dA  =  \frac{\tau^3}{3} \frac{\partial B}{\partial y} \Big\vert_0 (\sin \theta - \sin^2 \theta) \eeq
Now, we will add the integrals over $B(0,0)$ and over $B(x,y)-B(0,0)$ to obtain
\beq \int_{s_4 s_2 s'_1} B(x,y) dA - \int_{s_4 s_2 s_1} B(0,0) dA  =  - \tau^2 (1 - \cos \theta ) + \frac{\tau^3}{3} \frac{\partial B}{\partial y} \Big\vert_0 (\sin \theta - \sin^2 \theta) \label{variation}\eeq
Now, in order to find the "maximum" we have to differentiate it with respect to $\theta$ and equate derivative with zero:
\beq - \tau^2 B (0) \sin \theta + \frac{\tau^3}{3} \frac{\partial B}{\partial y} \Big\vert_0 ( \cos \theta  - 2 \sin \theta \cos \theta ) = 0 \eeq
After factoring out $-\tau^2$ from the above equation we obtain
\beq  B (0) \sin \theta - \frac{\tau}{3} \frac{\partial B}{\partial y} \Big\vert_0 ( \cos \theta  - 2 \sin \theta \cos \theta ) = 0 \label{sintheta} \eeq
We will now Taylor expand the above and solve it for $\theta$. Let us now see to what order we want to do that. The only purpose of knowing the value of $\theta$ is to substitute it into Equation \ref{variation}. Now, we want to compute the right hand side of Equation \ref{variation} up to $0(\tau^4)$. Now, the $0(\theta^2)$ term will lead to $0 (\tau^2 \theta^4)$ contribution in $\tau^2 (1 - \cos \theta)$ and $0 (\tau^3 \theta^2)$ contribution in $\tau^3 (\sin \theta - \sin^2 \theta)$. Since $0 (\theta) = 0 (\tau)$ these two contributions are $0 (\tau^6)$ and $0 (\tau^5)$ respectively. Since our calculation is up to $0 (\tau^4)$ both can be neglected. This means that we can neglect $0 (\theta^2)$ correction to $\theta$ and, therefore, write
\beq \theta = \frac{\tau}{3B(0)} \frac{\partial B}{\partial y} \Big\vert_0 \label{thetafirstorder} \eeq
Now, we will expand Equation \ref{variation} up to $0(\tau^4)$, 
\beq \int_{s_4 s_2 s'_1} B(x,y) dA - \int_{s_4 s_2 s_1} B(x,y) dA  = - \frac{\tau^2 \theta^2}{2} + \frac{\tau^3 \theta}{3} \frac{\partial B}{\partial y} \Big\vert_0 \eeq
and substitute the value of $\theta$ we have just found, to obtain
\beq \int_{s_4 s_2 s'_1} B(x,y) dA - \int_{s_4 s_2 s_1} B(x,y) dA  = \frac{\tau^4}{18 B(0)} \Big(\frac{\partial B}{\partial y} \Big\vert_0 \Big)^2 \eeq
The above equation pertains to the variation of the flux through a "triangle" $s_1 s_2 s_4$. We are interested, however, in the flux through the "square" $s_1 s_2 s_3 s_4$. But, one can clearly see that the flux through the square is equal to the sum of the fluxes through two triangles, $s_1 s_2 s_4$ and $s_3 s_2 s_4$. In light of the fact that $s_1$ is the only point that is being moved, the flux through $s_3 s_2 s_4$ will not change. Thus, the change of the flux through the square $s_1 s_2 s_3 s_4$ is identical to the change of the flux through the triangle $s_1 s_2 s_4$:
\beq \int_{s'_1 s_2 s_3 s_4} B(x,y) dA - \int_{s_1 s_2 s_3 s_4} B(x,y) dA  = \frac{\tau^4}{18 B(0)} \Big(\frac{\partial B}{\partial y} \Big\vert_0 \Big)^2 \eeq
Now, the above variation was only pertaining to the displacement of $s_1$. Now we will add the displacements of all four points. From symmetry it is clear that effects of displacement of $s_2$, $s_3$ and $s_4$ can be obtained by rewriting the above equation while replacing $y$ with $-x$, $-y$ and $+x$, respectively. Since the only place where $-x$ and $-y$ are present is derivative, squaring of that derivative will turn them into $+x$ and $+y$, respectively. Thus, we obtain 
\beq \int_{s'_1 s_2 s_3 s_4} B(x,y) dA - \int_{s_1 s_2 s_3 s_4} B(x,y) dA =  \frac{\tau^4}{18 B(0)} \Big(\frac{\partial B}{\partial y} \Big\vert_0 \Big)^2 \eeq
\beq \int_{s_1 s'_2 s_3 s_4} B(x,y) dA - \int_{s_1 s_2 s_3 s_4} B(x,y) dA =   \frac{\tau^4}{18 B(0)} \Big(\frac{\partial B}{\partial x} \Big\vert_0 \Big)^2 \eeq
\beq \int_{s_1 s_2 s'_3 s_4} B(x,y) dA - \int_{s_1 s_2 s_3 s_4} B(x,y) dA = \frac{\tau^4}{18 B(0)} \Big(\frac{\partial B}{\partial y} \Big\vert_0 \Big)^2 \eeq
\beq \int_{s_1 s_2 s_3 s'_4} B(x,y) dA - \int_{s_1 s_2 s_3 s_4} B(x,y) dA = \frac{\tau^4}{18 B(0)} \Big(\frac{\partial B}{\partial x} \Big\vert_0 \Big)^2 \eeq
Finally, in order to "shift" all four points by going from $\{s_1, s_2, s_3, s_4 \}$ to $\{s'_1, s'_2, s'_3, s'_4 \}$, we have to sum the effects of the shifts of any individual point (and we don't have to look at "interaction" between the shifts since that would be of higher order than we are concerned about). Thus, by suming the above four equations, we obtain
\beq \int_{s'_1 s'_2 s'_3  s'_4} B(x,y) dA - \int_{s_1 s_2 s_3 s_4} B(x,y) dA= \frac{\tau^4}{9B(0)} \Big( \Big( \frac{\partial B}{\partial x} \Big\vert_0 \Big)^2 + \Big( \frac{\partial B}{\partial y} \Big\vert_0 \Big)^2 \Big) \eeq
where $1/18$ changed to $1/9$ due to the presence of two identical copies of each term. We have not done yet. Apart from the "correction" due to the "shift" of the contour, there is also a correction over "any given" contour due to the derivatives of $B$, given by Equation \ref{noshift}:
\beq \int_{s_1 s_2 s_3 s_4} B(x,y) dA - \int_{s_1 s_2 s_3 s_4} B(0, 0) dA  = \frac{2}{3} \tau^4 \Big( \frac{\partial^2 B}{\partial x^2} \Big\vert_0 + \frac{\partial^2 B}{\partial y^2} \Big\vert_0 \Big) + 0 (\tau^6) \eeq
Now, both of the above correction is strictly due to $\partial_x B$ and $\partial_y b$. On the other hand, in \ref{vertical} we have found that the correction due to $\partial_z B$ is 
\beq \delta_z ({\rm Flux}) = \frac{\tau^4}{B (0)} \Big( \frac{\partial B}{\partial z} \Big\vert_z \Big)^2 \eeq
Thus, if we take all of these corrections into account, we will obtain the final equation for flux to be 
\beq {\rm Flux} = 2B (0) \tau^2 + \frac{\tau^4}{9B(0)} \Big( \Big( \frac{\partial B}{\partial x} \Big\vert_0 \Big)^2 + \Big( \frac{\partial B}{\partial y} \Big\vert_0 \Big)^2 \Big) + \nonumber \eeq
\beq + \frac{\tau^4}{B (0)} \Big( \frac{\partial B}{\partial z} \Big\vert_z \Big)^2 + \frac{2}{3} \tau^4 \Big( \frac{\partial^2 B}{\partial x^2} \Big\vert_0 + \frac{\partial^2 B}{\partial y^2} \Big\vert_0 \Big) \label{TwoZeroSix} \eeq
Now, the pre-Lagrangian is equal to the square of the flux, thus it is given by 
\beq {\cal J}_B = 4B^2(0) \tau^4 + \frac{4 \tau^6}{9} \Big( \Big( \frac{\partial B}{\partial x} \Big\vert_0 \Big)^2 + \Big( \frac{\partial B}{\partial y} \Big\vert_0 \Big)^2 \Big) + 4 \tau^6 \Big( \frac{\partial B}{\partial z} \Big)^2 + \frac{8 \tau^6 B(0)}{3} \Big( \frac{\partial^2 B}{\partial x^2} \Big\vert_0 + \frac{\partial^2 B}{\partial y^2} \Big\vert_0 \Big) \label{magneticlagrangian}\eeq
Let us now remind ourselves of the role that the above expression is playing. If we go back to scalar field, and if we don't care about the correction terms, then the pre-Lagrangian would be given by non-covariant expression
\beq {\cal J}_{\phi} = (\partial_0 \phi)^2 \eeq
We then write the above in a covariant form by introducing a vector $v_{\phi}^{\mu}$ given by 
\beq v_{\phi}^{\mu} = \delta^{\mu}_0 \eeq
where the subscript $\phi$ stands for field $\phi$ as opposed to coordinate index (thus, $v_{\phi}^{\mu}$ has \emph{one} coordinate index $\mu$). We then allow $v_{\phi}^{\mu}$ to rotate and rewrite the above expression as 
\beq {\cal J}_{\phi} = (v_B^{\mu} \partial_{\mu} \phi)^2, \eeq
 Such expression is \emph{not} invariant under rotation of $v_{\phi}^{\mu}$ and, therefore, can \emph{not} be identified with Lagrangian itself (which is why it is called "pre-Lagrangian"). In order to find the actual Lagrangian, we need to select $v_{\phi}^{\mu}$ that would \emph{minimize} the absolute value of pre-Lagrangian. Thus, in case of scalar field, we would have
\beq \partial^{\mu} \phi \partial_{\mu} \phi > 0 \Rightarrow v_{\rm scal}^{\mu} = \frac{\partial^{\mu} \phi}{\sqrt{\vert \partial^{\nu} \phi \partial_{\nu} \phi \vert}} \eeq
and then the substitution of the above $v_{\phi}^{\mu}$, indeed, gives us the \emph{actual} Lagrangian ${\cal L}_{\phi}$ (as opposed to "pre-Lagrangian ${\cal J}_{\phi}$):
\beq {\cal L} = (v^{\mu}_{\phi} \partial_{\mu} \phi)^2 = \partial^{\mu} \phi \partial_{\mu} \phi \eeq
We will now use similar concept for the electromagnetic field. So we would like to select a $t$-axis in such a way that the right hand side of Equation \ref{magneticlagrangian} is minimized. After that, we will substitute $v_B^{\mu}$ into the "pre-Lagrangian" given in \ref{magneticlagrangian}, in order to obtain the actual Lagrangian. Now, we already know from Section 6 that, up to finite order, $v_B^{\mu}$ is an eigenvector of $M^{\mu}_{\; \nu}$, where 
\beq M^{\mu}_{\; \nu} = F^{\mu}_{\; \rho} F^{\rho}_{\; \nu} \eeq
We would now like to find higher order corrections to $v_B^{\mu}$,
\beq v_B^{\prime \mu} = v_B^{\mu} + \delta v_B^{\mu} \eeq
 Now, in the frame where $\vec{E}$ and $\vec{B}$ are "exactly" parallel, the un-corrected expression for ${\cal J}_B$  is "exactly" minimized. Thus, the corrected expression is only "approximately" minimized in that frame. On the other hand, there is "another" frame where Lagrangian generator reaches "exact" minimum, but $\vec{E}$ and $\vec{B}$ are only approximately parallel. We will denote these two frames by $x^{\mu}$ and $x^{\prime \mu}$, respectively. Furthermore, we will identify the unit vectors in $t$ and $t'$ directions by $v_B^{\mu}$ and $v_B^{\prime \mu}$, respectively:
\beq x^0 = x^{\mu} v_{\mu} \; ; \; x^{\prime 0} = x^{\prime \mu} v'_{\mu} \eeq
Furthermore, we will select $z$-axis in such a way that, in $v_B^{\mu}$-frame, it coincides with a common direction of $\vec{E}$ and $\vec{B}$. If we denote $F_{\mu \nu}$ tensor in these two frames by $F_{\mu \nu}$ and $F^{\prime}_{\mu \nu}$, this means that 
\beq F_{13} = F_{23} = F_{01} = F_{02} = 0 \eeq
while the above statement will \emph{not} be true if we replace $F$ with $F^{\prime}$. Now, our task is to select $x^{\prime \mu}$-frame in such a way that $F^{\prime}_{12}$ is minimized. 

In order to make $F^{\prime}_{12}$ distinct from $F_{12}$ we have to "mix" it with something non-zero that has either index $1$ or index $2$. Given our constraints, the only non-zero element that meets the above description is $F_{12}$, itself, which prevents us from having first order variation. However, we \emph{can} produce second-order variation. For example, we can rotate $F_{03}$ to get $F''_{13}$ and then we can further rotate $F''_{13}$ to get $F'_{12}$. Another avenue is to first rotate $F_{12}$ to get either $F''_{02}$ or $F''_{13}$ and then we can further rotate the latter into $F'_{12}$. This last option would account to the rotation of $F_{12}$ through the $\Lambda^{\mu}_{\nu}$. However, all of the options we have just discussed will result in second order variation rather than first order (which is due to the fact that everything other than $F_{03}$ and $F_{12}$ is zero). 

Now, any kind of sequence of boosts can be produced through a single boost along the appropriately chosen direction. This means that we have to try to minimize the Lagrangian generator with respect to four degrees of freedom: three degrees of freedom tell us the direction of the boost, and the fourth tells us the magnitude of the boost. We can proceed by \emph{first} selecting a \emph{fixed} direction and minimizing with respect to magnitude, and \emph{after that} minimizing the produced "minimum" with respect to direction. Now, we have seen previously that the Lagrangian generator takes to form 
\beq {\cal J}_B = 4 \vert \vec{B} \vert ^2 \tau^4 +  \lambda \tau^6 \eeq
Now, from what we have just said, the variation of $\vert \vec{B} \vert$ is of the order of $(\delta v)^2$:
\beq \vert \vec{B}' \vert = \vert \vec{B} \vert + a (\delta v)^2 \eeq
This, however, might not be true for $\lambda$. After all, the above statement for $\vert \vec{B}' \vert$ was based on the assumption that $F_{12}$ and $F_{03}$ are the only non-zero components of the tensor. On the other hand, the coefficients next to $0 (\tau^6)$ terms are more complicated (for example, they have second derivatives, among other things); thus the "second derivatives" of things that were assumed to be zero are no longer zero. This means that the variation of $\lambda$ is of \emph{first} order:
\beq \lambda' = \lambda + b \delta v \eeq
This means that the modification of Lagrangian generator is given by 
\beq {\cal J}'_B = 4 (\vert \vec{B} \vert + a (\delta v)^2)^2 \tau^4 + (b+  d \delta v) \tau^6 \eeq
Up to $0 ((\delta v)^2)$ this becomes
\beq {\cal J}'_B = 4 \vert \vec{B} \vert^2 \tau^4 + 8 \vert \vec{B} \vert a (\delta v)^2 \tau^4 + d \tau^6 \delta v \eeq
In order to find the minimum, we have to equate the derivative of ${\cal J}'_B$ with respect to $\delta v$ to zero:
\beq 0 = \frac{\partial {\cal J}'_B}{\partial (\delta v)} = 16 \vert \vec{B} \vert a \tau^4 \delta v + d \tau^6 \eeq
which implies that 
\beq \delta v = \frac{d \tau^2}{16 \vert \vec{B} \vert a} \eeq
In other words, we have just shown that 
\beq \delta v = 0 (\tau^2) \eeq
This immediately implies that 
\beq (\delta v)^2 \tau^4 = 0 (\tau^8) \; ; \; \tau^6 \delta v = 0 (\tau^8) \eeq
Since we are computing up to $0 (\tau^6)$, this means that the impact of $\delta v$ can be neglected altogether. Therefore, we don't need to find out the value of $\delta v$ either. We can simply stick to the Equation \ref{magneticlagrangian} as our final expression both for Lagrangian generator \emph{as well as} actual Lagrangian, as far as $0 (\tau^6)$ is concerned. 

Let us now write the Equation \ref{magneticlagrangian} in a covariant form. First of all, in equations \ref{BisCovariant} and \ref{EisCovariant} we already found out a covariant expression for $\vert \vec{E} \vert^2$, $\vert \vec{B} \vert^2$ and $\vec{E} \cdot \vec{B}$. Thus, we can freely be using these three quantities. However, we can \emph{not} use un-contracted $B^{\mu}$, nor can we "contract" it with anything else, such as $v_B^{\mu} B_{\mu}$. Our task is to manipulate $\vert \vec{B} \vert^2$, $\vert \vec{E} \vert^2$ and $\vec{B} \cdot \vec{E}$ in such a way that we will arrive at the rest of the expressions we might need. In order to do it, we can utilize the constraints "at the origin", 
\beq B_x (0) = B_y(0) = 0 \; ; \; v_B^{\mu} (0) = \delta^{\mu}_0 \eeq
At the same time, these constraints no longer hold away from the origin:
\beq \partial_{\mu} B_x \neq 0 \; ; \; \partial_{\mu} B_y \neq 0 \; ; \; \partial_{\nu} v_B^{\mu} \eeq 
After all, we would like our coordinate system to be "rigid" in a sense that Christoffel's symbols are zero. This means that if $B_z$ "twists around" it can't possibly be identified with $z$-axis at more than one point. The most we can do is to rotate $z$-axis so that the two are identified at the origin. Furthermore, we recall from the previous discussion that $v_B^{\mu}$ is determined based on $\vec{E}$ and $\vec{B}$. This means that $v_B^{\mu}$ likewise "twists around". So, for the same exact reason, $v_B^{\mu}$ can be identified with $t$-axis \emph{only} at the origin, and not elsewhere: 
\beq v_B^{\mu} (0) = \delta^{\mu}_0 \; ; \; (\partial_{\nu} v_B^{\mu})(0) \neq 0 \eeq
Accordingly, at the origin, $B_z$ and $\vert \vec{B} \vert$ coincide; but their derivatives do not:
\beq B_z (0) = \vert \vec{B} (0) \vert \; ; \; \partial_{\mu} B_z (0) \neq \partial_{\mu} \vert \vec{B} (0) \vert \eeq
However, due to the fact that the derivatives of $B_x^2$ and $B_y^2$ are both zero, it is easy to see that the derivatives of $\vert \vec{B} \vert^2$ and $B_z^2$ coincide:
\beq \partial_z \vert \vec{B} \vert^2 = \partial_z B_z^2 = 2 B_z \partial_z B_z \eeq
However, as long as we take "squares" of the above quantities, their derivatives \emph{will} be zero:
\beq (\partial_{\mu} B_x^2)(0) = 2 B_x (0) \partial_{\mu} B_x = 0 \; ; \;  (\partial_{\mu} B_y^2)(0) = 2 B_y (0) \partial_{\mu} B_y = 0 \eeq
This also implies that 
\beq \partial_{\mu} \vert \vec{B} \vert^2 = \partial_{\mu} B_z^2 \eeq
Now by using 
\beq B_z (0) = \vert \vec{B} (0) \vert \eeq
the above equation evaluates to 
\beq \partial_{\mu} \vert \vec{B} \vert^2 = 2 B_z (0) \partial_{\mu} B_z = 2 \vert \vec{B} \vert \partial_{\mu} B_z \eeq
which implies that 
\beq \partial_{\mu} B_z = \frac{\partial_{\mu} \vert \vec{B} \vert^2}{2 \vert \vec{B} \vert} \label{SomethingMinor}\eeq
where $\vert \vec{B} \vert$ is given by Equation \ref{BisCovariant}. Now, we are not done yet: since we don't have a coordinate system, we can't have un-contracted index $\mu$. Let us discuss cases by cases how we get rid of it. Lets start with $\mu=3$; in other words, we want to produce $\partial_z B_z$. In light of the fact that $\vec{B}  (0)$ points along $z$-axis, we know that 
\beq B^{\mu} (0) \partial_{\mu} f = B^z (0) \partial_z f = \vert \vec{B} (0) \vert \partial_z f \eeq
This means that 
\beq \partial_z f = \frac{1}{\vert \vec{B} (0) \vert} B^{\mu} (0) \partial_{\mu} f \eeq
Now, in order to write a truly covariant expression, we have to write $B^{\mu}$ in terms of $F_{\mu \nu}$. Since $\vec{B}$ is parallel to $z$-axis, we know that 
\beq F_{01}=F_{02}=F_{13}=F_{23} =0 \eeq
Therefore we can rewrite $B^{\mu} \partial_{\mu} f$ as 
\beq B^{\mu} \partial_{\mu} f = B^3 \partial_3 f = F_{12} \partial_3 f \eeq
Now, we would like to come up with covariant expression that produces $F_{12}$ \emph{without} producing $F_{03}$. We will do the following trick. First we notice that $v_B^{\mu}$ is pointing along $z$ direction. This immediately implies that $\epsilon{\alpha \beta \gamma \delta} v_B^{\alpha}$ is non-zero only when $\beta$, $\gamma$ and $\delta$ are all non-zero at the same time. Thus, by contracting $F^{\mu \nu}$ with any two of these three indexes we will immediately "get rid" of $F_{03}$ and be left with $F_{12}$:
\beq \epsilon_{\alpha \beta \gamma \delta} v_B^{\alpha} F^{\beta \gamma} \partial^{\delta} f = - 2 F_{12} \partial_3 f = - 2B^3 \partial_3 f = - 2 B_z \partial_z f \eeq
where we have $(+, -, -, -)$ convention, and $B_z$ is identified with $B^3$ rather than $B_3$:
\beq F_{12} = B_z = B^3 = - B_3 \eeq
while $\partial_z$ is identified with $\partial_3$:
\beq \partial_z f = \partial_3 f \eeq
Now, by recalling that $B_z = \vert \vec{B} \vert$, we can rewrite the above expression as 
\beq \partial_z f = - \frac{1}{2 \vert \vec{B} \vert} \epsilon_{\alpha \beta \gamma \delta} v_B^{\alpha} F^{\beta \gamma} \partial^{\delta} f \label{TwoFortyThree} \eeq
Furthermore, immediately from the choice of the $t$-axis, we know that 
\beq \partial_t f = v_B^{\mu} \partial_{\mu} f \eeq
As far as $\partial_x$ and $\partial_y$ are concerned, the symmetry around $z$-axis prevents us from "separating" them from each other. But we can easily compute the sum of their squares by using 
\beq (\partial_x f)^2 + (\partial_y f)^2 = (\partial_0 f)^2 - (\partial_z f)^2 - \partial^{\mu} f \partial_{\mu} f \label{dxy}\eeq
By substituting the expressions for $\partial_0 f$ and $\partial_z f$ the above becomes
\beq (\partial_x f)^2 + (\partial_y f)^2 = (v_B^{\mu} \partial_{\mu} f)^2 - \frac{1}{4 \vert \vec{B} \vert^2} (\epsilon_{\alpha \beta \gamma \delta} v_B^{\alpha} F^{\beta \gamma} \partial^{\delta} f)^2 - \partial^{\mu} f \partial_{\mu} f \eeq
Now, let us apply it to $f=B_z$. We can  use Equation \ref{SomethingMinor}, 
\beq \partial_{\mu} B_z = \frac{\partial_{\mu} \vert \vec{B} \vert^2}{2 \vert \vec{B} \vert} \eeq
to do the following substitutions:
\beq (v_B^{\mu} \partial_{\mu} B_z)^2 = \Big( \frac{v_B^{\mu} \partial_{\mu} \vert \vec{B} \vert^2}{2 \vert \vec{B} \vert} \Big)^2  = \frac{1}{4 \vert \vec{B} \vert^2} (v_B^{\mu} \partial_{\mu} \vert \vec{B} \vert^2) \label{vdb}\eeq
\beq \frac{1}{4 \vert \vec{B} \vert^2} (\epsilon_{\alpha \beta \gamma \delta} v_B^{\alpha} F^{\beta \gamma} \partial^{\delta} B_z)^2 =\frac{1}{4 \vert \vec{B} \vert^2} \Big(\epsilon_{\alpha \beta \gamma \delta} v_B^{\alpha} F^{\beta \gamma} \frac{\partial^{\delta} \vert \vec{B} \vert^2}{2 \vert \vec{B} \vert} \Big)^2 \eeq
By combining factors of $2$ as well as powers of $\vert \vec{B} \vert$, the last equation can be further rewritten as 
\beq \frac{1}{4 \vert \vec{B} \vert^2} (\epsilon_{\alpha \beta \gamma \delta} v_B^{\alpha} F^{\beta \gamma} \partial^{\delta} B_z)^2= \frac{1}{16 \vert \vec{B} \vert^4} (\epsilon_{\alpha \beta \gamma \delta} v_B^{\alpha} F^{\beta \gamma} \partial^{\delta} \vert \vec{B} \vert^2 )^2 \label{epsilon}\eeq
Finally, $\partial^{\mu} f \partial_{\mu} f$ is now being replaced with $\partial^{\mu} B_z \partial_{\mu} B_z$, which can be rewritten as 
\beq \partial^{\mu} B_z \partial_{\mu} B_z = \frac{\partial^{\mu} \vert \vec{B} \vert^2}{2 \vert \vec{B} \vert}\frac{\partial_{\mu} \vert \vec{B} \vert^2}{2 \vert \vec{B} \vert} = \frac{1}{4 \vert \vec{B} \vert^2} \partial^{\mu} \vert \vec{B} \vert^2 \partial_{\mu} \vert \vec{B} \vert^2 \label{contractbz}\eeq
By substituting \ref{vdb}, \ref{epsilon} and \ref{contractbz} into \ref{dxy}, we obtain
\beq \Big(\frac{\partial B_z}{\partial x} \Big)^2 + \Big( \frac{\partial B_z}{\partial y} \Big)^2 = \frac{1}{4 \vert \vec{B} \vert^2} (v_B^{\mu} \partial_{\mu} \vert \vec{B} \vert^2 )^2 -  \label{TwoFiftyTwo} \eeq
\beq - \frac{1}{16 \vert \vec{B} \vert^4} (\epsilon_{\alpha \beta \gamma \delta} v_B^{\alpha} F^{\beta \gamma} \partial^{\delta} \vert \vec{B} \vert^2)^2 - \frac{1}{4 \vert \vec{B} \vert^2} \partial^{\mu} \vert \vec{B} \vert^2 \partial_{\mu} \vert \vec{B} \vert^2 \nonumber \eeq
Now, our final goal is to rewrite \ref{magneticlagrangian} in a covariant form. The only part of the above equation which we have not yet written in a covariant form is the second derivative terms. The only such expression present is $\partial_x^2 B_z + \partial_y^2 B_z$. Let us, therefore, find a covariant equivalent of the latter.  First, it is easy to see that the above expression can be written as
\beq \partial_x^2 B_z + \partial_y^2 B_z = \frac{1}{\vert \vec{B} \vert^2} F_k^{\; i} \partial_i \partial^j F_j^{\; k} \label{ijkcontraction}\eeq
The only problem is that if we will replace Latin symbols with Greek, we would obtain unwanted zero components. We would like to use the vector $v_B^{\mu}$ in order to "get rid" of them.  Suppose, for example, we want to write a covariant expression for $\vert \vec{w} \vert^2$ for some other vector $\vec{w}$. In the frame in which $t$-axis coincides with $v_B^{\mu}$, we can write it as 
\beq \vert \vec{w} \vert^2 = (w^0)^2 - w^{\mu} w_{\mu} = (w^{\mu} v_{\mu})^2 - w^{\mu} w_{\mu} \eeq
Now, in light of the fact that the expression we want to evaluate is a lot longer, we want our calculation to look as simple as possible. Thus, we will be using $\partial_0$ and $w_0$ in place of $v_B^{\mu} \partial_{\mu}$ and $v_B^{\mu} w_{\mu}$ and then we will replace all of the $0$-s with appropriate $v_B^{\mu}$-contractions at the very end. We will take care of one component at a time. Let us first "get rid" of $k$ in \ref{ijkcontraction}:
\beq \partial_x^2 B_z + \partial_y^2 B_z = \frac{1}{\vert \vec{B} \vert^2} F_{\rho}^{\; i} \partial_i \partial^j F_j^{\; \rho} - \frac{1}{\vert \vec{B} \vert^2} F_0^{\; i} \partial_i \partial^j F_j^{\; 0} \eeq
By comparing the right hand side of the above to the right hand side of \ref{ijkcontraction} one can see that the number of terms changed from $1$ to $2$ due to the "split" of space-alone term into covariant term together with time-alone one. In general, each time we get rid of a "Latin" index, the number of terms always doubles (thus, since we have to also get rid of $i$ and $j$ we will have $8$ terms at the end). Now, we will "get rid" of $i$ and obtain
\beq \partial_x^2 B_z + \partial_y^2 B_z = \frac{1}{\vert \vec{B} \vert^2} F_{\rho}^{\; 0} \partial_0 \partial^j F_j^{\; \rho} - \frac{1}{\vert \vec{B} \vert^2} F_{\rho}^{\mu} \partial_{\mu} \partial^j F_j^{\; \rho} - \frac{1}{\vert \vec{B} \vert^2} F_0^{\; 0} \partial_0 \partial^j F_j^{\; 0} + \frac{1}{\vert \vec{B} \vert^2} F_0^{\; \mu} \partial_{\mu} \partial^j F_j^{\; 0} \eeq
where, as usual, $F_0^{\; 0} =0$, but we are keeping that term in order to convince ourselves that we have the right number of terms and that this number "doubles" as expected. Finally, we will get rid of $j$:
\beq \partial_x^2 B_z + \partial_y^2 B_z = \frac{1}{\vert \vec{B} \vert^2} F_{\rho}^{\; 0} \partial_0 \partial^0 F_0^{\; \rho} - \frac{1}{\vert \vec{B} \vert^2} F_{\rho}^0 \partial_0 \partial^{\nu} F_{\nu}^{\; \rho}  - \frac{1}{\vert \vec{B} \vert^2} F_{\rho}^{\; \mu} \partial_{\mu} \partial^0 F_0^{\; \rho} + \frac{1}{\vert \vec{B} \vert^2} F_{\rho}^{\; \mu} \partial_{\mu} \partial^{\nu} F_{\nu}^{\; \rho} - \nonumber \eeq
\beq - \frac{1}{\vert \vec{B} \vert^2} F_0^{\; 0} \partial_0 \partial^0 F_0^{\; 0} + \frac{1}{\vert \vec{B} \vert^2} F_0^{\; 0} \partial_0 \partial^{\nu} F_{\nu}^{\; 0} + \frac{1}{\vert \vec{B} \vert^2} F_0^{\; \mu} \partial_{\mu} \partial^0 F_0^{\; 0} - \frac{1}{\vert \vec{B} \vert^2} F_0^{\mu} \partial_{\mu} \partial^{\nu} F_{\nu}^{\; 0} \eeq
The above equation has $8$ terms, as expected. Three of these $8$ terms involve $F_{00}$ and, therefore, are equal to zero. We can now get rid of these $3$ terms, and obtain $5$-term expression: 
\beq \partial_x^2 B_z + \partial_y^2 B_z = \frac{1}{\vert \vec{B} \vert^2} F_{\rho}^{\; 0} \partial_0 \partial^0 F_0^{\; \rho} - \frac{1}{\vert \vec{B} \vert^2} F_{\rho}^0 \partial_0 \partial^{\nu} F_{\nu}^{\; \rho}  - \label{secondder} \eeq
\beq - \frac{1}{\vert \vec{B} \vert^2} F_{\rho}^{\; \mu} \partial_{\mu} \partial^0 F_0^{\; \rho} + \frac{1}{\vert \vec{B} \vert^2} F_{\rho}^{\; \mu} \partial_{\mu} \partial^{\nu} F_{\nu}^{\; \rho} - \frac{1}{\vert \vec{B} \vert^2} F_0^{\mu} \partial_{\mu} \partial^{\nu} F_{\nu}^{\; 0} \nonumber \eeq
Finally, if we substitute $f= \vert \vec{B} \vert^2$ into Equation \ref{TwoFortyThree},
\beq \partial_z \vert \vec{B} \vert^2 = - \frac{1}{2 \vert \vec{B} \vert} \epsilon_{\alpha \beta \gamma \delta} v_B^{\alpha} F^{\beta \gamma} \partial^{\delta} \vert \vec{B} \vert^2 \label{TwoFortyThreeWithSubstitution} \eeq
and then substitute \ref{TwoFiftyTwo}, \ref{secondder} and \ref{TwoFortyThreeWithSubstitution} into \ref{TwoZeroSix}, and doing some simple combining of terms, we obtain
\beq {\cal L}_B = 4 \vert \vec{B} \vert^2 \tau^4 + \frac{\tau^6}{9 \vert \vec{B} \vert^2} (v_B^{\mu} \partial_{\mu} \vert \vec{B} \vert^2)^2 - \frac{2 \tau^6}{9 \vert \vec{B} \vert^4} (\epsilon_{\alpha \beta \gamma \delta} v_B^{\alpha} F^{\beta \gamma} \partial^{\delta} \vert \vec{B} \vert^2)^2 - \nonumber \eeq
\beq - \frac{\tau^6}{9 \vert \vec{B} \vert^2} \partial^{\mu} \vert \vec{B} \vert^2 \partial_{\mu} \vert \vec{B} \vert^2 + \frac{8 \tau^6}{3 \vert \vec{B} \vert} v_{\alpha} v_B^{\beta} \partial_{\gamma} \partial^{\delta} F_{\rho}^{\; \alpha} \partial_{\beta} \partial^{\gamma} F_{\delta}^{\; \rho}  - \frac{8 \tau^6}{3 \vert \vec{B} \vert} v_{\alpha} v_B^{\beta} F_{\rho}^{\; \alpha} \partial_{\beta} \partial^{\gamma} F_{\gamma}^{\; \rho} -  \eeq
\beq - \frac{8 \tau^6}{3B} v_{\beta} v_B^{\gamma} F_{\rho}^{\; \alpha} \partial_{\alpha} \partial^{\beta} F_{\gamma}^{\; \rho} + \frac{8 \tau^6}{3 \vert \vec{B} \vert} F_{\rho}^{\; \alpha} \partial_{\alpha} \partial^{\beta} F_{\beta}^{\; \rho} - \frac{8 \tau^6}{3 \vert \vec{B} \vert} v_B^{\rho} \partial_{\sigma} F_{\rho}^{\; \mu} \partial_{\mu} \partial^{\nu} F_{\nu}^{\; \sigma} \nonumber \eeq
where $B$ is given by Equation \ref{BisCovariant},
\beq \vert \vec{B} \vert^2 = \frac{1}{2} \Bigg(F^{\alpha \beta} F_{\alpha \beta} + (sgn(F^{\mu \nu}F_{\mu \nu})) \sqrt{(F^{\alpha \beta} F_{\alpha \beta})^2 - \frac{1}{16} (\epsilon_{\alpha \beta \gamma \delta} F^{\alpha \beta} F^{\gamma \delta})^2} \Bigg)  \eeq

\subsection*{7 Corrections to electric Lagrangian density}

Let us now turn to the "electric" Lagrangian. Again, we are considering a flux through a contour bounded by four points; but there is one important difference. In the "magnetic" case, all four points were lying on the equator of Alexandrov set. In the electric case, on the other hand, two of the four points lie on the equator, and the other two points coincide with the poles. This also has an impact on a particular ways the contour can be deformed. In magnetic case we were able to move all four points of the contour. In electric case, we can only move \emph{two} points that lie on equator; the other pair of points is constrained to coincide with poles and, therefore, can't be moved. 

Now, in the magnetic case, we were making a "rigid shift" of the contour in $z$ direction. This, of course, requires displacement of all four points. In the present situation, our inability to displace two of the points will prevent us from making such shifts. Furthermore, in the magnetic case we could slightly stretch or compress the length of the edges. In our present case, each of the edges is "constrained" to connect a point on a pole with a point on the equator. This ultimately constrains both its spacelike and timelike projections to coincide with $\tau$. As a result, the magnetic field has no contribution to flux through this contour, and the entire contribution comes solely from the electric field.

Let us denote the poles of Alexandrov set by $A$ and $C$, and denote the two points on equator by $B$ and $D$. In case of the constant electric field, points $B$ and $D$ will be on the opposite sides of the equator, chosen in such a way that the line connecting them is parallel to electric field. If, however, we introduce a derivative of electric field with respect to some perpendicular direction, then points $B$ and $D$ will shift in the direction of the derivative. At the same time, $A$ and $C$ will stayed "glued" to the poles. This will result in rotation of triangles $ABC$ and $ADC$ by an angle $\theta$ (which would be the same if we neglect the second derivatives). In light of the fact that the direction of electric field at $B$ is parallel to the one in $D$, it is easy to see that these two triangles will rotate "towards" each other. But the angle of rotation will be small.

Let us denote the square loop by $ABCD$, where $A$ and $C$ are the two poles of Alexandrov set and $B$ and $D$ are located at the equator. As usual, we will select $t$ axis in such a way that it passes through $A$ and $C$. Thus, 
\beq A = (- \tau, 0, 0, 0) \; , \; C = (\tau, 0, 0,0) \label{ElectricAC} \eeq
Furthermore, we will select $z$ axis to be the direction of the electric field at the origin. Thus, 
\beq E_x = \vec{r} \cdot \vec{\nabla} E_x \; , \;  E_y = \vec{r} \cdot \vec{\nabla} E_y \eeq
\beq E_z = E(0) + \vec{r} \cdot \vec{\nabla} E_z \eeq
If the derivative terms were zero, the flux would have been maximized by setting $B$ and $D$ on the $z$-axis. If the derivatives are non-zero, they will be shifted slightly from these locations. Thus, 
\beq B = (0, - \tau + 0 (\tau^2), 0 (\tau^2), 0 (\tau^4)) \; , \; D = (0, \tau + 0 (\tau^2), 0 (\tau^2), 0 (\tau^4)) \label{ElectricBD} \eeq
Here, $0 (\tau^2)$ terms correspond to the slight shift of points tht we have just discussed. The reason we used $0 (\tau^4)$ rather than $0 (\tau^2)$ in the $z$-coordinate, is that we are assuming that $B$ and $D$ lie \emph{exactly} on the surface of the sphere; in other words, the equation 
\beq B_x^2+B_y^2+B_z^2=D_x^2+D_y^2 + D_z^2 = \tau^2 \label{ElectricSurface} \eeq
holds \emph{exactly}. From the point of view of causal set theory, this is enforced by an assertion that all links are direct; that is, there is \emph{no} point $E$ satisfying $A \prec E \prec B$, $A \prec E \prec D$, $B \prec E \prec C$ or $D \prec E \prec C$. The combination of Equation \ref{ElectricSurface} with the fact that $x$- and $y$-coordinates deviate by $0 (\tau^2)$ implies that $z$-coordinate deviates by $0 (\tau^4)$. 

The fact that these links are direct is also the reason why we have not included the deviation of $t$-components. After all, if the point $B$ were to shift "upward" from the equator, then it would be "forced" to either be spacelike-separated from $C$ \emph{or} to have non-zero distance from $B$. In the former case we would get a "contradiction" with an "existence" of $BC$-link, while in the latter case we would get a contradiction with "direct nature" of $AB$-link. Similarly, if we were to shift $B$ downwards, we would either have to violate the existence of $AB$-link or direct nature of $BC$-link. Similar argument also prevents us from shifting point $D$ either upward or downward. At the same time, it should be acknowledged that  if we were to include gravity then $B$ and $D$ satisfying the above conditions would pick up $t$-component due to the bending of light cone. But, as we have stated earlier, we are postponing gravity to \cite{Gravity} and in this paper we are assuming flat spacetime. The assumption that curvature is \emph{exactly} zero corresponds to the assumption that the $t$-component is zero as well, up to the discreteness scale. 

Let us now discuss the order of magnitude up to which we will evaluate the spacelike $0 (\tau^2)$ displacement of these points. As we stated earlier, the unperturbed flux through the square loop is $0 (\tau^2)$. Now, the displacement has three possible effects: the derivative of the field with respect to the direction of displacement, the angle of the projection of flux, and the variation of area of the displaced contour. Now, if we were dealing with triangle, such as $ABC$, it is easy to see that the displacement between the centers of triangle and Alexandrov set is of $0 (\tau)$, leading to the variation of flux of $0 (\tau^2) \times 0 (\tau) = 0 (\tau^3)$. If, on the other hand, we consider the square instead of triangle, we would no longer have such displacement. Instead, the displacement distance would be of $0 (\tau^2)$ leading to the variation of flux of $0 (\tau^2) \times 0 (\tau^2) = 0 (\tau^4)$. As far as the projection goes, we expect the angle of the tilt of the contours to be $0 (\tau)$. This means that the \emph{cosine} of that angle is $1 - 0 (\tau^2)$. This would, again, lead to $0 (\tau^2) \times 0 (\tau^2)$ variation of the flux. 

Finally, in electric case, we expect the area variation to be zero, contrary to its $0 (\tau^2)$ value in magnetic case. After all, in magnetic case we had a contour "lifted" by all four points, while in electric case we have two triangles ($ABC$ and $ADC$) rotation. It is easy to see that the area changes in the former case but not the latter. At the same time, the area of the projection of "two triangles taken together" onto $yz$ plane will, in fact, change. But the effect of this variation of area is "absorbed" into the effect of the projection. In fact, one can re-think the "projection of the field onto tilted triangle" as "projection of the field onto the \emph{shaddow of} tilted triangle on the \emph{original} plane". If the field lines are perpendicular to the original plane, then any given line will pass through the tilted triangle if and only if it will also pass through its shadow. Thus, both fluxes get multiplied by $\cos \theta$ just for "different reasons" (in one case $\cos \theta$ comes from projection, in the other case it comes from area modification). 

In this respect we can re-think "projection onto triangles $ABC$ and $ADC$" in terms of "change of area of projection of contour $ABCD$ after shifting $B$ and $D$", while keeping in mind that said contour will no longer be square and, instead, it will be three-dimensional object. This new interpretation makes it more analogous to magnetic case: in magnetic case we were dealing with "area modification" due to "shift" of all four points ($A$, $B$, $C$, and $D$) while in electric case we are looking at area modification due to the shift of only two of the four ($B$ and $D$). In both cases the resulting variation of flux is $0 (\tau^2) \times 0 (\tau^2) = 0 (\tau^4)$, and in both cases source of \emph{one} of $0 (\tau^2)$ is $\cos \theta = 1 - 0 (\tau)$. In electric case we are using $\cos \theta$ in order to make a projection onto our two triangles while in magnetic case we need $\cos \theta$ in order to compute the modification of area of shifted rectangle. 

Be it as it may, the bottom line is that the original flux is $0 (\tau^2)$ and all of its higher order variations are of the order of $0 (\tau^2) \times 0 (\tau^2) = 0 (\tau^4)$. Now, the Lagrangian generator is the \emph{square} of the flux. Thus, if we compute flux up to leading order, the Lagrangian generator will take the form $(0 (\tau^2) + 0 (\tau^4))^2 = 0 (\tau^4) + 0 (\tau^6)$. Since this is the order of magnitude up to which we agreed to do our calculations, we do not need to include anything higher. In other words, we will compute the variation of the angle up to leading order. 

Let us now look more closely at the sources of the displacement. It is clear that linear variation of $E_z$ will shift the two points ($B$ and $D$) in the direction where the magnitude of $E_z$ is larger. Finite value of $ \partial E_z/ \partial x$ will result in $0 (\tau^4)$ variation to flux per the argument we just made; thus, the effects of second derivatives are of a higher order and can be thrown away. As far as $E_x$ and $E_y$ are concerned, their projections onto the original contour were zero (after all, our choice of coordinates implies that the original contour lies in $tz$-plane). Thus, in case of rotated contour their magnitude should be multiplied by $\sin \theta = 0 (\theta)$ (where $\theta$ is the angle of rotation \emph{away from} $tz$-plane). Apart from that, we were \emph{also} assuming that $\vec{E} (0)$ is parallel to $z$-axis. Thus, $E_x$ and $E_y$ are of the order of magnitude of shift "away from" the origin, which is $0 (\tau)$. Their flux, therefore, is $0 (\tau^2) \times 0 (\tau) \times 0 (\tau) = 0 (\tau^4)$. We notice that this is the \emph{main} contribution of $E_x$ and $E_y$ to flux. Thus, we do not need higher order terms. 

Let us now discuss the order of magnitude up to which we compute the angle. The value of the angle is based on the maximization of flux. Now, since we are assuming that the angle is small, we already know the value of flux up to $0 (\tau^2)$, which we treat as "constant". Furthermore, from our prior argument we know that the leading order correction is $0 (\tau^4)$; in other words, there are no $0 (\tau^3)$ terms. Finally, we also know that $0 (\theta) = 0 (\tau)$. Thus, we are interested in the terms of the form $\theta^k \tau^l$ where $k+l =4$. In light of scaling symmetry, every single term we produce needs to have a factor of $\tau^2$ which comes from the area of the contour. Thus, we are interested in the terms of the form $\tau^{2+k} \theta^l$ where $k+l =2$. Therefore, the most general correction that satisfies these conditions is 
\beq {\rm Correction} =  a \tau^4 + b \tau^3 \theta + c \tau^2 \theta^2 + 0 (\tau^3 \theta^2) + 0 (\tau^2 \theta^3) \eeq
In order to maximize the flux we have to equate the $\theta$-derivative of above with zero. In other words, 
\beq 0 = \partial_{\theta} (a \tau^4 + b \tau^3 \theta + c \tau^2 \theta^2 + 0 (\tau^3 \theta^2) + 0 (\tau^2 \theta^3)) = b \tau^3 + 2 c \theta \tau^2 + 0 (\tau^2 \theta) + 0 (\tau^2 \theta^2) \eeq
which implies that 
\beq \theta = - \frac{b}{2c} \tau + 0 (\tau^2) \label{ThetaElectric}\eeq
Thus, in order for $0 (\tau)$-term of Equation \ref{ThetaElectric} to be non-zero, we want $b$ to be non-zero as well. This means that the correction term needs to include something of the form $\theta \tau^3$. Now, as we said earlier, everything is to the order of $0 (\tau^2)$ or higher due to scaling symmetry. This means that we are only allowed an additional \emph{first} order in $\tau$. This means that we can assume that all fields are linear. 

Now, far as $\partial/ \partial t$ is concerned, we can split the triangle $ABC$ into triangles $AOB$ and $OBC$, where point $O$ is the center of Alexandrov set which we identify as the origin. If the electric field was linear in $t$, the contributions of these two triangles would have the same magnitude and opposite sign. This means that the correction of the flux through $ABC$ due to time derivative will be zero. In non-linear case, that correction will, of course, pick non-zero terms from second derivatives, but this would be of higher order than what we are interested in (after all, second derivatives will produce a multiple of $0 (\tau^2)$ which would be multiplied by $0 (\tau^2)$ coming from area thus leading to $0 (\tau^4)$ which, in combination with $\theta$-dependence would give $\theta \tau^4$ instead of $\theta \tau^3$ we are looking for). We can, similarly, split the triangle $ADC$ into $AOD$ and $ODC$ and argue in the similar way as above that $\partial/ \partial t$ does not lead to any flux through $ADC$ either, up to the order of magnitude we are interested in. To sum it up, we are making the following assumptions:

a) Equations \ref{ElectricAC} and \ref{ElectricSurface} hold \emph{exactly}

b) Equation \ref{ElectricBD} holds up to approximations given in the equation

c) All fields are linear

d) All fields are time-independent

We are now ready to perform the explicit calculation. In light of symmetry, we will restrict our calculation to triangle $ABC$, and the result will be easily generalizable to ADC and then the two answers will be added. The only non-trivial issue we will encounter is the presence of $\tau^3$ term in the case of triangle and its absence in case of rectangle. This is due to the fact that the center of triangle is displaced by $0 (\tau)$ from the origin leading to $0 (\tau^2) \times 0 (\tau) = 0 (\tau^3)$ whereas the center of rectangle is displaced by $0 (\tau^2)$ from the origin leading to $0 (\tau^2) \times 0 (\tau^2) = 0 (\tau^4)$ . As one would expect, these $0 (\tau^3)$ terms will cancel upon addition. 

Let us consider spherical coordinates. We will denote the angle between $OB$ and $z$-axis by $\theta$ and we will denote the angle between $0B$ and $xz$-plane by $\phi$. We will parametrize the $OB$ line by $\lambda$ which will go from $0$ to $\tau$. Thus, 
\beq x = \lambda \sin \theta \cos \phi \; , \; y = \lambda \sin \theta \sin \phi \; , \; z = \lambda \cos \theta \eeq
As we said previously, $\theta$ is assumed to be "small". At the same time, no such assumption is made regarding $\phi$. In fact, we treat $x$ and $y$ on equal footing and, therefore, $\phi$ can be anything between $0$ and $2 \pi$:
\beq \theta = 0 (\tau) \; , \; \phi \in [0, 2 \pi) \label{PhiLargerThanTheta} \eeq
The height of each "strip" is $2 (\tau - \lambda)$, where the factor of $2$ comes from the fact that half of the strip has positive $t$-coordinate and the other half has negative. Therefore, the area of the strip is $2 d \lambda (\tau - \lambda) $. Since we assume $t$-independence of fields, the flux through that strip is 
\beq d \Phi_{\rm ABC} = 2 d \lambda (\tau - \lambda) ( E_x \sin \theta \cos \phi + E_y \sin \theta \sin \phi + E_z \cos \theta) \label{ElectricFluxElement} \eeq
Furthermore, by using linearity and time-independance, we know that $E_z$, $E_x$ and $E_y$ are given by
\beq E_x = \frac{\partial E_x}{\partial z} \Big\vert_0 \lambda \cos \theta + \frac{\partial E_x}{\partial x} \Big\vert_0 \lambda \sin \theta \cos \phi + \frac{\partial E_x}{\partial y} \Big\vert_0 \lambda \sin \theta \sin \phi \nonumber \eeq
\beq E_y = \frac{\partial E_y}{\partial z} \Big\vert_0 \lambda \cos \theta + \frac{\partial E_y}{\partial x} \Big\vert_0 \lambda \sin \theta \cos \phi + \frac{\partial E_y}{\partial y} \Big\vert_0 \lambda \sin \theta \sin \phi \label{ElectricLinear} \eeq
\beq E_z = E_z (0) + \frac{\partial E_z}{\partial z} \Big\vert_0 \lambda \cos \theta + \frac{\partial E_z}{\partial x} \Big\vert_0 \lambda \sin \theta \cos \phi + \frac{\partial E_z}{\partial y} \Big\vert_0 \lambda \sin \theta \sin \phi \nonumber \eeq
By substituting Equation \ref{ElectricLinear} into Equation \ref{ElectricFluxElement} and taking the integral we obtain
\beq \Phi_{\rm ABC} = \int_0^{\tau}  d \lambda \; 2 (\tau - \lambda) \Big[ \Big( \frac{\partial E_x}{\partial z} \Big\vert_0 \lambda \cos \theta + \frac{\partial E_x}{\partial x} \Big\vert_0 \lambda \sin \theta \cos \phi + \frac{\partial E_x}{\partial y} \Big\vert_0 \lambda \sin \theta \sin \phi \Big) \sin \theta \cos \phi + \nonumber \eeq
\beq + \Big(\frac{\partial E_y}{\partial z} \Big\vert_0 \lambda \cos \theta + \frac{\partial E_y}{\partial x} \Big\vert_0 \lambda \sin \theta \cos \phi + \frac{\partial E_y}{\partial y} \Big\vert_0 \lambda \sin \theta \sin \phi  \Big) \sin \theta \sin \phi + \label{ElectricIntegral} \eeq
\beq + \Big( E_z (0) + \frac{\partial E_z}{\partial z} \Big\vert_0 \lambda \cos \theta + \frac{\partial E_z}{\partial x} \Big\vert_0 \lambda \sin \theta \cos \phi + \frac{\partial E_z}{\partial y} \Big\vert_0 \lambda \sin \theta \sin \phi \Big) \cos \theta \Big] \nonumber \eeq
Now, in light of the fact that integration goes from $0$ to $\tau$, it automatically adds extra $0 (\tau)$ to everything. Furhtermore, the factor $2 (\tau - \lambda)$ ands another $0 (\tau)$.  Apart from that, $E_x$ and $E_y$ terms are multiplied by $\sin \theta$ which is of $0 (\tau)$. Finally, we also notice that all of the terms, except for $E_z (0) \cos \theta$, come with extra factor $\lambda$ which is also of $0 (\tau)$. This means that in order to do calculation up to $0 (\tau^4)$ we have to compute $E_x$ and $E_y$ terms up to finite part (and replace the overall coefficients of $\sin \theta$ with $\theta$), and compute $E_z$-terms up to $0 (\tau)$, except for $E_z (0) \cos \theta$ term which will be computed up to $0 (\tau^2)$. This means that the integral simplifies to 
\beq \Phi_{\rm ABC} = \int_0^{\tau}  d \lambda \; 2 (\tau - \lambda) \Big( \frac{\partial E_x}{\partial z} \Big\vert_0 \lambda \theta \cos \phi + \frac{\partial E_y}{\partial z} \Big\vert_0 \lambda  \theta \sin \phi + \label{ElectricIntegral2} \eeq
\beq + E_z (0) \Big(1 - \frac{\theta^2}{2} \Big) + \frac{\partial E_z}{\partial z} \Big\vert_0 \lambda  + \frac{\partial E_z}{\partial x} \Big\vert_0 \lambda \theta \cos \phi + \frac{\partial E_z}{\partial y} \Big\vert_0 \lambda \theta \sin \phi  \Big) \nonumber \eeq
By evaluating the above integral and substituting the limits of integration $\lambda =0$ and $\lambda= \tau$ we obtain
\beq \Phi_{\rm ABC} = \tau^2 E_z (0) + \frac{\tau^3}{3} \frac{\partial E_z}{\partial z} \Big\vert_0 - \frac{\tau^2 \theta^2}{2} E_z (0) + \label{TwoVariableFlux} \eeq
\beq + \frac{\tau^3 \theta}{3} \cos \phi \Big( \frac{\partial E_z}{\partial x} \Big\vert_0 + \frac{\partial E_x}{\partial z} \Big\vert_0 \Big) + \frac{\tau^3 \theta}{3} \sin \phi \Big( \frac{\partial E_z}{\partial y} \Big\vert_0 + \frac{\partial E_y}{\partial z} \Big\vert_0 \Big) \nonumber \eeq 
The reason we have the $\tau^3$ term in the above expression despite our earlier remark that such terms drop out is that we are integrating over the triangle $ABC$ as opposed to the contour $ABCD$. When we have stated that variation of flux is of $0 (\tau^4)$ we were referring to the fact that the "displacement" of points is of $0 (\tau^2)$, leading to $0 (\tau^2) \times 0 (\tau^2) = 0 (\tau^2)$. However, in case of a triangle, the displacement of the center of triangle away from the center of Alexandrov set is of $0 (\tau)$ rather than $0 (\tau^2)$. Thus, our assumption was strictly referring to square loop and \emph{not} to the triangle. In case of square loop, on the other hand, the $\tau^3$ contribution will drop out, as expected. After all, $\partial E_z / \partial z$ will give \emph{positive} contribution to flux over triangle $ABC$ and \emph{negative} contribution over $ADC$, leading to cancellation. By inspection it is easy to see that $\partial E_z/ \partial z$ is the only source of $\tau^3$ which implies that the flux over $ABCD$ has no $0 (\tau^3)$ terms. We will first perform the calculation over $ABC$, thus carrying $\tau^3$ term as we go along. Then, after we are finished, we will "copy" our answer into the expression for flux over $ADC$ while changing the sign of $\tau^3$ term. After that, we will add the two answers and cancel $\tau^3$ (see Equation \ref{CancelCubic}). We will denote the fluxes through respective contours by $\phi_{ABC}$, $\phi_{ADC}$ and $\phi_{ABCD}$. 

Let us now maximize the flux $\phi_{\rm ABC}$ given in Equation \ref{TwoVariableFlux}. We will first assume fixed $\phi$ and maximize the above with respect to $\theta$. This will imply that $\theta$ is some function of $\phi$. By replacing the former with the latter we will re-express flux as a function of $\phi$ alone. \emph{After that} we will do maximization with respect to $\phi$ which will produce quantity independent of both of the angles. As far as $\theta$-maximization is concerned, we have
\beq 0 = \frac{\partial \Phi}{\partial \theta} = - \theta \tau^2 E_z (0) + \frac{\tau^3 }{3} \cos \phi \Big( \frac{\partial E_z}{\partial x} \Big\vert_0 + \frac{\partial E_x}{\partial z} \Big\vert_0 \Big) + \frac{\tau^3 }{3} \sin \phi \Big( \frac{\partial E_z}{\partial y} \Big\vert_0 + \frac{\partial E_y}{\partial z} \Big\vert_0 \Big) \eeq 
This implies that, at any given point that is \emph{potentially} a maximum, 
\beq \theta = \frac{\tau \cos \phi}{3 E_z (0)} \Big( \frac{\partial E_z}{\partial x} \Big\vert_0 + \frac{\partial E_x}{\partial z} \Big\vert_0 \Big) +  \frac{\tau \sin \phi}{3 E_z (0)} \Big( \frac{\partial E_z}{\partial y} \Big\vert_0 + \frac{\partial E_y}{\partial z} \Big\vert_0 \Big) \label{ThetaOfPhi} \eeq
Upon substituting Equation \ref{ThetaOfPhi} into Equation \ref{TwoVariableFlux} we obtain
\beq \Phi_{\rm ABC} = \tau^2 E_z (0) + \frac{\tau^3}{3} \frac{\partial E_z}{\partial z} \Big\vert_0 + \label{OneVariablePhi}\eeq
\beq + \frac{\tau^4}{18 E_z (0)} \Big( \Big( \frac{\partial E_z}{\partial x} \Big\vert_0 + \frac{\partial E_x}{\partial z} \Big\vert_0 \Big) \cos \phi + \Big( \frac{\partial E_z}{\partial y} \Big\vert_0 + \frac{\partial E_y}{\partial z} \Big\vert_0 \Big) \sin \phi \Big)^2  \nonumber \eeq
We would now like to maximize the Equation \ref{OneVariablePhi} with respect to $\phi$. In order to do that, we have to maximize $a \cos \phi + b \sin \phi$, where 
\beq a = \frac{\partial E_z}{\partial x} \Big\vert_0 + \frac{\partial E_x}{\partial z} \Big\vert_0 \; , \; b = \frac{\partial E_z}{\partial y} \Big\vert_0 + \frac{\partial E_y}{\partial z} \Big\vert_0 \label{abElectric} \eeq
In order to save ourselves a little bit of time, we can use geometry to maximize $a \cos \phi + b \sin \phi$. Consider a triangle $PQR$ that is rotated by the angle $\phi$ with respect to horizontal line, and suppose that point $P$ lies on that line. Furthermore, suppose that the segment $PQ$ is orthogonal to the segment $QR$. Finally let $M$ and $N$ be the respective projections of points $Q$ and $R$ onto the said horizontal line. From simple geometry, it is easy to see that 
\beq PM = a \cos \phi \; , \; MN = b \sin \phi \eeq
and, therefore, 
\beq PN = a \cos \phi + b \sin \phi \eeq
On the other hand, the orthogonality of $PQ$ and $QR$ implies that 
\beq PR = \sqrt{a^2 + b^2} \eeq
Therefore, by noticing that 
\beq PN = PR \cos NPR \eeq
we obtain 
\beq a \cos \phi + b \sin \phi = \sqrt{a^2 + b^2} \cos NPR \eeq
This implies that 
\beq \max (a \cos \phi + b \sin \phi) = \sqrt{a^2 + b^2} \label{MaxPythagorean} \eeq
By substituting Equations \ref{abElectric} and \ref{MaxPythagorean} into Equation \ref{OneVariablePhi}, we obtain
\beq \Phi_{\rm ABC}  = \tau^2 E_z (0) + \frac{\tau^3}{3} \frac{\partial E_z}{\partial z} \Big\vert_0 + \frac{\tau^4}{18 E_z (0)} \Big( \Big(\frac{\partial E_z}{\partial x} \Big\vert_0 + \frac{\partial E_x}{\partial z} \Big\vert_0 \Big)^2 + \Big( \frac{\partial E_z}{\partial y} \Big\vert_0 + \frac{\partial E_y}{\partial z} \Big\vert_0 \Big)^2 \Big) \label{PhiABC} \eeq
It is easy to see from symmetry argument that $\phi_{\rm ABD}$ takes the same form as $\phi_{\rm ABC}$ except that $\partial E_z / \partial z$ term has the opposite sign:
\beq \Phi_{\rm ADC}  = \tau^2 E_z (0) - \frac{\tau^3}{3} \frac{\partial E_z}{\partial z} \Big\vert_0 + \frac{\tau^4}{18 E_z (0)} \Big( \Big(\frac{\partial E_z}{\partial x} \Big\vert_0 + \frac{\partial E_x}{\partial z} \Big\vert_0 \Big)^2 + \Big( \frac{\partial E_z}{\partial y} \Big\vert_0 + \frac{\partial E_y}{\partial z} \Big\vert_0 \Big)^2 \Big) \label{PhiADC} \eeq
Finally, by adding Equation \ref{PhiABC} to Equation \ref{PhiADC} we obtain the flux through the loop $ABCD$:
\beq \Phi_{\rm ABCD}  = 2 \tau^2 E_z (0) + \frac{\tau^4}{9 E_z (0)} \Big( \Big(\frac{\partial E_z}{\partial x} \Big\vert_0 + \frac{\partial E_x}{\partial z} \Big\vert_0 \Big)^2 + \Big( \frac{\partial E_z}{\partial y} \Big\vert_0 + \frac{\partial E_y}{\partial z} \Big\vert_0 \Big)^2 \Big) \label{CancelCubic} \eeq
The reason the Equation \ref{CancelCubic} lacks $\tau^3$ term is that this term comes with opposite sign in Equation \ref{PhiABC} and Equation \ref{PhiADC}, leading to cancellation. This is, in fact, what we expect: after all, back in the qualitative discussion we have argued that the flux through the rectangle is of $0 (\tau^2) + 0 (\tau^4)$ and, therefore, lacks $0 (\tau^3)$ term. In our argument we have assumed that the displacement is of $0 (\tau^2)$. This assumption is not correct for the case of rectangle: after all, the center of triangle is displaced from the origin by $0 (\tau)$. Thus, our argument regarding the "absence of $0 (\tau^3)$ fails in the case of triangle but stands in the case of rectangle. This implies that the $0 (\tau^3)$ terms coming from the two triangles cancel out. This, in fact, is what we have just seen. 

Now, just like in the the magnetic case, our goal is to arive at covariant expression. We will use the same trick as before. First, we will restore spacelike rotational covariance by replacing $\hat{z}$ with $\vec{E}/ \vert \vec{E} \vert$. After that, we will restore Lorentz covariance by replacing $\delta^{\mu}_0$ with $v^{\mu}$ and $E^k$ with $v^{\mu} F_{\mu \nu}$. Let us start with space covariance. For our convenience, let us rewrite  Equation \ref{CancelCubic} as 
\beq \Phi_{\rm ABCD}  = 2 \tau^2 E_z (0) + \frac{\tau^4}{9 E_z (0)} \Big( \Big(\frac{\partial E_z}{\partial x} \Big\vert_0 \Big)^2 + \Big( \frac{\partial E_z}{\partial y} \Big\vert_0 \Big)^2 + \label{NonCovariantFlux} \eeq
\beq + \Big( \frac{\partial E_x}{\partial z} \Big\vert_0 \Big)^2 + \Big(\frac{\partial E_y}{\partial z} \Big\vert_0 \Big)^2 +  2 \frac{\partial E_z}{\partial x} \Big\vert_0 \frac{\partial E_x}{\partial z} \Big\vert_0 +  2 \frac{\partial E_z}{\partial y} \Big\vert_0 \frac{\partial E_y}{\partial z} \Big\vert_0 \Big) \nonumber \eeq
Let us first find the rotationally-symmetric expression for $(\partial_x E_z)^2 + (\partial_y E_z)^2$. We can first add and subtract $(\partial_z E_z)^2$:
\beq (\partial_x E_z)^2 + (\partial_y E_z)^2 = (\partial_x E_z)^2 + (\partial_y E_z)^2 + (\partial_z E_z)^2 - (\partial_z E_z)^2 \eeq
which we can further rewrite as 
\beq (\partial_x E_z)^2 + (\partial_y E_z)^2 = \partial_k E_z \partial_k E_z - (\partial_z E_z)^2 \label{Electricxy1} \eeq
Now, since we have assumed that $E$ is parallel to $z$-axis,
\beq E^k = E \delta^k_3 \eeq
we obtain
\beq \partial_k E_z \partial_k E_z = \frac{E^i E^j \partial_k E^i \partial_k E^j}{E^l E^l} \; ; \; \partial_z E_z = \frac{E^i E^j \partial_i E^j}{E^k E^k} \label{Electricxy2}\eeq
By substitution of Equation \ref{Electricxy2} into the Equation \ref{Electricxy1}, we obtain
\beq (\partial_x E_z)^2 + (\partial_y E_z)^2 = \frac{E^i E^j \partial_i E^k \partial_j E^k}{E^l E^l} - \Big(\frac{E^i E^j \partial_i E^j}{E^k E^k}\Big)^2 \label{Electricxy3} \eeq
Of course, the above expression is only rotationally covariant, but it is \emph{not} Lorentz covariant (as evident by the presence of spacelike indexes and absence of timelike ones).  We will first deal with spacelike rotational covariance of the rest of the terms and we will come back and make things Lorentz covariant after we are done. Let us, therefore, move to the rotational covariance of $(\partial_z E_x)^2 + (\partial_z E_y)^2$. Similarly to what we have done before, we will add and subtract $(\partial_z E_z)^2$ term: 
\beq (\partial_z E_x)^2 + (\partial_z E_y)^2 = (\partial_z E_x)^2 + (\partial_z E_y)^2 + (\partial_z E_z)^2 - (\partial_z E_z)^2 \eeq
We can now rewrite it as 
\beq (\partial_z E_x)^2 + (\partial_z E_y)^2 = \partial_3 E^k \partial_3 E^k  - (\partial_z E_z)^2 \label{Electricxy4} \eeq
After that, we can again use 
\beq E^k = E \delta_3^k \eeq
to rewrite the above two terms as 
\beq \partial_3 E^k \partial_3 E^k = \frac{E^i E^j \partial_i E^k \partial_j E^k}{E^l E^l} \; ; \; \partial_z E_z = \frac{E^i E^j \partial_i E^j}{E^k E^k} \label{Electricxy5}\eeq
Finally, by substituting Equation \ref{Electricxy5} into Equation \ref{Electricxy4}, we obtain
\beq (\partial_z E_x)^2 + (\partial_z E_y)^2 = \frac{E^i E^j \partial_i E^k \partial_j E^k}{E^l E^l} - \Big(\frac{E^i E^j \partial_i E^j}{E^k E^k} \Big)^2 \label{Electricxy6} \eeq
Finally, let us evaluate $\partial_x E_z \partial_z E_x + \partial_y E_z \partial_z E_y$. Again, we add and subtract $(\partial_z E_z)^2$. But, for our convenience, we will express $+ (\partial_z E_z)^2$ as $\partial_z E_z \partial_z E_z$, while leaving $- (\partial_z E_z)^2$ in the original form:
\beq \partial_x E_z \partial_z E_x + \partial_y E_z \partial_z E_y = \partial_x E_z \partial_z E_x + \partial_y E_z \partial_z E_y +\partial_z E_z \partial_z E_z - (\partial_z E_z)^2 \eeq
We can now rewrite it as 
\beq \partial_x E_z \partial_z E_x + \partial_y E_z \partial_z E_y = \partial_k E^3 \partial_3 E^k  - (\partial_z E_z)^2 \label{Electricxy7}\eeq
We can now use 
\beq E^k = E \delta^k_3 \eeq
in order to rewrite the two terms on the right hand side as 
\beq \partial_k E^3 \partial_3 E^k = \frac{E^i E^j \partial_k E^i \partial_j E^k}{E^l E^l} \; , \; \partial_z E_z = \frac{E^i E^j \partial_i E^j}{E^k E^k} \label{Electricxy8}\eeq
By substituting Equation \ref{Electricxy8} into Equation \ref{Electricxy7}, we obtain 
\beq \partial_x E_z \partial_z E_x + \partial_y E_z \partial_z E_y = \frac{E^i E^j \partial_k E^i \partial_j E^k}{E^l E^l}  - \Big(\frac{E^i E^j \partial_i E^j}{E^k E^k} \Big)^2 \label{Electricxy9}\eeq
We can now substitute Equations \ref{Electricxy3}, \ref{Electricxy6} and \ref{Electricxy9} into equation \ref{NonCovariantFlux} to obtain
\beq \Phi_{\rm ABCD}  = 2 \tau^2 \sqrt{E^k E^k} + \frac{\tau^4}{9 \sqrt{E^k E^k}} \Big( \frac{E^i E^j \partial_i E^k \partial_j E^k}{E^l E^l} - \Big(\frac{E^i E^j \partial_i E^j}{E^k E^k}\Big)^2 + \label{Electricxyz1} \eeq
\beq + \frac{E^i E^j \partial_i E^k \partial_j E^k}{E^l E^l} - \Big(\frac{E^i E^j \partial_i E^j}{E^k E^k} \Big)^2  +  \frac{2 E^i E^j \partial_k E^i \partial_j E^k}{E^l E^l}  - 2 \Big(\frac{E^i E^j \partial_i E^j}{E^k E^k} \Big)^2 \Big) \nonumber \eeq
Now, the Lagrangian generator  is given by 
\beq {\cal J}_E = \Phi_{\rm ABCD}^2 \label{DefinitionElectricFlux}\eeq
By substituting Equation \ref{Electricxyz1} into Equation \ref{DefinitionElectricFlux} and evaluating it up to $0(\tau^6)$, we obtain
\beq {\cal J}_E = 4 \tau^4 E^k E^k + \frac{4 \tau^6}{9} \Big( \frac{E^i E^j \partial_i E^k \partial_j E^k}{E^l E^l} - \Big(\frac{E^i E^j \partial_i E^j}{E^k E^k}\Big)^2 + \label{Electricxyz2} \eeq
\beq + \frac{E^i E^j \partial_i E^k \partial_j E^k}{E^l E^l} - \Big(\frac{E^i E^j \partial_i E^j}{E^k E^k} \Big)^2  +  \frac{2 E^i E^j \partial_k E^i \partial_j E^k}{E^l E^l}  - 2 \Big(\frac{E^i E^j \partial_i E^j}{E^k E^k} \Big)^2 \Big)^2 \nonumber \eeq
Finally, let us rewrite the above in covariant form. Like we were saying earlier, $t$-axis is parallel to a "vector field" $v^{\mu}$ which is associated with the direction of "smallest variation" of $A^{\mu}$. Thus, 
\beq v^{\mu} = \delta^{\mu}_0 \eeq
We can now rewrite $E^k$ as $F_{k0}$ and subsequently rewrite the latter as $v^{\mu} F_{\mu \nu}$ where "Latin" index $k$ now turns into "Greek" index $\nu$:
\beq E^k = F_{k0} \longrightarrow v^{\mu} F_{\mu \nu} \eeq
We will, however, leave $\vert \vec{E} \vert^2$ in that seemingly non-covariant form, mainly because we already have an expression for $\vec{E}$,  
\beq \vert \vec{E} \vert^2 = \frac{1}{2} \Bigg(F^{\alpha \beta} F_{\alpha \beta} - (sgn(F^{\mu \nu}F_{\mu \nu})) \sqrt{(F^{\alpha \beta} F_{\alpha \beta})^2 - \frac{1}{16} (\epsilon_{\alpha \beta \gamma \delta} F^{\alpha \beta} F^{\gamma \delta})^2} \Bigg) \eeq
which we "like" better since it has no reference to $v^{\mu}$. Thus, we can rewrite Equation \ref{Electricxyz2} as 
\beq {\cal J}_E = 4 \tau^4 \vert \vec{E} \vert^2 + \frac{4 \tau^6}{9} \Big( \frac{v_{\alpha} v^{\beta} v_{\rho} v_{\sigma} F^{\rho \mu} F^{\sigma \nu} \partial_{\mu} F^{\alpha \gamma} \partial_{\nu} E_{\beta \gamma}}{\vert \vec{E} \vert^2} - \nonumber \eeq
\beq - \Big(\frac{v_{\rho} v_{\sigma} v_{\eta} F^{\rho \mu} F^{\sigma \nu} \partial_{\mu} F^{\eta \nu} }{\vert \vec{E} \vert^2}\Big)^2 + \frac{v_{\alpha} v^{\beta} v_{\rho} v_{\sigma} F^{\rho \mu} F^{\sigma \nu}\partial_{\mu} F^{\alpha \eta} \partial_{\nu} F_{\beta \eta}}{\vert \vec{E} \vert^2} \label{ElectricCovariant} \eeq
\beq  - \Big(\frac{v_{\rho} v_{\sigma} v^{\eta} F^{\rho \mu} F^{\sigma \nu} \partial_{\mu} F_{\nu \eta}}{\vert \vec{E} \vert^2} \Big)^2  +  \frac{2 v_{\rho} v_{\sigma} F^{\rho \mu} F^{\sigma \nu}  \partial_{\eta} E^{\mu} \partial_{\nu} E^{\eta}}{\vert \vec{E} \vert^2}  - 2 \Big(\frac{v_{\rho} v_{\sigma} v_{\eta} F^{\mu \rho} F^{\nu \sigma} \partial_{\mu} F^{\nu \eta}}{\vert \vec{E} \vert^2} \Big)^2 \Big)^2 \nonumber \eeq
Now we have to ask ourselves whether or not we have to rotate the axis of Alexandrov set in order to minimize $\cal J$. The $0 (\tau^4)$ term is already minimized with the Alexandrov set chosen the way it was. This means that the $0 (\tau^4)$ term behaves like $\theta^2$, where $\theta$ is the angle by which we tilt. On the other hand, the $0 (\tau^6)$ term has completely different structure from $0 (\tau^4)$ one. So we have no reason to believe its derivative with respect to $\theta$ is small. Thus, we would expect it to have large linear term in $\theta$. Therefore, $\cal J$ evolves as 
\beq {\cal J}_E = {\cal J}_{E0} + a \theta^2 \tau^4 + b \theta \tau^6 \eeq
This implies that it reaches the minimum at 
\beq 0 = \frac{d {\cal J}}{d \theta} = 2a \theta \tau^4  + b \tau^6 \eeq
Therefore
\beq \theta = - \frac{b \tau^2}{2a} \eeq
The $\cal L$ will be equal to $\cal J$ \emph{at} that specific $\theta$. In other words, 
\beq {\cal L}_E = {\cal J}_0 + a \Big(- \frac{b \tau^2}{2a}\Big)^2 \tau^4 + b \Big(- \frac{b \tau^2}{2a}\Big) \tau^6 = {\cal J}_0 - \frac{b^2 \tau^8 }{4a} \eeq
Thus, $\cal L$ deviates from ${\cal J}_0$ by $0 (\tau^8)$ term. Since our calculation is up to $0 (\tau^6)$, we can ignore that term. Thus, we can blindly identify $\cal L$ with the value of $\cal J$ for the Alexandrov set that has \emph{not} beeen rotated. Thus, our final answer is
\beq {\cal L}_E = 4 \tau^4 \vert \vec{E} \vert^2 + \frac{4 \tau^6}{9} \Big( \frac{v_{\alpha} v^{\beta} v_{\rho} v_{\sigma} F^{\rho \mu} F^{\sigma \nu} \partial_{\mu} F^{\alpha \gamma} \partial_{\nu} E_{\beta \gamma}}{\vert \vec{E} \vert^2} - \nonumber \eeq
\beq - \Big(\frac{v_{\rho} v_{\sigma} v_{\eta} F^{\rho \mu} F^{\sigma \nu} \partial_{\mu} F^{\eta \nu} }{\vert \vec{E} \vert^2}\Big)^2 + \frac{v_{\alpha} v^{\beta} v_{\rho} v_{\sigma} F^{\rho \mu} F^{\sigma \nu}\partial_{\mu} F^{\alpha \eta} \partial_{\nu} F_{\beta \eta}}{\vert \vec{E} \vert^2} \label{ElectricCovariant2} \eeq
\beq  - \Big(\frac{v_{\rho} v_{\sigma} v^{\eta} F^{\rho \mu} F^{\sigma \nu} \partial_{\mu} F_{\nu \eta}}{\vert \vec{E} \vert^2} \Big)^2  +  \frac{2 v_{\rho} v_{\sigma} F^{\rho \mu} F^{\sigma \nu}  \partial_{\eta} E^{\mu} \partial_{\nu} E^{\eta}}{\vert \vec{E} \vert^2}  - 2 \Big(\frac{v_{\rho} v_{\sigma} v_{\eta} F^{\mu \rho} F^{\nu \sigma} \partial_{\mu} F^{\nu \eta}}{\vert \vec{E} \vert^2} \Big)^2 \Big)^2 \nonumber \eeq
where $\vert \vec{E} \vert^2$ is given by Equation \ref{EisCovariant},
\beq \vert \vec{E} \vert^2 = \frac{1}{2} \Bigg(F^{\alpha \beta} F_{\alpha \beta} - (sgn(F^{\mu \nu}F_{\mu \nu})) \sqrt{(F^{\alpha \beta} F_{\alpha \beta})^2 - \frac{1}{16} (\epsilon_{\alpha \beta \gamma \delta} F^{\alpha \beta} F^{\gamma \delta})^2} \Bigg) \eeq

\subsection*{8 Corrections to total electromagnetic Lagrangian}

Let us summarize what we have found so far. In Section 5 we have found the the first order expressions for "magnetic" and "electric" Lagrangians. These were defined by merely squaring $\vert \vec{B} \vert$ and $\vert \vec{E} \vert$ respectively. The latter are given by "covariant" expressions, 
\beq \vert \vec{B} \vert = \frac{1}{2} \Bigg(\sqrt{F^{\alpha \beta} F_{\alpha \beta} + \frac{1}{4} \epsilon_{\alpha \beta \gamma \delta} F^{\alpha \beta} F^{\gamma \delta}} + (sgn(F^{\mu \nu}F_{\mu \nu})) \sqrt{F^{\alpha \beta} F_{\alpha \beta} - \frac{1}{4} \epsilon_{\alpha \beta \gamma \delta} F^{\alpha \beta} F^{\gamma \delta}}  \Bigg) \eeq
\beq \vert \vec{E} \vert = \frac{1}{2} \Bigg(\sqrt{F^{\alpha \beta} F_{\alpha \beta} + \frac{1}{4} \epsilon_{\alpha \beta \gamma \delta} F^{\alpha \beta} F^{\gamma \delta}} -(sgn(F^{\mu \nu}F_{\mu \nu})) \sqrt{F^{\alpha \beta} F_{\alpha \beta} - \frac{1}{4} \epsilon_{\alpha \beta \gamma \delta} F^{\alpha \beta} F^{\gamma \delta}}  \Bigg) \eeq
Then, in Section 6, we have found a correction to "magnetic" Lagrangian. The "corrected" expression is given by
\beq {\cal L}_B = 4 \vert \vec{B} \vert^2 \tau^4 + \frac{\tau^6}{9 \vert \vec{B} \vert^2} (v_B^{\mu} \partial_{\mu} \vert \vec{B} \vert^2)^2 - \frac{2 \tau^6}{9 \vert \vec{B} \vert^4} (\epsilon_{\alpha \beta \gamma \delta} v_B^{\alpha} F^{\beta \gamma} \partial^{\delta} \vert \vec{B} \vert^2)^2 - \nonumber \eeq
\beq - \frac{\tau^6}{9 \vert \vec{B} \vert^2} \partial^{\mu} \vert \vec{B} \vert^2 \partial_{\mu} \vert \vec{B} \vert^2 + \frac{8 \tau^6}{3 \vert \vec{B} \vert} v_{\alpha} v_B^{\beta} \partial_{\gamma} \partial^{\delta} F_{\rho}^{\; \alpha} \partial_{\beta} \partial^{\gamma} F_{\delta}^{\; \rho}  - \frac{8 \tau^6}{3 \vert \vec{B} \vert} v_{\alpha} v_B^{\beta} F_{\rho}^{\; \alpha} \partial_{\beta} \partial^{\gamma} F_{\gamma}^{\; \rho} -  \eeq
\beq - \frac{8 \tau^6}{3B} v_{\beta} v_B^{\gamma} F_{\rho}^{\; \alpha} \partial_{\alpha} \partial^{\beta} F_{\gamma}^{\; \rho} + \frac{8 \tau^6}{3 \vert \vec{B} \vert} F_{\rho}^{\; \alpha} \partial_{\alpha} \partial^{\beta} F_{\beta}^{\; \rho} - \frac{8 \tau^6}{3 \vert \vec{B} \vert} v_B^{\rho} \partial_{\sigma} F_{\rho}^{\; \mu} \partial_{\mu} \partial^{\nu} F_{\nu}^{\; \sigma} \nonumber \eeq
Finally, in Section 7, we have found a correction for "electric" Lagrangian, and the corrected expression was given by
\beq {\cal L}_E = 4 \tau^4 \vert \vec{E} \vert^2 + \frac{4 \tau^6}{9} \Big( \frac{v_{\alpha} v^{\beta} v_{\rho} v_{\sigma} F^{\rho \mu} F^{\sigma \nu} \partial_{\mu} F^{\alpha \gamma} \partial_{\nu} E_{\beta \gamma}}{\vert \vec{E} \vert^2} - \nonumber \eeq
\beq - \Big(\frac{v_{\rho} v_{\sigma} v_{\eta} F^{\rho \mu} F^{\sigma \nu} \partial_{\mu} F^{\eta \nu} }{\vert \vec{E} \vert^2}\Big)^2 + \frac{v_{\alpha} v^{\beta} v_{\rho} v_{\sigma} F^{\rho \mu} F^{\sigma \nu}\partial_{\mu} F^{\alpha \eta} \partial_{\nu} F_{\beta \eta}}{\vert \vec{E} \vert^2}\eeq
\beq  - \Big(\frac{v_{\rho} v_{\sigma} v^{\eta} F^{\rho \mu} F^{\sigma \nu} \partial_{\mu} F_{\nu \eta}}{\vert \vec{E} \vert^2} \Big)^2  +  \frac{2 v_{\rho} v_{\sigma} F^{\rho \mu} F^{\sigma \nu}  \partial_{\eta} E^{\mu} \partial_{\nu} E^{\eta}}{\vert \vec{E} \vert^2}  - 2 \Big(\frac{v_{\rho} v_{\sigma} v_{\eta} F^{\mu \rho} F^{\nu \sigma} \partial_{\mu} F^{\nu \eta}}{\vert \vec{E} \vert^2} \Big)^2 \Big)^2 \nonumber \eeq
Now, if we combine the two Lagrangians by using 
\beq {\cal L} = {\cal L}_B - {\cal L}_E \eeq
we obtain
\beq {\cal L} = 4 \tau^4 (\vert \vec{B} \vert^2  - \vert \vec{E} \vert^2) + \frac{\tau^6}{9 \vert \vec{B} \vert^2} (v_B^{\mu} \partial_{\mu} \vert \vec{B} \vert^2)^2 - \frac{2 \tau^6}{9 \vert \vec{B} \vert^4} (\epsilon_{\alpha \beta \gamma \delta} v_B^{\alpha} F^{\beta \gamma} \partial^{\delta} \vert \vec{B} \vert^2)^2 - \nonumber \eeq
\beq - \frac{\tau^6}{9 \vert \vec{B} \vert^2} \partial^{\mu} \vert \vec{B} \vert^2 \partial_{\mu} \vert \vec{B} \vert^2 + \frac{8 \tau^6}{3 \vert \vec{B} \vert} v_{\alpha} v_B^{\beta} \partial_{\gamma} \partial^{\delta} F_{\rho}^{\; \alpha} \partial_{\beta} \partial^{\gamma} F_{\delta}^{\; \rho}  - \frac{8 \tau^6}{3 \vert \vec{B} \vert} v_{\alpha} v_B^{\beta} F_{\rho}^{\; \alpha} \partial_{\beta} \partial^{\gamma} F_{\gamma}^{\; \rho} -  \eeq
\beq - \frac{8 \tau^6}{3B} v_{\beta} v_B^{\gamma} F_{\rho}^{\; \alpha} \partial_{\alpha} \partial^{\beta} F_{\gamma}^{\; \rho} + \frac{8 \tau^6}{3 \vert \vec{B} \vert} F_{\rho}^{\; \alpha} \partial_{\alpha} \partial^{\beta} F_{\beta}^{\; \rho} - \frac{8 \tau^6}{3 \vert \vec{B} \vert} v_B^{\rho} \partial_{\sigma} F_{\rho}^{\; \mu} \partial_{\mu} \partial^{\nu} F_{\nu}^{\; \sigma} \nonumber \eeq
  \beq - \frac{4 \tau^6}{9} \Big( \frac{v_{\alpha} v^{\beta} v_{\rho} v_{\sigma} F^{\rho \mu} F^{\sigma \nu} \partial_{\mu} F^{\alpha \gamma} \partial_{\nu} E_{\beta \gamma}}{\vert \vec{E} \vert^2} - \nonumber \eeq
\beq - \Big(\frac{v_{\rho} v_{\sigma} v_{\eta} F^{\rho \mu} F^{\sigma \nu} \partial_{\mu} F^{\eta \nu} }{\vert \vec{E} \vert^2}\Big)^2 + \frac{v_{\alpha} v^{\beta} v_{\rho} v_{\sigma} F^{\rho \mu} F^{\sigma \nu}\partial_{\mu} F^{\alpha \eta} \partial_{\nu} F_{\beta \eta}}{\vert \vec{E} \vert^2} \eeq
\beq  - \Big(\frac{v_{\rho} v_{\sigma} v^{\eta} F^{\rho \mu} F^{\sigma \nu} \partial_{\mu} F_{\nu \eta}}{\vert \vec{E} \vert^2} \Big)^2  +  \frac{2 v_{\rho} v_{\sigma} F^{\rho \mu} F^{\sigma \nu}  \partial_{\eta} E^{\mu} \partial_{\nu} E^{\eta}}{\vert \vec{E} \vert^2}  - 2 \Big(\frac{v_{\rho} v_{\sigma} v_{\eta} F^{\mu \rho} F^{\nu \sigma} \partial_{\mu} F^{\nu \eta}}{\vert \vec{E} \vert^2} \Big)^2 \Big)^2 \nonumber \eeq
From Equations \ref{BisCovariant} and \ref{EisCovariant} one can easilly see that $\epsilon_{\alpha \beta \gamma \delta} F^{\alpha \beta} F^{\gamma \delta}$ cancel out when one computes $B^2 - E^2$. Thus, as one would expect,
\beq B^2 - E^2 = F^{\mu \nu} F_{\mu \nu} \eeq
This allows us to rewrite the above expression as 
\beq {\cal L} = 4 \tau^4 F^{\mu \nu} F_{\mu \nu}  + \frac{\tau^6}{9 \vert \vec{B} \vert^2} (v_B^{\mu} \partial_{\mu} \vert \vec{B} \vert^2)^2 - \frac{2 \tau^6}{9 \vert \vec{B} \vert^4} (\epsilon_{\alpha \beta \gamma \delta} v_B^{\alpha} F^{\beta \gamma} \partial^{\delta} \vert \vec{B} \vert^2)^2 - \nonumber \eeq
\beq - \frac{\tau^6}{9 \vert \vec{B} \vert^2} \partial^{\mu} \vert \vec{B} \vert^2 \partial_{\mu} \vert \vec{B} \vert^2 + \frac{8 \tau^6}{3 \vert \vec{B} \vert} v_{\alpha} v_B^{\beta} \partial_{\gamma} \partial^{\delta} F_{\rho}^{\; \alpha} \partial_{\beta} \partial^{\gamma} F_{\delta}^{\; \rho}  - \frac{8 \tau^6}{3 \vert \vec{B} \vert} v_{\alpha} v_B^{\beta} F_{\rho}^{\; \alpha} \partial_{\beta} \partial^{\gamma} F_{\gamma}^{\; \rho} -  \eeq
\beq - \frac{8 \tau^6}{3B} v_{\beta} v_B^{\gamma} F_{\rho}^{\; \alpha} \partial_{\alpha} \partial^{\beta} F_{\gamma}^{\; \rho} + \frac{8 \tau^6}{3 \vert \vec{B} \vert} F_{\rho}^{\; \alpha} \partial_{\alpha} \partial^{\beta} F_{\beta}^{\; \rho} - \frac{8 \tau^6}{3 \vert \vec{B} \vert} v_B^{\rho} \partial_{\sigma} F_{\rho}^{\; \mu} \partial_{\mu} \partial^{\nu} F_{\nu}^{\; \sigma} \nonumber \eeq
  \beq - \frac{4 \tau^6}{9} \Big( \frac{v_{\alpha} v^{\beta} v_{\rho} v_{\sigma} F^{\rho \mu} F^{\sigma \nu} \partial_{\mu} F^{\alpha \gamma} \partial_{\nu} E_{\beta \gamma}}{\vert \vec{E} \vert^2} - \nonumber \eeq
\beq - \Big(\frac{v_{\rho} v_{\sigma} v_{\eta} F^{\rho \mu} F^{\sigma \nu} \partial_{\mu} F^{\eta \nu} }{\vert \vec{E} \vert^2}\Big)^2 + \frac{v_{\alpha} v^{\beta} v_{\rho} v_{\sigma} F^{\rho \mu} F^{\sigma \nu}\partial_{\mu} F^{\alpha \eta} \partial_{\nu} F_{\beta \eta}}{\vert \vec{E} \vert^2} \label{ElectricCovariant} \eeq
\beq  - \Big(\frac{v_{\rho} v_{\sigma} v^{\eta} F^{\rho \mu} F^{\sigma \nu} \partial_{\mu} F_{\nu \eta}}{\vert \vec{E} \vert^2} \Big)^2  +  \frac{2 v_{\rho} v_{\sigma} F^{\rho \mu} F^{\sigma \nu}  \partial_{\eta} E^{\mu} \partial_{\nu} E^{\eta}}{\vert \vec{E} \vert^2}  - 2 \Big(\frac{v_{\rho} v_{\sigma} v_{\eta} F^{\mu \rho} F^{\nu \sigma} \partial_{\mu} F^{\nu \eta}}{\vert \vec{E} \vert^2} \Big)^2 \Big)^2 \nonumber \eeq
Thus, the leading order term coincides with usual Lagrangian. But then highly non-trivial effects come up in higher order corrections.

\subsection*{9. Conclusion}

In this paper we have proposed a way of defining Lagrangian on a causal set. This required us to find a "geometric" way of defining Lagrangians so that we would no longer have to refer to coordinate system. Our intention was to design a geometric constructions in such a way that the resulting Lagrangians match their analytic definition in the coordinate case. However, in light of the discrete nature of causal set theory, these geometric constructions have finite size. As a result, they are bound to take contributions from higher order derivatives. Yet, analytic solutions as we know it are only functions of first derivatives of the fields. This means that the geometric definition of Lagrangian no longer matches the analytic one at higher order. Furthermore, it is quite clear that the specific mismatch between geometric and analytic solution will be different for different choices of "geometry". This allows us to vary non-linear effects at will by varying the geometric definition of Lagrangians. 

In many cases, non linearity is introduced under the motive of "grasping at the straws"  in attempts of explaining a phenomena that can't be explained linearly. One example that comes to mind is quantum measurement theory. But, of course, there might be other, more ''mundane'' examples where non-linear effects are hypothesized without direct observations. While none of these issues were explored in this paper, it is conceivable that this paper will be useful in reframing these situations some time down the road. In particular, one can attempt to consider that the hypothesized non-linearities came from the "geometric" introduction of Lagrangian as opposed to anything extra that had been "put by hand". In fact, one can even use 'scientific method" by looking at different ''geometrical constructions'', investigating predicted non-linearities of each one, and seeing which of the ''non-linearities'' match what we see in the lab. this paper, however, shows a natural way of introducing non-linearity. 

One very hypothetical example of this is the explanation of OPERA experiment. It is conceivable that non-linear interaction between neutrino and background fields would alter the speed of the former. While we have not attempted to tackle neutrino field, it was still demonstrated on the example of Klein Gordon and Maxwell fields, that non-linear effects can be vastly different in case of different Lagrangians. This implies that it is conceivable that one field (such as neutrino) is affected more than other fields. But, of course, we can't make this claim until we actually work out the neutrino directly. In fact, it should be emphasized that this paper made no attempt to study neutrino, nor did it present any argument that the ''non-linear effects'' would lead to simple deviation of speed of signal propagation; in fact, it seems more reasonable to believe that the non linear effects are a lot more complex.  Nevertheless, it might still be worth it to try to tackle neutrino with different geometric constructions before dismissing the possibility. 

Independently of points made above, I believe that the prediction of non-linear effects is important from the point of view of predictive power of causal set theory. Due to the fact that causal set does not have regular structure (such as lattice) it is very difficult to perform any kind of analytic calculations on a causal set, and so far most of the work has been numeric (with a notable exception involving an ''analytic'' calculation of cosmological constant, \cite{Cosmological}). This paper, on the other hand, provides an example of an \emph{analytic} prediction of causal set theory by proposing a set of differential equations that deviates from the one typically accepted. At the same time, however, there is a serious shortcoming. In particular, the deviations lead to highly non linear terms, and, therefore, none of the usual methods of evaluating path integral will work. This means that a serious investigation into alternative ways of evaluating path integrals is in order.


\begin{thebibliography}{77}

\bibitem{Johnston1} Johnston S.``Particle propagators on discrete spacetime'' {\sl  Class. Quantum Grav 4143-4149}  (2008)  and {\tt arXiv:0806.3083}.

\bibitem{Johnston2} Johnston S. ``The Feynman propagator for a Free Scalar Field on a Causal Set'' {\sl  Phys. Rev. Lett. 103, 180401}  (2009)  and {\tt arXiv:0909.0944 }.

\bibitem{SverdlovBombelli} R Sverdlov and L. Bombelli 2008 “Introduction of Bosonic Fields into Causal Set Theory”, Workshop on Continuum and Lattice Approaches to Quantum Gravity, University of Sussex, United Kingdom, PoS(CLAQG08)014

\bibitem{BombelliSverdlov} R. Sverdlov and L. Bombelli, Dynamics for causal sets with matter fields: A Lagrangian-based
approach, contribution to the Proceedings of the DICE2008 Conference, J. Phys.: Conf. Ser. 174
(2009) 012019 [arXiv:0905.1506].

\bibitem{Grassmann1} R.Sverdlov `` Novel definition of Grassmann numbers and spinor fields" (2008) {\tt arXiv:. arXiv:0808.0756}.

\bibitem{Grassmann2} R.Sverdlov ``An attempt to resolve apparent paradoxes in definitions of Grassmann numbers and spinor fields 
" (2009) {\tt arXiv:. arXiv:0908.2605}.

\bibitem{Gravity} R.Sverdlov ``Non-linear corrections to Lagrangians predicted by causal set theory: Effects of curvature" (in preparation)

\bibitem{Fermions} R.Sverdlov ``Non-linear corrections to Lagrangians predicted by causal set theory: Fermionic case" (in preparation)

\bibitem{Cosmological} M. Henneaux and C. Teitelboim, “The Cosmological Constant and General Covariance”,
Phys. Lett. B 222 : 195 (1989)

\end{thebibliography}
\end{document}